\documentclass[twocolumn]{aastex6}
\usepackage{graphicx}
\usepackage{amsmath}

\newcommand{\vsini}{v \sin{i}}
\newcommand{\dnu}{\Delta \nu}
\newcommand{\dnuo}{\Delta \nu_0}
\newcommand{\dnuscl}{\Delta \nu_{\rm scl}}
\newcommand{\numax}{\nu_{\rm max}}
\newcommand{\msun}{{\rm M_\odot}}
\newcommand{\rsun}{{\rm R_\odot}}
\newcommand{\feh}{{\rm [Fe/H]}}
\newcommand{\mh}{{\rm [M/H]}}
\newcommand{\mrho}{\langle \rho \rangle}
\newcommand{\mrhosun}{\langle \rho_\odot\rangle}
\newcommand{\teff}{T_{\rm eff}}
\newcommand{\logg}{\log{g}}
\newcommand{\tsun}{\rm T_{eff,\odot}}
\newcommand{\gsun}{\rm g_\odot}
\newcommand{\numaxsun}{\nu_{\rm max,\odot}}
\newcommand{\dnusun}{\dnu_\odot}
\newcommand{\yr}{Y$_{\rm R}$}

\newcommand{\gbmr}{\texttt{BeS/G-$\dnuo$}}

\begin{document}
\title{The First APOKASC Catalog of Kepler Dwarf and Subgiant Stars}
\author{Aldo Serenelli\altaffilmark{1}}
\author{Jennifer Johnson\altaffilmark{2}}
\author{Daniel Huber\altaffilmark{3,4,5,6}}
\author{Marc Pinsonneault\altaffilmark{2}}
\author{Warrick H. Ball\altaffilmark{7,8,9,6}}
\author{Jamie Tayar\altaffilmark{2}}
\author{Victor Silva Aguirre\altaffilmark{6}}
\author{Sarbani Basu\altaffilmark{10}}
\author{Nicholas Troup\altaffilmark{11}}
\author{Saskia Hekker\altaffilmark{8,6}}
\author{Thomas Kallinger\altaffilmark{12}}
\author{Dennis Stello\altaffilmark{13,4,6}}
\author{Guy R. Davies\altaffilmark{9,6}}
\author{Mikkel N. Lund\altaffilmark{9,6}}
\author{Savita Mathur\altaffilmark{14}}
\author{Benoit Mosser\altaffilmark{15}}
\author{Keivan G. Stassun\altaffilmark{16,17}}
\author{William J. Chaplin\altaffilmark{9,6}}
\author{Yvonne Elsworth\altaffilmark{9,6}}
\author{Rafael A. Garc\'ia\altaffilmark{18,19}}
\author{Rasmus Handberg\altaffilmark{6}}
\author{Jon Holtzman\altaffilmark{20}}
\author{Fred Hearty\altaffilmark{21}}
\author{D. A. Garc\'ia-Hern\'andez\altaffilmark{22,23}}
\author{Patrick Gaulme\altaffilmark{24}}
\author{Olga Zamora\altaffilmark{22,23}}

\altaffiltext{1}{Institute of Space Sciences (IEEC-CSIC), Campus UAB, Carrer de Can Magrans S/N, E-08193, Barcelona, Spain}
\altaffiltext{2}{Department of Astronomy, The Ohio State University, Columbus, OH 43210, USA}
\altaffiltext{3}{Institute for Astronomy, University of Hawai`i, 2680 Woodlawn Drive, Honolulu, HI 96822, US}
\altaffiltext{4}{Sydney Institute for Astronomy, School of Physics, University of Sydney, NSW 2006, Australia}
\altaffiltext{5}{SETI Institute, 189 Bernardo Avenue, Mountain View, CA 94043, USA}
\altaffiltext{6}{Stellar Astrophysics Centre, Department of Physics and Astronomy, Aarhus University, Ny Munkegade 120, DK-8000 Aarhus C, Denmark}
\altaffiltext{7}{Institut f\"ur Astrophysik, Georg-August-Universit\"at G\"ottingen, Friedrich-Hund-Platz 1, D-37077, G\"ottingen, Germany}
\altaffiltext{8}{Max-Planck-Institut f\"ur Sonnensystemforschung, Justus-von-Liebig-Weg 3, D-37077 G\"ottingen, Germany}
\altaffiltext{9}{School of Physics and Astronomy, University of Birmingham, Edgbaston Park Road, West Midlands, Birmingham, B15 2TT UK}
\altaffiltext{10}{Department of Astronomy, Yale University, P.O. Box 208101, New Haven, CT 06520-8101, USA}
\altaffiltext{11}{Department of Astronomy, University of Virginia, Charlottesville, VA 22904-4325, USA}
\altaffiltext{12}{Institute of Astrophysics, University of Vienna, T\"urkenschanzstrasse 17, 1180, Vienna, Austria}
\altaffiltext{13}{School of Physics, University of New South Wales, NSW 2052, Australia}
\altaffiltext{14}{Space Science Institute, 4750 Walnut St. Suite 205, Boulder, CO 80301, USA}
\altaffiltext{15}{LESIA, Observatoire de Paris, PSL Research Univ., CNRS, Univ. Pierre et Marie Curie, Universit\'e Paris Diderot, 92195, Meudon, France}
\altaffiltext{16}{Vanderbilt University, Department of Physics \& Astronomy, 6301 Stevenson Center Lane, Nashville, TN 37235, USA}
\altaffiltext{17}{Fisk University, Department of Physics, 1000 17th Avenue N., Nashville, TN 37208, USA}
\altaffiltext{18}{IRFU, CEA, Universit\'e Paris-Saclay, F-91191 Gif-sur-Yvette, France}
\altaffiltext{19}{Universit\'e Paris Diderot, AIM, Sorbonne Paris Cit\'e, CEA, CNRS, F-91191 Gif-sur-Yvette, France}
\altaffiltext{20}{New Mexico State University, Las Cruces, NM 88003, USA}
\altaffiltext{21}{Department of Astronomy and Astrophysics, Institute for Gravitation and the Cosmos, The Pennsylvania State University, University Park, PA 16802, USA}
\altaffiltext{22}{Instituto de Astrof\'isica de Canarias (IAC), V\'ia L\'actea s/n, E-38200 La Laguna, Tenerife, Spain}
\altaffiltext{23}{Departamento de AstrofÃ\'isica, Universidad de La Laguna (ULL), E-38206 La Laguna, Tenerife, Spain}
\altaffiltext{24}{Apache Point Observatory, P.O. Box 59, Sunspot, NM 88349}

\email{aldos@ice.csic.es}

\begin{abstract}

We present the  first APOKASC catalog of spectroscopic  and asteroseismic data
for dwarfs  and subgiants. Asteroseismic  data for  our sample of  415 objects
have been obtained by the \emph{Kepler}  mission in short, 58.5~s, cadence and
lightcurves  span  from  30  up  to more  than  1000~days.  The  spectroscopic
parameters are based on spectra taken  as part of the Apache Point Observatory
Galactic Evolution  Experiment (APOGEE) and  correspond to Data Release  13 of
the  Sloan Digital  Sky  Survey. We  analyze our  data  using two  independent
$\teff$ scales, the spectroscopic values from DR13 and those derived from SDSS
\emph{griz} photometry.   We use the  differences in our results  arising from
these choices as a test of systematic temperature uncertainties, and find that
they   can  lead   to   significant  differences   in   the  derived   stellar
properties.  Determinations   of  surface  gravity  ($\logg$),   mean  density
($\mrho$), radius  ($R$), mass ($M$),  and age  ($\tau$) for the  whole sample
have  been carried  out by  means of  (stellar) grid-based  modeling. We  have
thoroughly assessed random  and systematic error sources  in the spectroscopic
and asteroseismic data, as well as in the grid-based modeling determination of
the stellar quantities provided in  the catalog. We provide stellar properties
determined for each of the two $\teff$ scales. The median combined (random and
systematic)  uncertainties  are  2\% (0.01~dex; $\logg$),  3.4\%  ($\mrho$),  2.6\%
($R$), 5.1\%  ($M$), and 19\% ($\tau$)  for the photometric $\teff$  scale and
2\% ($\logg$),  3.5\% ($\mrho$),  2.7\%  ($R$),  6.3\% ($M$),  and  23\%
($\tau$)  for the  spectroscopic scale.  We present  comparisons with  stellar
quantities in the asteroseismic catalog by \citet{chaplin:2014} that highlight
the  importance of  having  metallicity measurements  for determining  stellar
parameters accurately. Finally, we compare  our results with those coming from
a variety of sources, including  stellar radii determined from TGAS parallaxes
and asteroseismic  analyses based  on individual frequencies.  We find  a very
good  agreement  for  all  inferred  quantities.  The  latter  comparison,  in
particular, gives a strong support  to the determination of stellar quantities
based on  global seismology,  a relevant  result for  future missions  such as
\emph{TESS} and \emph{PLATO}.  
\end{abstract}


\maketitle

\section{Introduction}

The  advent  of  space-borne  asteroseismology  of  stars  showing  solar-like
oscillations has  been instrumental  in the  dawning of  the era  of precision
stellar astrophysics. First \emph{CoRoT} \citep{michel:2008,deridder:2009} and
then   \emph{Kepler}   \citep{gilliland:2010,bedding:2010,chaplin:2010}   have
provided  a  novel  way  to  determine  precise  stellar  parameters  --  most
importantly  mean density  ($\mrho$), surface  gravity ($\logg$),  mass ($M$),
radius ($R$), and age ($\tau$) -- for large numbers of stars. 

The capability that  asteroseismology offers for the  determination of stellar
parameters requires  knowledge of effective temperature  ($\teff$) but stellar
ages in particular  are sensitive to the metallicity, and  to some extent even
to the  mixture of heavy  elements. In  turn, composition information  is also
necessary to enable stellar and galactic studies. Stellar spectroscopy is thus
essential       for      realizing       the      full       potential      of
asteroseismology.  \citet{pinsonneault:2014}  produced  the first  catalog  of
asteroseismically  derived stellar  parameters for  red giant  stars based  on
asteroseismic  data from  \emph{Kepler}  and from  the high-resolution  Apache
Point Observatory  Galactic Evolution Experiment  (APOGEE), part of  the Sloan
Digital Sky Survey (SDSS). This first  APOKASC catalog included more than 1916
stars observed as part of  the SDSS-DR10 \citep{ahn:2014}. When completed, the
full APOKASC catalog will contain about 15,000 stars.  

Other surveys focused on the  \emph{Kepler} or \emph{CoRoT} fields include the
SAGA  survey  \citep{casagrande:2014,casagrande:2016}  that has  observed  the
\emph{Kepler}  field with  Str\"omgren  photometry with  DR1 containing  about
1,000  stars  and CoRoGEE,  the  combination  of  APOGEE and  \emph{CoRoT},  a
combined        dataset       of        606       red        giant       stars
\citep{chiappini:2015,anders:2017}. These  efforts have  now been  extended to
include   the    fields   of    the   \emph{Kepler}   extended    mission   K2
\citep{howell:2014,lund:2016b,stello:2017},   with   large-scale   observation
campaigns  being carried  out  by APOGEE  (Zasowski et  al.  in prep.),  Galah
(\citealp{desilva:2015}, Sharma  et al.  in prep.),  LAMOST \citep{zhao:2012},
and RAVE \citep{valentini:2017}.

Most of the  activity related to large  scale surveys is focused  on red giant
stars, as their larger oscillation amplitudes and brightness, and their longer
oscillation periods, make them much easier targets for asteroseismology. Joint
asteroseismic and  spectroscopic analysis  of the  smaller dwarf  and subgiant
samples, however,  remains important. First,  the spatial distribution  of the
dwarf sample is  much more confined to the  solar neighborhoold ($\sim$300~pc)
than the giant sample ($\sim$2000~pc \citealp{huber:2014} or even further than
4000~pc for the  faint giants \citealp{mathur:2016}). Properties,  such as the
age-metallicity  and age-velocity  disperson  relations, can  then be  readily
compared   to    such   pioneering   work   as    \citet{edvarsson:1993}   and
\citet{nordstrom:2004}.   The sources  of  systematic error  can therefore  be
investigated; hence bridging results out to  the regions sampled by the giants
from stars where we have the most accurate independent observations (parallax,
interferometry, etc). 

Stellar parameters  of dwarf and  subgiant stars  can be determined  best from
asteroseismic  measurements when  frequencies of  many individual  oscillation
modes   can   be  measured   precisely   and   matched  by   stellar   models.
\citet{lebreton:2014}  discusses  in  detail different  approaches  for  using
frequency data. Recent examples  based on  \emph{Kepler} targets  include: 42
main  sequence and  subgiant stars  \citep{appourchaux:2012,metcalfe:2014}, 33
exoplanet host stars \citep{davies:2016b,victor:2015}. More recently, \citep{lund:2017,victor:2017}
have analyzed the LEGACY sample, composed by 66 stars that have  \emph{Kepler} 
  light    curves    longer    than   1    year. These samples overlap. 
  The LEGACY sample includes some of the stars in \citet{metcalfe:2014}, while works by  
  \citet{bellinger:2016,creevey:2017} are based on  seismic data from \citep{davies:2016b}
  and the LEGACY  sample.
  In contrast,  global seismic
parameters, namely the  large frequency separation ($\dnu$)  and the frequency
of maximum power  ($\numax$) that are closely related to  $\mrho$ and $\logg$,
are available for a much larger  number of dwarfs and subgiants. \emph{Kepler}
has provided $\dnu$ and $\numax$ for  more than 500 main sequence and subgiant
stars  and determinations  of stellar  properties were  initially reported  by
\citet{chaplin:2011}, based  on the  Kepler Input  Catalog $\teff$  and $\feh$
values. The  first catalog for  this sample  including stellar ages,  based on
more  accurate  and  precise  photometric $\teff$  determinations,  but  still
lacking $\feh$  measurements for the whole  sample, was later on  presented in
\citet[hereafter C14]{chaplin:2014}.

A  subset of  our sample  allows  us to  critically compare  the results  from
detailed modeling of individual frequencies  with those from scaling relations
relative to the Sun.  We also,  crucially, have detailed abundance data on the
stars in our sample, which was not  true in C14, and we have $\teff$ estimates
from photometry  and spectroscopy  for the  whole sample  for the  first time.
Both  of  these advantages  justify  the  presentation  of a  revised  catalog
including this new APOGEE data. 

In   this  work,   we   present   the  extension   of   the  APOKASC   catalog
\citep{pinsonneault:2014} to include 415 dwarf  and subgiant stars that form a
homogeneous sample  with the  previously released and  also future  red giants
catalog.  Spectroscopic parameters  are  determined from  spectra released  in
SDSS-III  DR10 but  with  data extracted  using the  methods  of SDSS-IV  DR13
(\citealt{albareti:2017},  Holtzmann  et  al.  2017  in  prep.).  The  catalog
includes   newly   determined   global  seismic   parameters   obtained   from
\emph{Kepler} short-cadence  light curves,  improving over  those used  in C14
both because  of longer  duration lightcurves and  in the  analysis techniques
\citep{handberg:2014}. In Section~\ref{sec:sample} we  describe the sample and
the     spectroscopic     and      asteroseismic     data     and     analysis
methods. Section~\ref{sec:gbm} presents the grid-based modeling (GBM) approach
used    to   determine    stellar   parameters    for   the    catalog   while
Section~\ref{sec:core} is devoted  to the determination of  errors and central
values  for the  catalog. Section~\ref{sec:catalog}  presents the  information
contained  in the  catalog and  comparison with  previous work.  A summary  is
presented  in  Section~\ref{sec:summary}.   Finally,  two  appendices  include
detailed information about stellar models  and extensive comparisons among GBM
pipelines used in this work.

\section{The Sample}  \label{sec:sample}

The  catalog  consists  of  stars with  detected  solar-like  oscillations  in
\emph{Kepler} short-cadence data as reported in \citet{chaplin:2011a} and with
spectroscopic      observations       from      the       APOGEE-1      survey
\citep{majewski:2015}. APOGEE-1 observed  415 out of the  $\sim\,600$ stars in
\citet{chaplin:2011a}.  Stars with  photometric  temperature estimates  higher
than 6500~K  were de-prioritized because the  IR spectra from APOGEE  were not
expected  to be  informative  for  such hot  targets.  There  were also  other
targeting  constraints   as  summarized  by  \citet{zasowski:2013}.   For  our
purposes, the  most important were that  the APOGEE field of  view is slightly
smaller than  a \emph{Kepler} CCD module,  therefore objects that fell  on the
corners of a \emph{Kepler} CCD module  were not observed. In addition, targets
that fell within  $\sim$70 arcseconds of another target could  be discarded in
favor  of the  higher priority  giant  target. Finally,  APOGEE-1 had  already
completed its observations of certain parts of the Kepler field when the dwarf
targets were added.  

The APOGEE survey  is a bright-time component of the  Sloan Digital Sky Survey
III   \citep{eisenstein:2011}.   Using    the   Sloan   Foundation   Telescope
\citep{gunn:2006} at Apache Point 
Observatory,  APOGEE  observed 230  science  targets  simultaneously across  a
7~deg$^2$  field of  view in  the standard  observing mode.  An additional  35
fibers were devoted to standard stars and 35 fibers to sky. 300 fibers fed the
APOGEE spectrograph  \citep{wilson:2012}. APOGEE  spectra cover  the $H$-band,
between 1.51\,$\mu$m and 1.7\,$\mu$m, at a resolution R=22,500.

\subsection{Spectroscopic analysis}  \label{sec:spectro}

APOGEE data is reduced in a  three-step process. Initially, observations of an
individual plate on  an individual night are reduced taking  into account: the
calibration of the detector, detection  of bad pixels, wavelength calibration,
dither shifts between exposures in a visit, correction of individual exposures
and determination of  radial velocity of each object by  using a best-matching
stellar template. The  second step combines multiple individual  visits to the
same objects: it  corrects for each visit-specific radial  velocity and coadds
the spectra. The final spectra are resampled onto a fixed wavelength grid with
constant   dispersion   in   $\log{\lambda}$.   Details  can   be   found   in
\citet{nidever:2015}.  

The final step,  extraction of stellar parameters and  chemical abundances, is
done  by  the  APOGEE  Stellar Parameters  and  Chemical  Abundances  pipeline
(ASPCAP).  Here we summarize its main characteristics.  

ASPCAP  performs $\chi^2$  minimization  over grids  of precomputed  synthetic
spectra  (e.g. \citealt{zamora:2015}).  The  basic parameter  space search  is
performed over  $\teff$, $\logg$, \mh, ${\rm  \alpha/M}$, to which C,  N, line
broadening and/or microturbulence can be  added. The optimization is done with
\texttt{FERRE} \citep{allende:2006}, which performs a $\chi^2$ minimization by
comparing fluxes as  a function of wavelength with weights  that come directly
from   the    uncertainties   in    the   fluxes   determined    during   data
reduction. Synthetic spectra across the grids are constructed by interpolation
of model fluxes  for spectra computed at  the nodes of the grid.  This is more
robust  than  interpolating  atmospheric  structures and  then  computing  the
synthetic spectra from the interpolated structure \citep{meszaros:2013}.  
A   detailed   description   of   how   ASPCAP   works   can   be   found   in
\citet{garciaperez:2016}.  

Once ASPCAP has determined the best-fit stellar parameters to the
grid of synthetic spectra, these parameters are compared to external
values of high accuracy (''ground truth'') and, if necessary, a calibration relation
is applied to bring the ASPCAP values into agreement with ground
truth. For this work, seismic measurements provide accurate gravities,
but the calibration of the ASPCAP temperatures and metallicities
are relevant. For $\teff$, the comparision values are based
on $J-K$ colors and the color-Teff relation of \citet{gonzalez:2009}. 
For metallicity, the comparison values are
the mean metallicities of well-studied clusters. Details
about how the calibration samples were constructed are
available in \citet{holtzman:2015} for DR12 and Holtzman
et al. (2017, in preparation) for DR13.


\subsubsection{Preliminary analysis with modified DR12 parameters}

The APOGEE stellar parameters and abundances released prior to Data Release 13
(DR13)  were  derived   assuming  that  stellar  rotation   was  a  negligible
contribution   to   line-broadening   at   the  resolution   of   the   APOGEE
spectra. Therefore, the synthetic spectra  grid used for $\chi^2$ minimization
was  not  convolved with  stellar  rotation  broadening profiles.  While  this
assumption is  usually appropriate for giants,  which are the majority  of the
stars observed  by APOGEE, warmer  dwarfs ($\teff \gtrsim$ 6000K)  and younger
dwarfs of all  temperatures can still show noticeable rotation  even at APOGEE
resolution. This  was realized at an  early stage of this  project, during the
preparation of DR12, so a modified  pipeline was developed to correct for this
effect  by adding  rotation  velocity  as another  dimension  in  the grid  of
atmosphere models. When rotation is  included in the optimization procedure, C
and N are fixed to solar values, i.e. ${\rm [C/Fe]=0}$ and ${\rm [N/Fe]=0}$, 
to keep the computational needs constrained.  

The  DR12-rot   spectroscopic  parameters   were  used  in   combination  with
asteroseismic  data  (Sect.\,\ref{sec:seismicparam})  by  all  the  Grid-Based
Modeling pipelines  (GBMs; see Sect.\,\ref{sec:gbm}) to  carry out comparisons
among the codes. The offsets and  different standard deviations found in these
comparisons  do not  depend  on whether  we  adopt the  DR12-rot  or the  DR13
spectroscopic values.  

\subsubsection{DR13}

The final  ASPCAP spectroscopic parameters  used in the catalog  correspond to
DR13  \citep{albareti:2017}.  DR13  includes  by  default  a  grid  of  models
including rotation  for dwarfs,  an improved relation  between microturbulence
and  gravity, and  a relation  for macroturbulence.  Some of  the improvements
brought about  by DR13 are  discussed in \citep{albareti:2017} and  in greater
extent by  Holtzmann et  al. (2017  in prep.).  Here we  focus on  $\teff$ and
metallicity \mh\ for the APOKASC sample.

\subsubsection{Adopted Effective Temperatures} \label{sec:teffscales}

\begin{figure}
\includegraphics[scale=.42]{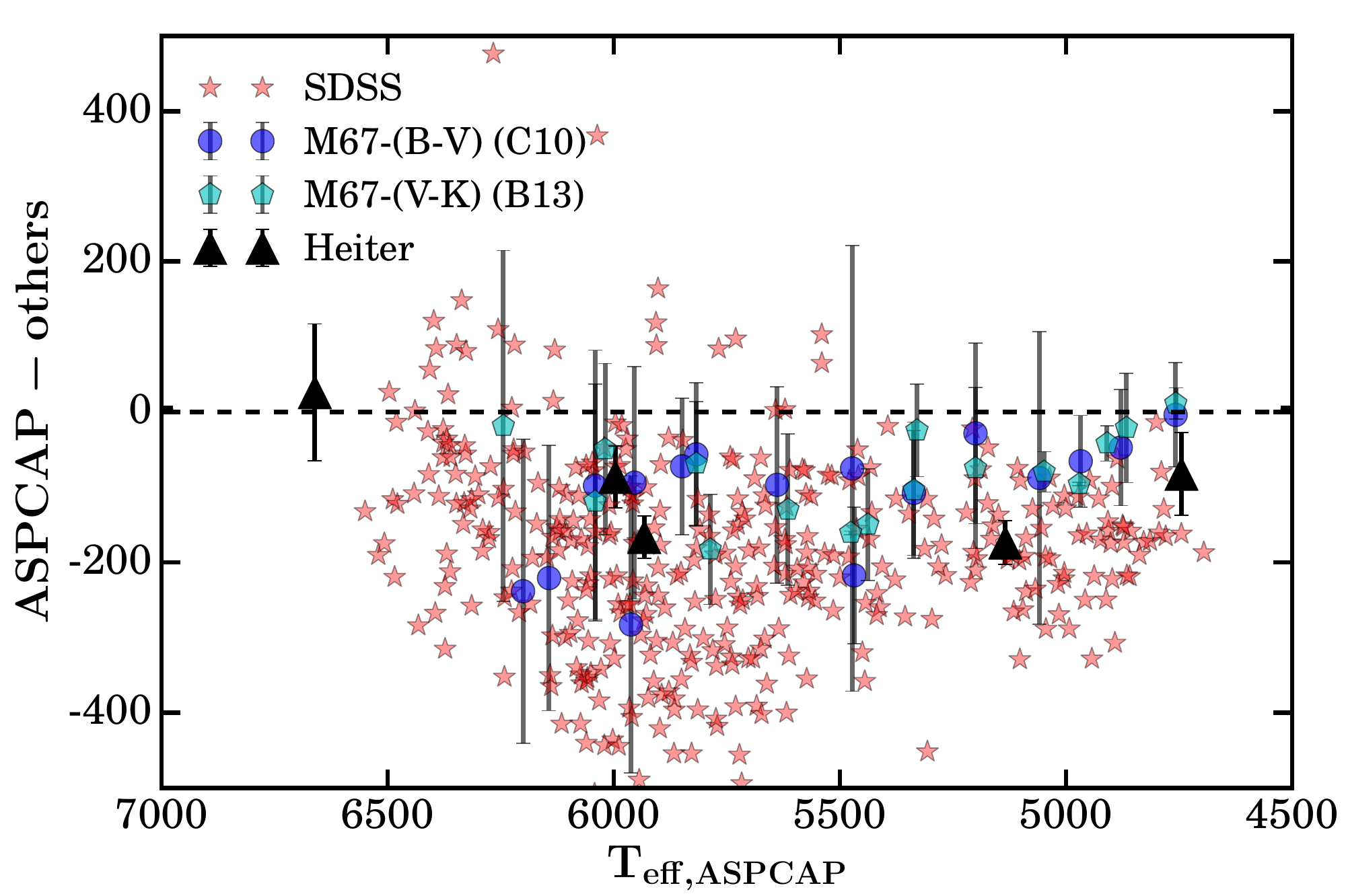}
\caption{Differences between  the spectroscopic  ASPCAP $\teff$ scales  and: 1) the
  SDSS photometric $\teff$  scale for our full sample of  stars; 2) median binned
  differences (100~K bins)  between the ASPCAP and two photometric  $\teff$ scales of M67
  stars (C10: \citealt{casagrande:2010}, B13: \citealt{boyajian:2013}). Error
  bars denote the dispersion of the binned temperatures. Typical uncertainties
  in the photometric $\teff$ of individual M67 stars are $< 80$\,K; 3) 
  five    Gaia    benchmark   stars   \citep{heiter:2015} for which error bars
  are $\teff$ uncertainties for each individual star as given in \citet{heiter:2015}. \label{fig:comp_phot_aspcap}} 
\end{figure}

We present  results in the catalog  using two different sets  of temperatures:
photometric  temperatures  from measurements  in  the  SDSS \emph{griz}  bands
\citep{pinsonneault:2012} and  ASPCAP spectroscopic temperatures from  DR13. 
A
comparison  between photometric  and spectroscopic  temperature scales  in the
\emph{Kepler}   field    is    shown   in
Figure~\ref{fig:comp_phot_aspcap}.   For  \emph{Kepler}  targets, there  is  a
systematic offset (defined as spec-phot)  of $-165$~K between 5000-6000 K with
a 134~K dispersion. At hotter temperatures the offset is $-220$~K with a 163~K
dispersion. The formal $\teff$ uncertainty returned by ASPCAP is 70~K, whereas
the median uncertainty for the SDSS $\teff$ scale is 62~K with a dispersion of
27~K.  Two  concerns  with photometric  temperatures are:  1) the  zero
  point shifts  from using different photometric  color-$\teff$ relations and,
  2) the uncertainties  caused by reddening corrections. The latter are small, 
  with E($B-V$) median and dispersion of 0.023 and 0.015 respectively for the
  \emph{Kepler} dwarfs and subgiants \citep{huber:2017}. Concerning the impact 
  of the KIC extinction values used in \citet{pinsonneault:2012}, this has been 
  tested by \citet{huber:2017} who  found that using improved reddening estimates
  over the original KIC values and stellar metallicities (\citealt{pinsonneault:2012} 
  assumed a constant 
  $\feh=-0.2$ value for the whole sample, as $\feh$ information was not available 
  at the time for the whole \emph{Kepler} sample) leads to a $\approx -20$~K zero 
  point shift of the SDSS $\teff$ scale.

To test the temperature scale  further, we calculated photometric temperatures
for stars observed by  APOGEE in the open cluster M67,  which has a well-known
solar  metallicity,   [Fe/H]=$-$0.01  \citep{jacobson:2011},   and  reddening,
E($B-V$)=0.04 mag \citep{taylor:2007}, as well as accurate optical $B$ and $V$
band photometry \citep{sandquist:2004}. To investigate the effect of different
color-$\teff$  relations  we  used \citet{casagrande:2010}  to  convert  $B-V$
colors and \citet{boyajian:2013} to convert  $V-K$ colors to temperatures. The
uncertainties in  the $\teff$  from photometric  errors are  $<$ 80K. 
These comparisons
are also included in Fig.~\ref{fig:comp_phot_aspcap}.We also
compared the ASPCAP  temperatures to the temperatures for  five Gaia benchmark
stars  \citep{heiter:2015},  which  are  based on  interferometric  radii  and
parallax  measurements.  The  random uncertainties  in these  measurements are
$<\,$2\%.  Figure~\ref{fig:comp_phot_aspcap} also  shows that  the photometric
M67 temperatures and Gaia benchmark star temperatures agree well with the SDSS
temperature scale, particularly below 6250\,K. This agrees with the conclusion
of \citet{pinsonneault:2012}  that there was good  agreement among photometric
scales  for  stars  in  this   temperature  range.  Therefore, our  preferred
temperature scale  for the \emph{Kepler}  field stars is the  SDSS \emph{griz}
temperature scale. For the spectroscopic temperature scale, we include a
systematic uncertainty component in the seismically determined stellar parameters to account
the systematic offset with respect to the photometric scale. This is done by 
varying the ASPCAP temperature scales by $\pm100$\,K and quantifying the impact
in the final estimated stellar parameters ($g$, $\mrho$, $R$, $M$,  and $\tau$). 
Details are given in Sect.\,\ref{sec:systteff}.

At temperatures  greater than 6250\,K, there  are causes for concern  for both
temperature  scales. The  near-infrared H-band  spectra of  hotter stars  have
fewer  absorption  features,  decreasing   their  sensitivity  to  temperature
(Holtzman et al 2017 in prep.).  \citet{pinsonneault:2012} noted discrepancies
among the photometric  and spectroscopic measurements when  they compared with
their  SDSS-color  based  temperatures  in this  temperature  range.  Possible
reasons  include:   (1)  rapid   rotation  producing   non-isothermal  surface
temperatures, which violates an assumption of the infrared flux method or, (2)
a  lack   of  calibrators   for  color-$\teff$   relations  at   these  hotter
temperatures.   Currently   ongoing   efforts   to   carefully   measure   and
cross-validate interferometric  angular diameters with  the CHARA array  for a
range  of spectral  types (Karovicova  et al.,  in prep.)  as well  as planned
detailed  investigation   of  individual   hot  G-stars  will   contribute  to
understanding this discrepancy.

\subsubsection{Adopted Metallicities}

\begin{figure}
\includegraphics[scale=.42]{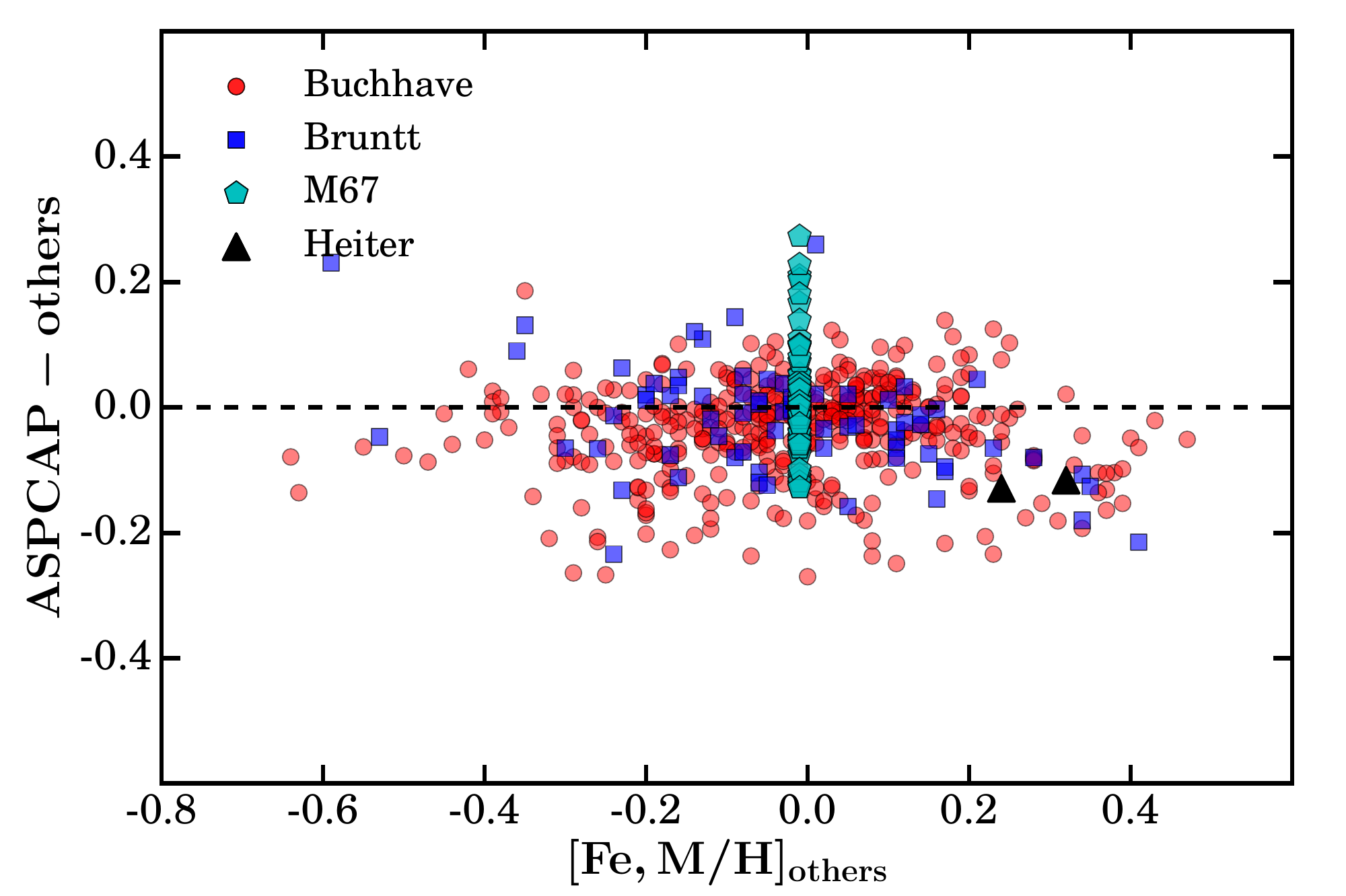}
\caption{Differences  between ASPCAP  metallicities and  results from  optical
  spectroscopy  \citet{buchhave:2015} and  \citet{bruntt:2012}.  For M67,  our
  fiducial   value   is   the   mean   cluster   value   [Fe/H]=$-$0.01   from
  \citep{jacobson:2011}. \label{fig:feh}}
\end{figure}

The  metallicities  $\mh$   we  use  throughout  this  work   are  the  ASPCAP
metallicities corresponding  to DR13. Formal uncertainties  returned by ASPCAP
have   a  median   value  of   0.025~dex   with  a   very  small,   0.004~dex,
dispersion. These estimates of uncertainty  are based on the small measurement
scatter in members of star clusters. Figure~\ref{fig:feh} shows the comparison
of ASPCAP $\mh$ values with results from optical spectroscopy including $\feh$
measurements for  71 stars in  common with the \citet{bruntt:2012}  sample and
$\mh$  for  400  stars  in  common  with  \citet{buchhave:2015}.  The  overall
agreement is very good in both cases. The median and median absolute deviation\footnote{
For a sample $\left\{x_i\right\}_{i=1,...,N}$, the median absolute deviation is defined 
as: $\hbox{median}\left[|x_i - \hbox{median}(\left\{x_i\right\})|\right]$. It is a 
measurement of dispersion less sensitive to outlayers than the standard deviation.} 
of   the  $\mh$   difference  between   ASPCAP  and   \citet{bruntt:2012}  are
$-0.019$~dex and 0.052~dex respectively. With respect to \citet{buchhave:2015}
these values  are $-0.028$ and  0.046~dex respectively. The figure  also shows
ASPCAP results  for the calibration cluster  M67, for which we  adopt the mean
cluster value  $\feh=-0.01\pm 0.05$~dex  reported by  \citet{jacobson:2011} as
our fiducial  value. Only stars  in the range $\logg=3.3$-$4.5$\,dex  
appropriate for
our sample are shown. The ASPCAP dispersion for stars in this gravity range is
0.08~dex. Finally,  two Gaia benchmark subgiants  from \citet{heiter:2015} are
also shown.

Based on these  comparisons, we do not find significant  offsets in the ASPCAP
$\mh$  scale,  but  typically  the   dispersion  is  larger  than  the  formal
uncertainties returned  by ASPCAP, which  highlights the possibility  that the
true uncertainty is underestimated. In  lightof the dispersion that we observe
in  external measurements  of  subgiant stars  relative to  APOGEE,  we add  a
0.1~dex uncertainty in quadrature to  the ASPCAP formal uncertainty, and adopt
these as our final $\mh$ uncertainties for the catalog. Note that in 
constructing the catalog, we use the same $\mh$ values in combination with both
temperature scales. This is formally inconsistent and might be a systematic source of 
uncertainty. But, as shown in this section and particularly with the comparison against 
\citet{buchhave:2015} results, ASPCAP $\mh$ determinations are robust so 
combining them with either temperature scale is a safe procedure.

\subsection{Input asteroseismic data}\label{sec:seismicparam}

Stars in the APOKASC sample have been observed by \emph{Kepler}, in cadence of
58.5\,s. The sample  includes 284 stars observed  for up to 40  days, 31 stars
between 40 and 100 days, 57 between 100  and 900 days and 55 stars between 900
and     1055    days.     Data    were     downloaded    from     the    KASOC
database\footnote{\url{kasoc.phys.au.dk}}  and  corrected   using  the  method
presented in  \citet{handberg:2014}. This method computes  two median-filtered
versions of  the time series with  different filter window widths.  A weighted
combination is  made of  the two  based on their  relative differences  to one
another  ---  thereby both  short  and  long-term instrumental  and  transient
features are corrected. We refer  to \citet{handberg:2014} for further details
on the operations of the filter.

The duration  of the timeseries and  the brightness of the  star are important
quantities in  determining the quality of  the seismic data. The  amplitude of
solar-like oscillations, however, depends  strongly on stellar parameters such
as the  luminosity $L$,  mass $M$, and  $\teff$. \citet{huber:2011}  has found
that oscillation amplitudes in the \emph{Kepler} photometric band $K_P$ behave
as   $A_{K_P}   \propto   L^{0.838}    M^{-1.32}   \teff^{-1.8}$   or,   using
Eq.\,\ref{eq:numax}   (Sect.~\ref{sec:gbm},   below),  as   $A_{K_P}   \propto
\numax^{-1}  R^{-0.324}   M^{-0.3}  \teff^{1.552}$.   In  the   short  cadence
\emph{Kepler} sample $\numax$ varies by more than one order of magnitude while
the  other quantities  in  the  relation above  either  vary  by much  smaller
fractions (e.g. $\teff$)  or induce milder dependences (e.g. $R$).  It is also
known  that  stellar  activity  may lead  to  smaller  oscillation  amplitudes
\citep{chaplin:2011,garcia:2010,kiefer:2017}.

A measure  of the quality of  the seismic detection is  given by the
height-to-background ratio (HBR), defined  as the peak power excess divided by
the  total power  coming from  granulation and  shot noise  at $\numax$ (see e.g. \citealt{mosser:2012}).   HBR
higher than  1 indicates  a secure detection.  We include HBR  as part  of the
APOKASC catalog presented in this work.


Global  seismic  parameters $\dnu$  and  $\numax$  were extracted  using  five
independent automated  seismic pipelines. This gives us the possibility
to  of identifying potential issues that might cause systematic offsets in the determination
of global seismic parameters or in the determination of uncertainties that could 
go unnoticed if only one seismic set of results were available. Moreover, this 
allows have a quantitative estimate of uncertainties linked to the methods used in the
seismic pipelines. The  procedures employed by  each of them are described below.

\texttt{A2Z}: $\dnu$ is obtained by computing  the power spectrum of the power
spectrum (PS2).  The background is fit  with a photon noise  component and two
Harvey-like granulation models \citep{harvey:1985} with  a slope fixed to 4. A
Gaussian  function is  used to  fit the  envelope of  the p-mode  region in  a
smoothed power  spectrum \citep{mathur:2010}. The uncertainties  in $\dnu$ and
$\numax$ were computed with the weighted  centroids method and are known to be
quite  conservative.  The  A2Z  pipeline   was  applied  to  the  KADACS  data
\citep{garcia:2011},  allowing  us  to  compare  the  results  from  different
calibrated data.

 \texttt{COR}: $\dnu$ is measured with  the autocorrelation of the oscillation
 signal. In  practice, this autocorrelation  signal is delivered by  the power
 spectrum    of    the    filtered    power    spectrum    of    the    signal
 \citep{mosser:2009}. The  properties of  the filter make  use of  the seismic
 scaling relations, in order to optimize the seismic signature. A more precise
 value  of  the large  separation  is  then  obtained using  the  second-order
 asymptotic expansion \citep{mosser:2013}. The resulting uncertainty in $\dnu$
 is  often better  than 0.1\,\%.  $\numax$ and  the background  parameters are
 provided by a local analysis of the excess power. The uncertainty in $\numax$
 is a fraction of $\dnu$, typically of about $\dnu/5$.

\texttt{FITTER}: Determination  of $\numax$ is  done by performing  a Bayesian
background fit following \citet{davies:2016}. This approach fits a 2 component
Harvey model in addition  to a Gaussian p-mode hump, all  modified by the sinc
squared apodization function.   A white noise background is added  to this. To
determine $\dnu$  a squared  Gaussian plus  a flat background  is fitted  to a
narrow region in  the critically sampled power spectrum of  the power spectrum
(PS2)   around   half   the   initial  estimate   of   the   large   frequency
spacing.  Estimated  values and  uncertainties  for  $\numax$ and  $\dnu$  are
determined  as  the  median  and  the  standard  deviation  of  the  posterior
probability  distributions  for  the   central  frequencies  of  the  Gaussian
components in both methods.

\texttt{OCT}: The background  model includes a 2-component Harvey  model and a
white noise component.  After background subtraction, the  spectrum is heavily
smoothed and a $\numax$ is obtained from  a Gaussian fit to the residual power
spectrum. $\dnu$  is determined from  a weighted  average of the  $\dnu/2$ and
$\dnu/4$ features  extracted from the  PS2.  The uncertainty is  determined as
the standard deviation of grouped data following Eq.~6 in \citet{hekker:2010},
where more detailes can be found.

\texttt{SYD}:   Data   are   analyzed   using   the   methods   described   in
\citet{huber:2009},  using  on  average  a frequency  range  between  100-7500
$\mu\hbox{Hz}$. The background  is modeled using a two  component Harvey model
with  the white  noise  component fixed  to the  mean  value measured  between
7300-7500 $\mu\hbox{Hz}$. Uncertainties on $\dnu$ and $\numax$ were calculated
using Monte-Carlo  simulations as described in  \citet{huber:2011}. The median
uncertainties in the sample are ~2\% in  $\dnu$ and ~5\% in $\numax$ for stars
with less than 100 days of data,  and ~0.1\% and ~0.7\% for stars observed for
the entire mission.

\begin{figure*}
\centering\includegraphics[scale=.45]{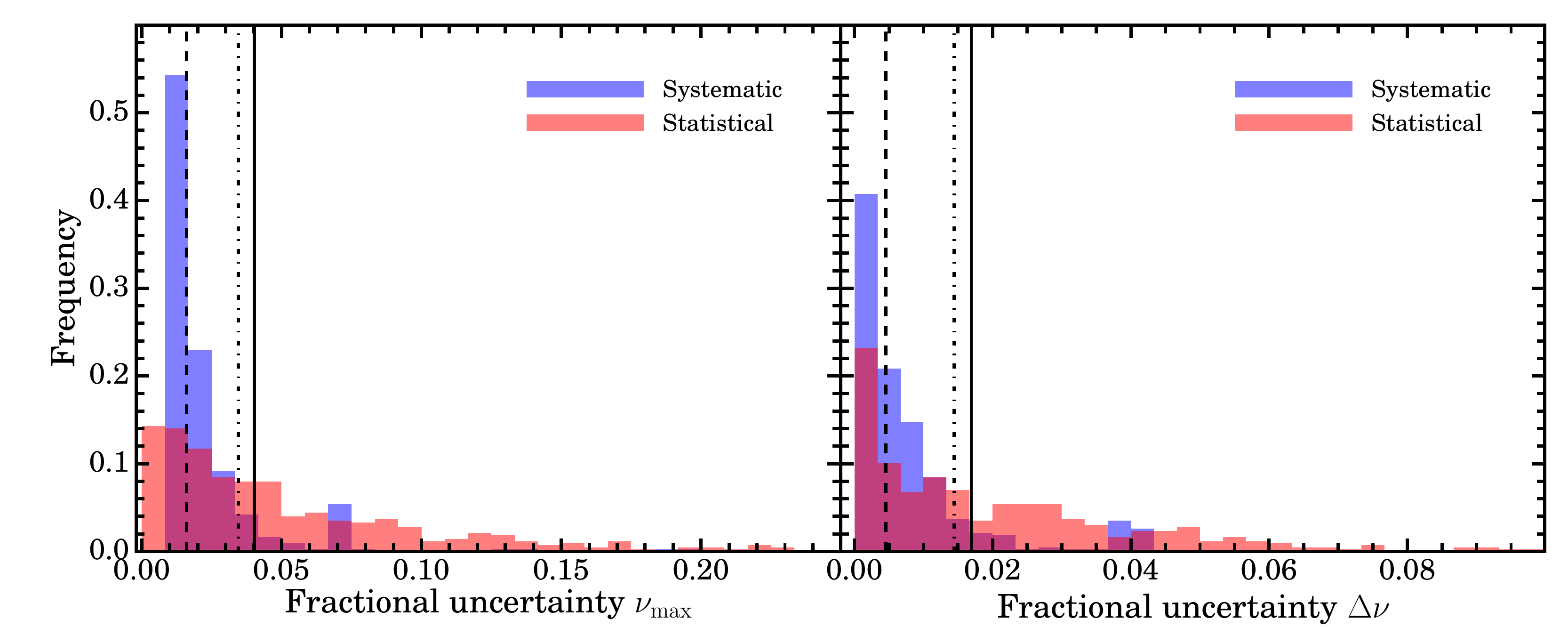}
\caption{Distribution of  errors for global seismic  quantities $\numax$ (left
  panel) and  $\dnu$ (right panel).  Vertical dashed, dotted-dashed  and solid
  lines are the  median values of the systematic, statistical  and total error
  distributions. \label{fig:pipes}}
\end{figure*}

Note that  uncertainties returned by  each pipeline  are formal, in  the sense
that  they  do not  take  into  account  systematic  differences in  e.g.  the
definition of $\dnu$ and $\numax$. For this reason, results from all pipelines
have been  used to assess the  systematic component contributing to  the total
uncertainties of the global seismic  parameters (for early work on comparisons
among seismic pipelines see e.g \citealt{verner:2011,hekker:2011}).

For  the central  values for  $\numax$ and  $\dnu$ we  adopt results  from the
\texttt{SYD} pipeline.  This choice  is made  because \texttt{SYD}  yields the
smallest average  deviation from the median  values for the sample  as a whole
and they are available for the  complete sample. In addition, it is consistent
with that in C14, where the asteroseismic fundamental properties of stars were
presented after the initial 10 months of \emph{Kepler} observations. We define
here the  formal error given by  the \texttt{SYD} pipeline as  the statistical
uncertainty for each quantity.

In  order to  estimate systematic  uncertainties  we proceed  as follows.  For
$\numax$ and  $\dnu$ separately,  we compute $\delta  Q_{ij}= (Q_{i,j}-Q_{{\rm
    SYD},j})/Q_{{\rm SYD},j}$,  where $Q$  is either  $\numax$ or  $\dnu$, $j$
runs through all stars in the sample and $i$ over all the pipelines other than
\texttt{SYD}. We  define clear  outliers as  measurements $Q_{i,j}$  for which
$|\delta  Q_{ij}| >  0.2$. There are 71 individual measurement, i.e. combinations of 
$(i,j)$ from all non-SYD pipelines that fall in this condition for either $\numax$ or $\dnu$. 
This is equivalent to about 4\% of the measurements for all non-SYD pipelines. 
These outliers are  removed from  the estimate  of
systematic uncertainties.  The resulting distributions of  $\{\delta Q_{ij}\}$
are characterized by mean values of  0.1\% and 0.02\% for $\numax$ and $\dnu$,
respectively,  and standard  deviations of  $\sigma_{\rm sys}(\nu_{\rm  max})=
3.3\%$ and  $\sigma_{\rm sys}(\Delta\nu)=2\%$.  The mean values  indicate that
there are  no strong  systematic deviations among  pipelines. Then,  we define
thresholds  of  7\%  and  4\%,  equal to  twice  the  values  of  $\sigma_{\rm
  sys}(\nu_{\rm max})$  and $\sigma_{\rm  sys}(\Delta\nu)$. For each  star, we
now  compute  the median  of  a  given  seismic  quantity and  remove  results
discrepant from  this median  by more  than the  threshold defined  above. The
process  is iterated  until no  outliers are  found or  only results  from two
pipelines  remain.  The 1~$\sigma$  systematic  uncertainty  for each  seismic
quantity and star  is then defined as the standard  deviation of the remaining
values.   \texttt{SYD} results  are always  kept because  they constitute  the
central seismic values  used in this work. If only  the \texttt{SYD} result is
available,  then a  systematic uncertainty  equal  to the  threshold value  is
assigned.

The  final   distributions  of   statistical  (\texttt{SYD})   and  systematic
uncertainties  are  shown  for  $\numax$  and $\dnu$  in  the  two  panels  of
Fig.\,\ref{fig:pipes}. For both quantities statistical uncertainties have much
more  extended distributions,  i.e.  they  dominate the  total  error for  the
majority of the sample. In the case of the systematic distributions, two small
bumps  are seen  at 7\%  and  4\% for  $\numax$ and  $\dnu$ respectively  that
correspond to  stars with results just  from \texttt{SYD}. For each  star, the
total error is computed by adding in quadrature the systematic and statistical
uncertainties.  Vertical   lines  show   the  median  values   of  systematic,
statistical  and total  errors in  increasing order.  The latter  are 4\%  for
$\numax$ and 1.7\% for $\dnu$.

sec:systgbm\section{Grid-based modeling} \label{sec:gbm}

Stellar mass  and radius can be  estimated from the global  seismic parameters
$\numax$  and $\dnu$,  and $\teff$.  If $g$  is the  stellar surface  gravity,
$\langle \rho  \rangle$ the mean  density, $M$ the  mass, and $R$  the radius,
then the scaling relations \citep{ulrich:1986,brown:1991,kjeldsen:1995}
\begin{align}
 \frac{\numax}{\numaxsun}  & \simeq 
\left(\frac{g}{\gsun} \right) \left(\frac{\teff}{\tsun} \right)^{-1/2}  \label{eq:numax}\\
 \frac{\dnu}{\dnusun} & \simeq \left(\frac{\mrho}{\mrhosun}\right)^{1/2} \label{eq:dnu}
\end{align}
can be readily inverted such that stellar mass and radius are approximately given by
\begin{align}
\frac{M}{\msun} & \simeq \left(\frac{\numax}{\numaxsun}\right)^3 \left(\frac{\dnu}{\dnusun}\right)^{-4} \left(\frac{\teff}{\tsun}\right)^{3/2} \label{eq:mass} \\
\frac{R}{\rsun} & \simeq \left(\frac{\numax}{\numaxsun} \right) \left(\frac{\dnu}{\dnusun} \right)^{-2} \left(\frac{\teff}{\tsun} \right)^{1/2}. \label{eq:radius}
\end{align}

Direct determination  of $M$ and $R$  from the equations above  can be
  qualitatively improved by constructing  grids of stellar evolutionary tracks
  that include $\numax$ and $\dnu$.  A  more refined analysis is possible when
  information about  the composition of  the star  is also available,  e.g. by
  means of $\mh$. Then, a given set of 
observables is compared against a  large set of stellar models from which
best fitting values  and uncertainties or even  full statistical distributions
for  the stellar  quantities of  interest  can be  determined. This  so-called
grid-based modeling  (GBM) allows  the estimation  of other  quantities beyond
mass and radius, most importantly stellar age $\tau$.

\begin{figure}[!ht]
\centering
\includegraphics[scale=.42]{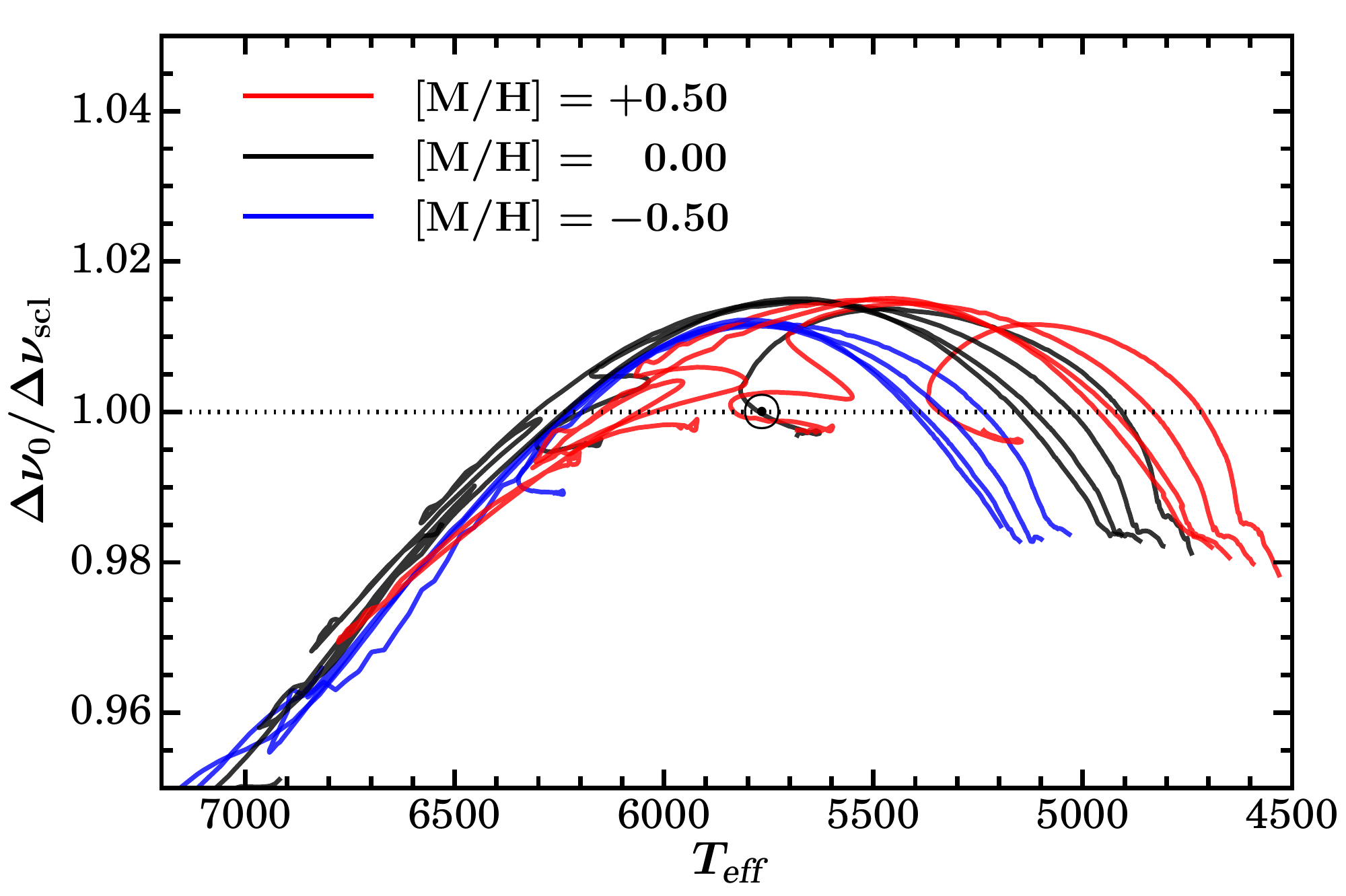}
\caption{Ratio between $\dnuo$ and $\dnuscl$ for stellar models with 1, 1.2, 1.4, 
and 1.6~M$_\odot$ (from right to left at the low $\teff$ end) 
and $\mh=-0.5, 0.0, +0.5$ (blue, black and red respectively). 
At low $\teff$ tracks are truncated at $\logg=3.2$ because all APOKASC sample
has higher $\logg$. 
 \label{fig:dnucorr}}
\end{figure}

Scaling  relations (Eqs.\,\ref{eq:numax}-\ref{eq:dnu})  applied to  dwarfs and
subgiants      are     accurate      to     within      a     few      percent
\citep{stello:2009,huber:2012,victor:2012,white:2013,coelho:2015,huber:2017}. The
first relation is qualitatively understood by theory of stochastically excited
oscillations,  but   its  accuracy  still  defies   a  definite  understanding
\citep{belkacem:2011}.  On the  other hand,  the  relation for  $\dnu$ can  be
tested  by stellar  models, at  least  up to  the point  allowed by  adiabatic
oscillation  frequencies or  the poor  modeling of  surface effects.  In fact,
Eq.\,\ref{eq:dnu} can be dropped altogether in  GBM if, for each stellar model
in  the  grid,  $\dnu$ is  computed  from  an  appropriate  fit to  the  model
frequencies of radial ($\ell=0$) modes. \citet{white:2011} found that for main
sequence and subgiant stars $\dnu$ derived from model frequencies deviate from
Eq.\,\ref{eq:dnu} by a few percent, with clear trends depending on the stellar
$\teff$.  In  the remainder  of  this  work we  denote  by  $\dnuo$ the  large
frequency separation  determined in stellar  models by using  frequencies from
$\ell=0$ modes and  by $\dnuscl$ the large frequency  separation obtained when
Eq.\,\ref{eq:dnu}     is     used     to     compute     it     in     stellar
models. Figure~\ref{fig:dnucorr} shows the relative difference between $\dnuo$
and $\dnuscl$ as a function  of $\teff$ for representative evolutionary tracks
based on a calibration that reproduces  the global seismic solar properties. The 
mass range of the tracks shown is between 1 and 1.6~M$_\odot$ and tracks cover all 
evolutionary phases of interest for our catalog, from ZAMS up to evolved subgiant 
phase (defined here for illustration purposes as $\logg=3.2$). 
As shown in Fig.\,\ref{fig:comp_phot_aspcap}, $\teff$ in our APOKASC sample range
from $\sim 6500$ down to $\sim 4700$\,K. Then, from Fig.\,\ref{fig:dnucorr} we 
see that $\dnuo$ departures from the  pure $\dnuscl$ scaling
relation can vary between approximately $-3$\% and 2\% depending on the stellar
mass, $\teff$ and $\mh$. In later sections 
we discuss in detail the impact of these departures in the determination of 
fundamental stellar parameters.

For this  catalog, determination of stellar  parameters is based upon  GBM. By
construction,  unlike  the plain  use  of  scaling relations,  GBM  introduces
dependencies on stellar  models and also on the  statistical approach employed
to characterize  stellar quantities of interest.  In a similar manner  as done
with asteroseismic  pipelines, we use  a number  of GBM pipelines  that employ
different stellar models  and/or statistical methods. In this way,  we aim not
only  at determining  stellar  parameters  but also  at  obtaining a  sensible
quantification of  the systematic  uncertainties involved.  In total,  we have
employed seven different GBM pipelines but some  of them have been run in more
than  one  \emph{mode},  i.e.  with   different  sets  of  stellar  models  or
assumptions  regarding calculation  of $\dnu$,  leading  to a  grand total  of
twelve different sets of GBM results.

Here  we  include a  summary  of  the  characteristics  of the  different  GBM
calculations,  sorted  by pipelines.  A list of the
combinations of grids of stellar models,  GBM pipelines, and choice of $\dnuo$
or  $\dnuscl$   used  in   this  work   is  given   for  quick   reference  in
Table~\ref{tab:gbms}. A  description  of the  physical  inputs
adopted  in   each  of   the  grids   of  stellar   models  is   presented  in
Appendix~\ref{app:models}. Additional  details about the numerical  methods of
different pipelines are given in the  respective references.

\texttt{BASTA:}   For    this   work,    the   BAyesian    STellar   Algorithm
\citep{victor:2015}  uses   stellar  models  computed   with  \texttt{GARSTEC}
\citep{weiss:2008}.  The  mass and  $\feh$  ranges  covered  by the  grid  are
M/$\msun =\left\{0.7\,,1.80\right\}$ with $\Delta M = 0.01\,\msun$ and $\feh =
\left\{-0.65\,,+0.50\right\}$~dex    with    $\Delta    \feh$    =    0.05~dex
respectively. Determination of stellar parameters is performed with an adapted
version       of      the       Bayesian      approach       described      in
\citet{serenelli:2013}.  \texttt{BASTA} has  been  run both  with $\dnuo$  and
$\dnuscl$,     that    we     identify    as     \texttt{BAS/G-$\dnuo$}    and
\texttt{BAS/G-$\dnuscl$}.  For  each quantity,  its central  value and  $\pm 1
\sigma$ limits correspond to the median and the $\pm 34.1\%$ limits around the
median respectively.

\texttt{BeSPP}: The Bellaterra Stellar Parameters Pipeline uses stellar models
computed with \texttt{GARSTEC} but with some differences with respect to those
used     in     \texttt{BASTA}.     The      grid     covers     the     range
M/$\msun$=$\left\{0.6\,,3.0\right\}$ with a step  $\Delta {\rm M}/\msun= 0.02$
and  \feh=$\left\{-3.0\,,+0.6\right\}$~dex  with  $\Delta \feh=  0.1$~dex  for
$\feh$ between  $-$3.0 and  0.0~dex and  $\Delta \feh=  0.05$~dex for  $\feh >
0.0$~dex.  The determination  of  stellar parameters  is done  by  means of  a
Bayesian  approach based  on the  method described  in \citet{serenelli:2013},
extended to  include asteroseismic  inputs. \texttt{BeSPP}  has been  run both
with      $\dnuo$      and     $\dnuscl$:      \texttt{BeS/G-$\dnuo$}      and
\texttt{BeS/G-$\dnuscl$}.  For  each quantity,  its central  value and  $\pm 1
\sigma$ limits correspond to the median and the $\pm 34.1\%$ limits around the
median respectively.

 \texttt{GOE}: Two grids  of stellar models have been used,  one computed with
 \texttt{CESTAM2k} \citep{marques:2013}  and the other one  with \texttt{MESA}
 \citep{paxton:2013}.     Both      grids     span     the      mass     range
 M/$\msun$=$\left\{0.6\,,2.0\right\}$  with  $\Delta  M  =  0.02\,\msun$.  The
 metallicity range  Z$=\left\{0.003\,,0.040\right\}$ with a 0.003  step, which
 is  roughly  equivalent  to  $\feh  =  \left\{-1.3\,,+0.40\right\}$~dex.  The
 \texttt{CESTAM2k}  grid has  been  run  with $\dnuo$  and  $\dnuscl$ and  the
 \texttt{MESA}    grid    just   with    $\dnuscl$:    \texttt{GOE/C-$\dnuo$},
 \texttt{GOE/C-$\dnuscl$}, and \texttt{GOE/M-$\dnuscl$}.  The determination of
 stellar parameters is based on an independent implementation of \texttt{SEEK}
 \citep{quirion:2010}, but  adopting Bayesian priors  only to account  for the
 inhomogeneity    sampling     of    the     grids    of     stellar    models
 \citep{hekker:2014}. For each quantity, its  central value and $\pm 1 \sigma$
 limits correspond to the median and the $\pm 34.1\%$ limits around the median
 respectively.
 
\texttt{MPS}: This pipeline  is based on an independent  implementation of the
likelihood method described in \citet{basu:2010}. It is based on a Monte Carlo
method in which observed parameters are sampled and, for each realization, the
likelihood  of all  models in  the grid  is computed.  All of  this likelihood
distributions are then  combined to obtain the  global likelihood distribution
from   which    stellar   parameters   and   uncertainties    are   determined
\citep{hekker:2014}. Stellar models are the canonical models (no overshooting)
of  \texttt{BaSTI} \citep{pietrinferni:2004},  span the  mass and  metallicity
ranges M/$\msun$=$\left\{0.5\,,4.5\right\}$ with $\Delta  M = 0.05\,\msun$ and
Z$=\left\{0.0001\,,0.040\right\}$,  and  use  $\dnuscl$.  We  identify  it  as
\texttt{MPS/B-$\dnuscl$}.
 
\texttt{RADIUS}:     Stellar     models     are    computed     with     ASTEC
\citep{astec:2008}. The grid  covers M/$\msun$=$\left\{0.5\,,4.0\right\}$ with
a        step       $\Delta        {\rm       M}/\msun=        0.01$       and
\feh=$\left\{-1.27\,,+0.47\right\}$~dex  with   $\Delta  \feh=   0.1$~dex  and
employs  $\dnuscl$. We  name  it \texttt{RAD/A-$\dnuscl$}.  Central values  of
stellar  parameters  are  determined  from  the best  fitting  model  and  the
1$\sigma$ uncertainties are estimated by  considering the full range of values
of a given parameter  spanned by the models that are  within 3~$\sigma$ of all
the input parameters, and assuming it  represents a $\pm 3\sigma$ range of the
output parameters \citep{stello:2009b}.

\texttt{SFP}: Seismic  Fundamental Parameters. Stellar  models in SFP  are the
\texttt{BaSTI} canonical models \citep{pietrinferni:2004}. The grid covers the
range                 M/$\msun$=$\left\{0.7\,,4.5\right\}$                 and
Z$=\left\{0.001\,,0.040\right\}$  and  it has  been  interpolated  to offer  a
denser representation  of the parameter  space so that $\Delta  {\rm M}/\msun=
0.02$    and    $\Delta    \rm{Z}=    0.001$.    In    this    implementation,
\texttt{SFP/B-$\dnuscl$}, $\dnuscl$ is used,  and the determination of stellar
parameters   is   done    with   a   Bayesian   method,    as   described   in
\citet{kallinger:2010}  and \citet{kallinger:2012}.  The central  value quoted
for a  given stellar quantity is  the most probable value  in the distribution
and the $\pm 1 \sigma$ values correspond to the limits containing $\pm 34.1\%$
of the probability distribution around the most probable value. 

\texttt{YB}: The Yale-Birmingham  pipeline \citep{basu:2010,gai:2011} has been
run with  two different grids  of stellar models.  One grid has  been computed
with  \texttt{YREC2} and  covers  M/$\msun$=$\left\{0.8\,,3.0\right\}$ with  a
step  $\Delta {\rm  M}/\msun= 0.02$  and \feh=$\left\{-0.6\,,+0.6\right\}$~dex
with $\Delta \feh= 0.05$~dex. The other  one is formed by the canonical models
from \texttt{BaSTI}  \citep{pietrinferni:2004} (see \texttt{MPS}  above). Both
grids  use $\dnuscl$  and  we identify  them  as \texttt{YB/YR-$\dnuscl$}  and
\texttt{YB/B-$\dnuscl$}.  Stellar parameters  are determined  through a  Monte
Carlo approach, where input data is used to generate large Gaussian samples of
input parameters. The stellar parameters  of the model with highest likelihood
is assigned to each  realization of the sample. These models  are then used to
build the probability  distribution. For each quantity, its  central value and
$\pm 1  \sigma$ limits correspond  to the median  and the $\pm  34.1\%$ limits
around the median respectively.

\floattable
\begin{deluxetable}{lccl|lccl}
\tabletypesize{\footnotesize}
\tablewidth{8cm}
\tablecaption{Summary of GBM\label{tab:gbms}}
\tablehead{GBM & Stellar code & Mode & \multicolumn{1}{c}{Short name} & GBM & Stellar code & Mode & \multicolumn{1}{c}{Short name}}
\startdata
\texttt{BASTA} & \texttt{GARSTEC} & $\dnuscl$ & \texttt{BAS/G-$\dnuscl$} & \texttt{GOE} & \texttt{MESA} & $\dnuscl$ & \texttt{GOE/M-$\dnuscl$} \\ 
\texttt{BASTA} & \texttt{GARSTEC} & $\dnuo$ & \texttt{BAS/G-$\dnuo$} &\texttt{MPS} & \texttt{BASTI} & $\dnuscl$ & \texttt{MPS/B-$\dnuscl$} \\ 
\texttt{BeSPP} & \texttt{GARSTEC} & $\dnuscl$ & \texttt{BeS/G-$\dnuscl$} & \texttt{RADIUS} & \texttt{ASTEC} & $\dnuscl$ & \texttt{RAD/A-$\dnuscl$} \\ 
\texttt{BeSPP} & \texttt{GARSTEC} & $\dnuo$ & \texttt{BeS/G-$\dnuo$} & \texttt{SFP} & \texttt{BASTI} & $\dnuscl$ & \texttt{SFP/B-$\dnuscl$} \\ 
\texttt{GOE}   & \texttt{CESTAM2k} & $\dnuscl$ & \texttt{GOE/C-$\dnuscl$} & \texttt{YB} & \texttt{YREC2} & $\dnuscl$ & \texttt{YB/YR-$\dnuscl$} \\ 
\texttt{GOE}   & \texttt{CESTAM2k} & $\dnuo$ & \texttt{GOE/C-$\dnuo$} & \texttt{YB} & \texttt{BASTI} & $\dnuscl$ & \texttt{YB/B-$\dnuscl$} \\ 
\enddata
\tablecomments{$\dnuscl$ stands for $\dnu$ as obtained directly from Eq.\,(\ref{eq:dnu}) and $\dnuo$ for $\dnu$ obtained from the slope of a linear fit to radial model frequencies (Sect.\,\ref{sec:gbm}).}
\end{deluxetable}

\section{Determination of stellar parameters and uncertainties} \label{sec:core}

\subsection{Choice of central values}  \label{sec:central}

The combination of  spectroscopic and asteroseismic data is used  in this work
to produce a catalog that includes determinations of the following quantities:
$g$, $\mrho$,  $M$, $R$,  and $\tau$.  The central  values in  the catalog
correspond  to  the  determination  of  these  quantities  obtained  with  the
combination \texttt{BeSPP,GARSTEC,$\dnuo$},  or \gbmr\ for short.  This choice
is based on the following reasons. 

The grid of stellar models used in \texttt{BeSPP} is the only one in this work
that includes microscopic diffusion (see Appendix~\ref{app:models} for details
on the  implementation). This  is a relevant  physical process  for solar-like
stars. In  the particular case of  the Sun, it  is necessary to include  it in
order  to  obtain  a  good   agreement  with  helioseismic  measurements  (see
\citealt{bahcall:1995,jcd:1996}  among many  others).  Additional evidence  of
microscopic diffusion  can be  found in high  precision spectroscopic  work on
clusters that show depletion of metals of up to 0.2~dex in the very metal poor
clusters  NGC6397 and  M30  \citep{nordlander:2012,gruyters:2016}  and a  more
subtle depletion,  below a 0.1~dex  level, in the higher  metallicity clusters
NGC 6752 and M67 \citep{gruyters:2014,onehag:2014}. 

\begin{figure}[!ht]
\includegraphics[scale=.42]{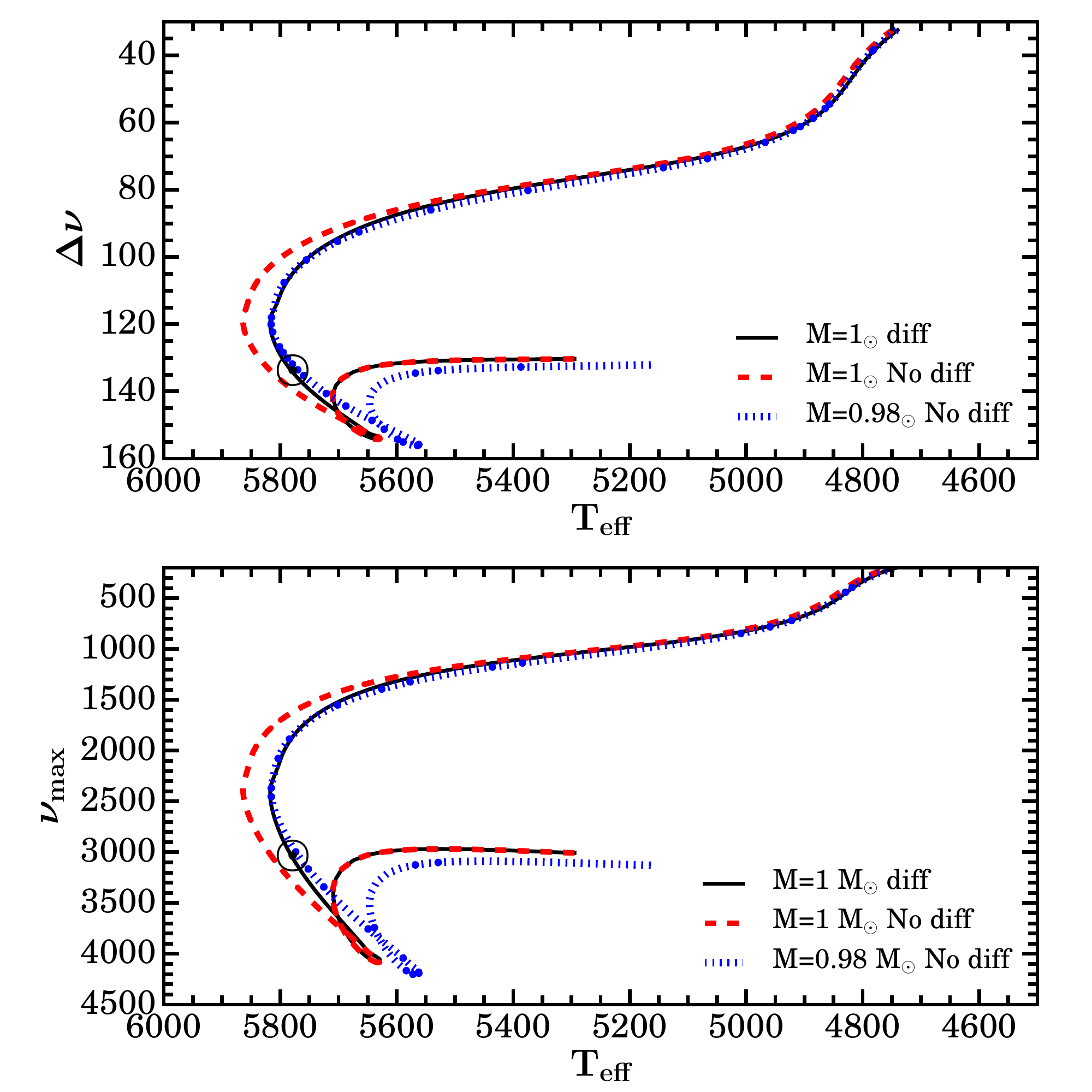}
\caption{Comparison  of 1~$\msun$  evolutionary  tracks using  the same  solar
  calibrated  mixing length  and same  initial composition,  with and  without
  microscopic diffusion. The  top (bottom) panel shows $\dnu$  ($\numax$) as a
  function  of  $\teff$.  For  given  observables  around  the  turn-off,  the
  difference in mass is about 0.02~$\msun$, as shown by the 0.98~$\msun$ track
  without diffusion. Circle denotes the position of the Sun. \label{fig:solar}}
\end{figure}

An  additional reason  is that  the solar  calibrations performed  to fix  the
mixing length  parameter $\alpha_c$  are typically done  including microscopic
diffusion. But  then, when the grids  are computed neglecting this  process, a
mismatch is  produced between the  calibrated $\teff$  scale set by  the solar
calibrated $\alpha_c$ and  the actual $\teff$ scale in the  grid. In practice,
this implies  that grids  computed in  this way  do not  contain a  model that
reproduces the basic solar properties, i.e. solar radius and luminosity at the
present-day solar  age.  To illustrate this,  Figure~\ref{fig:solar} shows, in
the  $\teff-\dnu$  and $\teff-\numax$  planes,  the  evolutionary track  of  a
1~M$_\odot$ solar calibrated track including  diffusion (solid black line) and
the corresponding track using the same calibrated mixing length $\alpha_c$ but
without diffusion (dashed  red line).  It is apparent from  these figures that
solar seismic properties  cannot be reproduced by the  evolutionary model that
does  not include  diffusion. The  mismatch corresponds,  approximately, to  a
0.02~M$_\odot$ difference, as  seen by the overlap  between the 0.98~M$_\odot$
evolutionary track  without microscopic diffusion  (blue dotted line)  and the
solar calibrated  track.  This  might not seem  as a large  effect, but  as we
discuss in later sections, this is  a sizeable number compared to typical mass
uncertainties determined from asteroseismic  data. The mismatch between tracks
with or  without diffusion  depends on  the evolutionary  stage; it  builds up
starting  from  the ZAMS,  it  is  maximum at  the  turn  off, and  it  almost
completely vanishes as  stars evolve off the main sequence  and the convective
envelope deepens, restoring the initial surface composition.  

The final  reason for  using $\gbmr$  results as the  reference values  in the
catalog  relates to  the  calculation of  $\dnuo$.  There  are  three sets  of
results  based  on  GBMs   that  rely  upon  $\dnuo$,  \texttt{BAS/G-$\dnuo$},
\texttt{BeS/G-$\dnuo$}, and  \texttt{GOE/C-$\dnuo$}. The advantage  of relying
on $\dnuo$ is that it captures  deviations from the pure scaling $\dnuscl$ due
to       detailed      structure       of       stellar      models       (see
e.g.  \citealt{belkacem:2013}).   But,  its  determination   from  theoretical
frequencies  is affected  by  poor  modeling of  stellar  atmospheres and  the
neglect  of   non-adiabatic  effects   in  the   outermost  layers   of  stars
\citep{rosenthal:1999}. In the \gbmr\ grid  for example, the solar model gives
$\dnuo=136.3\,\mu$Hz,   about  1\%   larger  than   the  reference   $\dnusun=
135.1\pm0.1\,\mu$Hz determined  from the  \texttt{SYD} seismic  pipeline. This
difference implies that grids relying on $\dnuo$ will not be able to reproduce
a solar  model unless they  are rescaled so that  $\dnuo$ for the  solar model
matched  $\dnusun$.  Such  correction  is  applied in  \gbmr\  by computing  a
calibration factor  $f_{\dnu}= \dnusun/\dnu_{0,\rm  SM}$ (here SM  means solar
model) and  rescaling $\dnuo$ in the  whole grid according to  it. Other grids
based on $\dnuo$ do not correct for this effect and therefore do not reproduce
solar  properties  when fed  with  solar  seismic quantities.   Applying  this
correction factor  to the whole  grid of  stellar models carries  the implicit
assumption that the  surface effect produces a fractional  variation in $\dnu$
that is constant for all stars, analogous to using a fixed $\alpha_c$ based on
a solar calibration. Work is  ongoing in developing surface effect corrections
(e.g. \citealt{sonoi:2015,ball:2014}) and more sophisticated model atmospheres
to have more realistic calculations of  frequencies and, in the context of the
present  work, corrections  to  $\dnuo$  that do  not  rely  on solar  scaling
\citep{ball:2016,trampedach:2017}.  

\begin{figure*}[ht!]
\centering
\includegraphics[scale=.38]{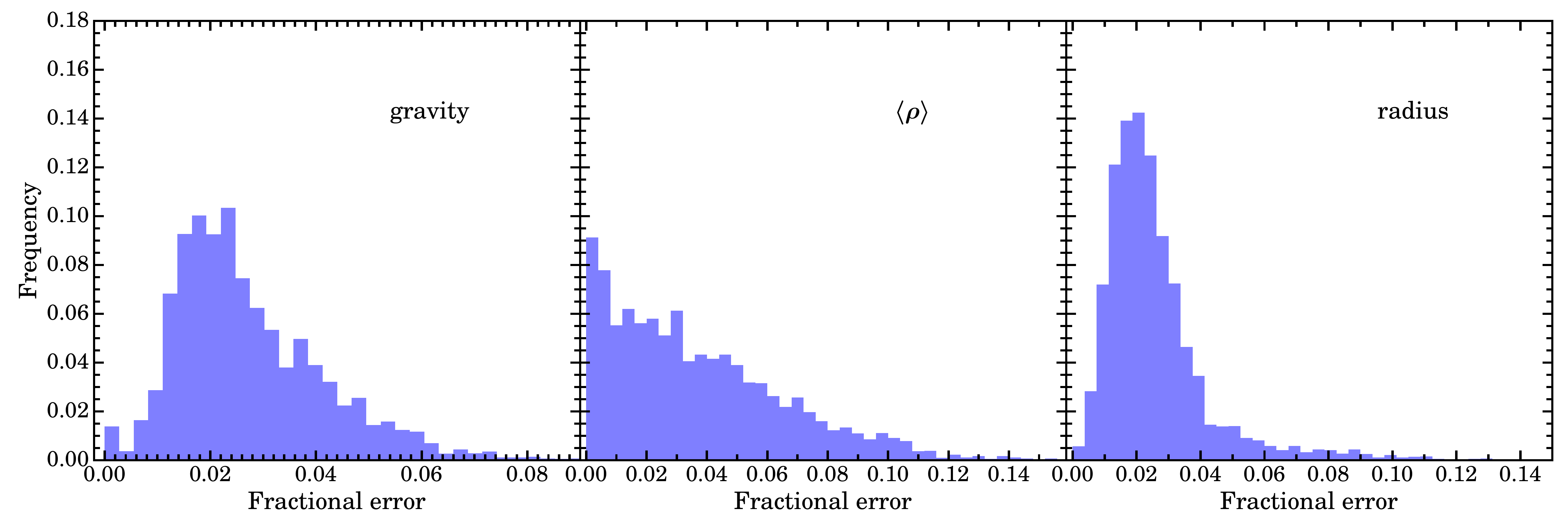}
\includegraphics[scale=0.38]{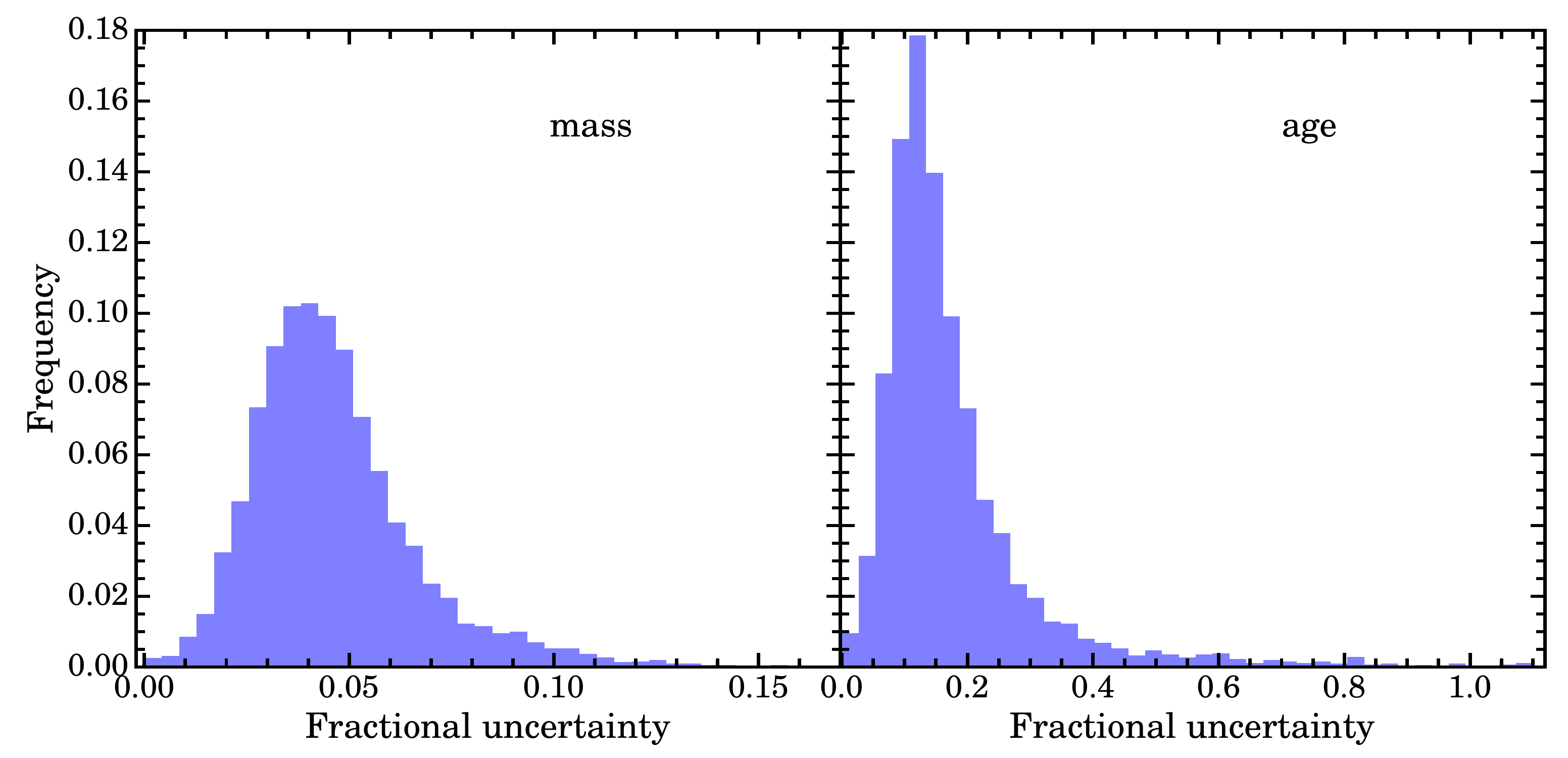}
\caption{Statistical errors for all stars and all pipelines (see Table~\ref{tab:gbms} 
for a full list of pipelines). \label{fig:stat}}
\end{figure*}

The underlying differences  between \gbmr\ used to provide  the central values
in  the catalog  and other  GBMs are  in fact  a positive  aspect in  our work
because they allow a more robust determination of systematic uncertainties, as
discussed later in Section~\ref{sec:systgbm}.

\subsection{Statistical uncertainties}  \label{sec:stat}

The  adopted  formal uncertainties  from  ASPCAP  are  69~K for  $\teff$  and,
typically, 0.10~dex  for $\mh$ (Sect.\,\ref{sec:spectro}). In  the SDSS scale,
$\teff$ errors are characterized by a mean of 67~K, very similar to the ASPCAP
formal error, with  a dispersion of 25~K. Metallicities are  those from ASPCAP
so their uncertainties are treated in the way described above.

Each GBM  pipeline returns central  values and  uncertainties for each  of the
stellar parameters included in the  catalog, i.e. $M$, $R$, $\mrho$, $g$
and  $\tau$. The  uncertainty returned  by  each pipeline  is the  statistical
uncertainty associated  with each parameter,  a measure of the  precision with
which stellar  parameters can be  determined from the  available spectroscopic
and  asteroseismic  data. For  each  stellar  parameter, the  distribution  of
statistical   uncertainties   from  all   twelve   GBM   sets  is   shown   in
Fig.\,\ref{fig:stat}.  This figure  shows  results obtained  using the  ASPCAP
spectroscopic parameters  but, due  to the  similarity in  $\teff$ statistical
uncertainties between the ASPCAP and  the SDSS temperature scales, results and
the discussion that follows are very similar for both scales. 

\begin{figure}
\centering
\includegraphics[scale=.42]{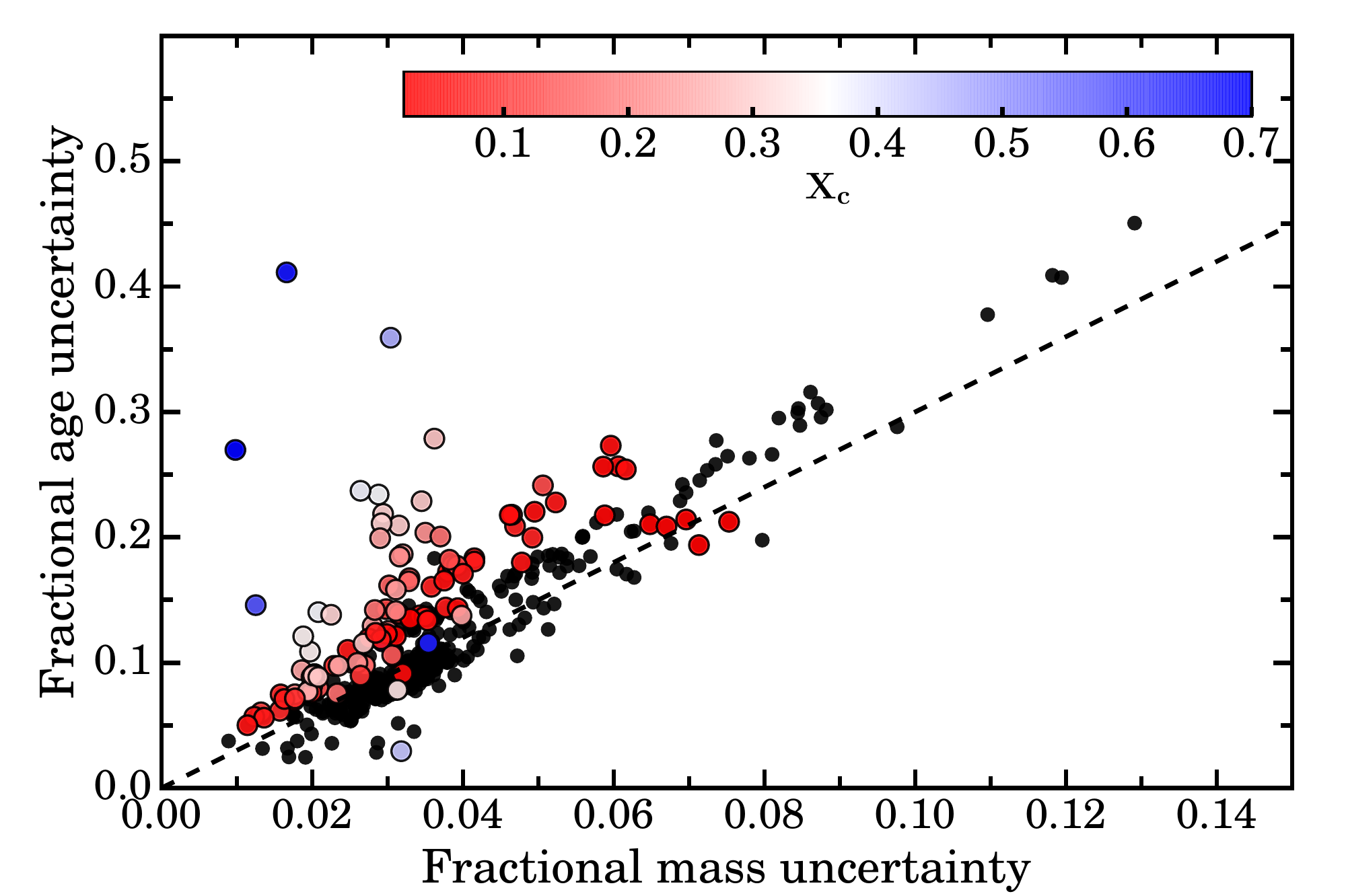}
\caption{Relation between mass and  age statistical uncertainties. Color coded
  is the central  hydrogen mass fraction $X_c$. Stars with  $X_{\rm c} < 0.02$
  are shown  as black symbols. Dashed  line represents a 3:1  relation between
  age and mass uncertainties.\label{fig:maxc}} 
\end{figure}

Interpreting the distribution of statistical uncertainties for $g$ and $\mrho$
is straight  forward from the scaling  relations. For $g$, the  dominant error
source comes from  the linear dependence of $g$ with  $\numax$. The additional
dependence  on $\sqrt{\teff}$  is  small because  $\delta \teff/\teff$  ranges
between 1  and 1.5\% for  the whole  sample, so this  propagates at most  as a
0.7\%  ($  <  0.01$~dex)  uncertainty  into $g$.  For  $\mrho$,  the  dominant
uncertainty source is  $\dnu$, augmented by the corresponding factor  of 2. To
first order,  there is no explicit  dependence of $\mrho$ on  other quantities
than  $\dnu$, and  given that  the $\dnu$  uncertainty distributions  does not
peak, neither does the distribution of $\mrho$ statistical uncertainties. 

For radius, this  discussion is more interesting  and it is a  good example of
the advantages of using GBM to determine stellar parameters instead of relying
simply  on  the scaling  relations.  Let  us  consider the  median  fractional
uncertainties  of $\numax$  and $\dnu$  shown in  Fig.\,\ref{fig:pipes} as  an
example. These are  4\% and 1.7\% respectively, and assume  a 1.3\% fractional
uncertainty  for   $\teff$  (corresponding   to  a  typical   $\teff=  5400$~K
value).  Using Eq.\,\ref{eq:radius}  to propagate  these errors  we obtain  an
estimate  of  the  median  fractional   uncertainty  of  $\delta  R/R  \approx
5\%$. However,  the distribution in Fig.\,\ref{fig:stat}  shows a distribution
that peaks around 2\% and, in fact, 89\% of the GBM results have $\delta R/R <
4.5\%$. The reason  for this apparent discrepancy is that  $\numax$ and $\dnu$
are  not independent  quantities in  stars, as  they both  depend on  the same
intrinsic  quantities. Stellar  evolution models  used in  GBM incorporate  by
construction their  correlation. Based on stellar  models, \citet{stello:2009}
have  found the  simple relation  $\dnu \propto  \numax^{0.77}$ between  these
quantities. Using this  relation, we can go back  to Eq.\,\ref{eq:radius} and,
because typically  $\numax$ has a  larger fractional uncertainty  than $\dnu$,
replace $\dnu$ by  its dependence on $\numax$ to find  an approximate relation
$R  \propto  (\teff/\numax)^{1/2}$.  Now,  using the  median  uncertainty  for
$\numax$ and 1.3\% for $\teff$ as before we obtain $\delta R/R \approx 2.2\%$,
well in agreement  with the distribution seen  in Fig.\,\ref{fig:stat}. Linear
propagation of errors from Eq.~\ref{eq:radius} without accounting for physical
correlation  between  $\numax$  and  $\dnu$  leads  to  an  overestimation  of
uncertainties  in the  inferred  stellar  parameters. \cite{gai:2011}  already
discussed  that  GBM  leads  to  smaller errors  than  direct  application  of
Eq.~\ref{eq:radius}, but  the simple explanation related  to the $\numax-\dnu$
relation being imprinted in stellar models was not discussed in that work. 

There  is an  analogous  discussion  for the  statistical  uncertainty in  the
determination  of  stellar   masses.   Use  of  scaling   relations  leads  to
overestimation of the  uncertainty because there is no  information on stellar
structure  and evolution  in those  relations. Using  median uncertainties  as
above, we derive an expected $\delta  M/M \approx 14\%$, more than three times
the median  value in  the distribution of  statistical uncertainties  shown in
Fig.\,\ref{fig:stat}. Again,  accounting for the correlation  between $\numax$
and $\dnu$  leads to a  much smaller  value of $\sim  2\%$. Note that  in this
§estimate  the fractional  uncertainty  of  $\dnu$ is  assumed  to  be that  of
$\numax$ scaled by the factor 0.77. It is a crude estimate, but good enough to
guide interpretation of uncertainties  in mass determinations. The uncertainty
distribution of masses shown in Fig.\,\ref{fig:stat} shows, however, a maximum
around 3.5-4\%,  larger than  the above  estimate. This is  mostly due  to the
effect of  the metallicity  uncertainty, which  is not  taken into  account in
scaling relations. In fact, there is a correlation between metallicity
  and seismic  mass determinations  such that a  0.1~dex uncertainty  in $\mh$
  corresponds to about 2.5-3\%  uncertainty in mass (Sect.~\ref{sec:chap14}).
This  uncertainty,  combined with  the  2\%  mass uncertainty  estimated  from
seismic  quantities and  $\teff$ gives  a  good understanding  of the  typical
statistical mass uncertainty we have found in our GBM results.

The discussion  above is simplified, for  example in that the  role of stellar
metallicity is  not accounted  for, but it  shows that using  GBM is  not only
qualitatively better than employing  scaling relations because it incorporates
knowledge on stellar structure, but in  fact leads to more precise estimations
of   stellar  parameters   because  models   naturally  incorporate   existing
correlations between physical quantities. Typically,  we can conclude that GBM
leads to  a precision  in mass and  radius that  is at least  a factor  of two
better than  pure scaling relations. Moreover,  it is so due  to good physical
reasons.

Finally, interpretation  of statistical uncertainties of  stellar ages depends
on the  evolutionary state of the  star. The total evolutionary  lifetime of a
star on the main sequence scales approximately with its mass as $\tau_{\rm MS}
\propto  M^{-\alpha}$, with  $\alpha  \approx  3 -  3.5$  for  low mass  stars
\citep{serenelli:2007}.  Therefore, for  stars at  the  end or  past the  main
sequence, it  is expected that the  age uncertainty is roughly  $\alpha$ times
the mass uncertainty. Fig.\,\ref{fig:maxc} shows the relation between mass and
age uncertainties obtained  with \texttt{BeSPP}, and color  coded according to
the central hydrogen  abundance, a measure of the evolutionary  status of each
star. For  stars that have  depleted hydrogen  (black symbols), which  in fact
comprise most of the sample, the linear relation is clearly visible. For stars
that are still on  the main sequence, the correlation between  mass and age is
naturally lost because the stellar mass does not determine its current age. In
practice, results shown in Fig.\,\ref{fig:maxc}  allow to establish, as a rule
of thumb,  that a lower  bound for the  fractional age uncertainty  is $\delta
\tau / \tau \gtrsim 3 \cdot \delta M / M$ (see also \citealt{davies:2016c}).

\subsection{Systematic uncertainties}\label{sec:syst}

\subsubsection{GBM peculiarities}\label{sec:systgbm}

The  second  major  source  of  uncertainties   has  its  origin  in  the  GBM
pipelines.  Several  aspects need  to  be  considered: the  stellar  evolution
tracks,  the  statistical  method  to  extract  stellar  parameters,  and  the
calculation of $\dnu$ in evolutionary tracks.

As  described in  Sect.\,\ref{sec:gbm}, different  pipelines typically  employ
different grids  of stellar  models. Stellar evolution  theory still  has some
uncertainty; there is not a unique  choice of what different modelers consider
is the \emph{best  physics} that should be employed in  computation of stellar
models. In the realm of low mass stars, different possibilities are available,
among  others, for  nuclear reaction  rates, radiative  opacities (atomic  and
molecular), treatment  of microscopic diffusion, implementation  and amount of
overshooting  among other  physical inputs  or processes.  Moreover, different
numerical  implementations of  similar physics  and numerical  aspects of  the
integration of the equations of stellar  structure and evolution can also lead
to differences in the computed stellar models. Detailed comparisons of stellar
models for dwarf stars using different  physics but the same code or different
codes  but the  same underlying  physics, have  been object  of some  studies,
although                   of                  limited                   scope
(\citealt{lebreton:2008,marconi:2008,stancliffe:2016}                    among
others). Additionally, GBM pipelines can  implement differently a given set of
stellar tracks  (e.g. interpolating stellar tracks  to a finer grid)  and also
employ  different statistical  approaches  to extract  stellar parameters  and
their uncertainties so  even using the same underlying grid  of stellar models
does not imply that different GBM pipelines will lead to the same results.

\begin{figure*}[ht!]
\centering \includegraphics[scale=.22]{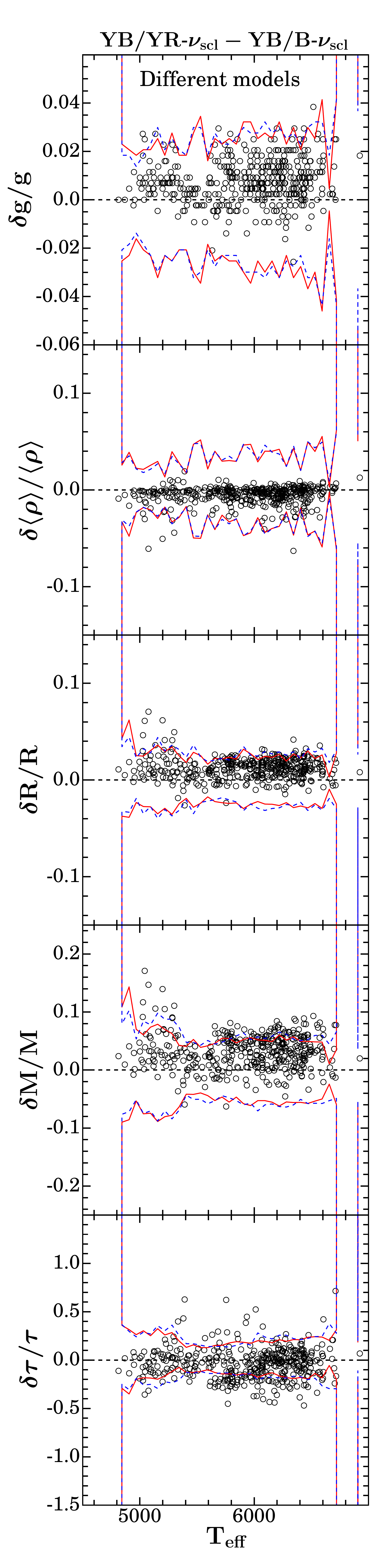}\includegraphics[scale=.22]{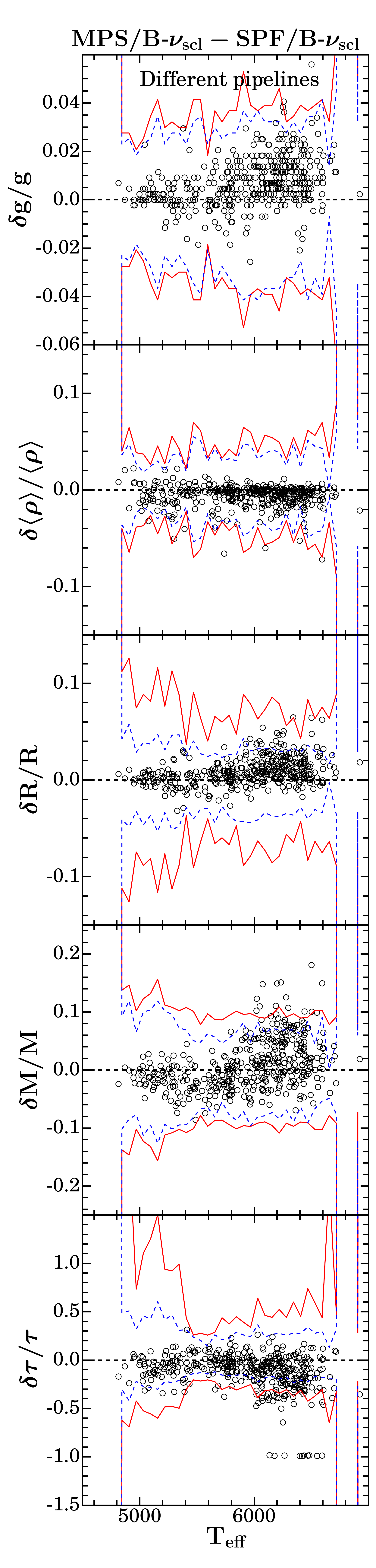}\includegraphics[scale=.22]{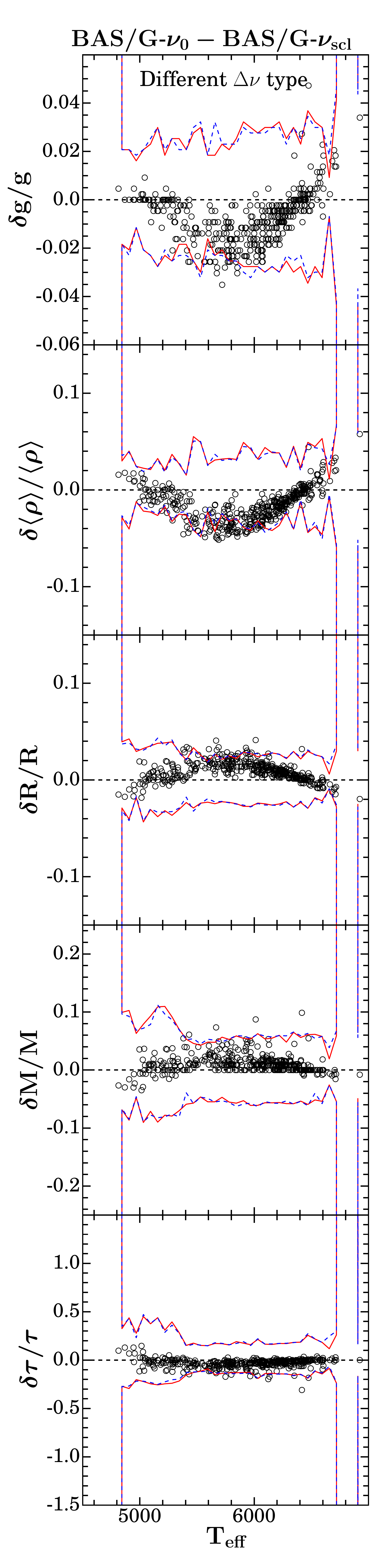}\includegraphics[scale=.22]{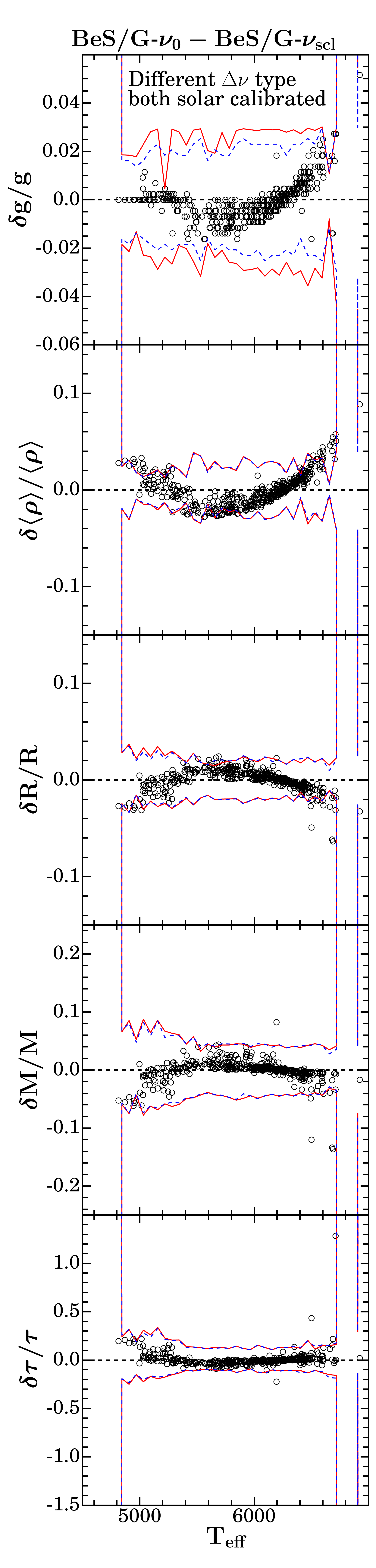}
\caption{Fractional  differences (except  for $\logg$,  given in  dex) in  the
  estimated stellar  parameters from  different GBM  pipelines. Pairs  of GBMs
  used in each  column and sense in which differences  are computed are listed
  at   the    top   using    the   form:   Pipeline/Grid-$\dnu$    type  (see
  Sect.~\ref{sec:gbm} and Table~\ref{tab:gbms} for  naming convention). 
  The  columns are from  left to
  right: (1)  Same pipeline but  different models, (2) different  pipeline but
  the same models, (3) same pipeline and models but different $\dnu$ type and,
  (4)  same pipeline  and  model  but different  $\dnu$  type  both scaled  to
  solar.  Red  solid and  blue  dashed  lines show  statistical  uncertainties
  returned  by  the  first  and  second pipelines  respectively,  in  bins  of
  60~K. \label{fig:gbms_main}}
\end{figure*}

C14 has  considered this  problem and performed  several comparisons  aimed at
disentangling  the effects  of statistical  methods and  the use  of different
grids of stellar models. Here, we have also carried out such comparisons based
on   results    from   the    twelve   GBM    sets   of    results   available
(Table~\ref{tab:gbms}). Extensive GBM one-to-one  comparisons are presented in
Appendix~\ref{app:gbm}.  In  this section,  we  use  a  subset of  results  to
illustrate the  typical systematic  uncertainties arising from  four different
sources:  different  evolutionary  tracks, different  statistical  methods  to
determine stellar parameters,  the use of $\dnuo$ or $\dnuscl$,  and the solar
calibration of $\dnuo$ (Sect.\,\ref{sec:central}).

The typical  relevance of each  uncertainty source  is shown, for  all stellar
parameters,  in Figure~\ref{fig:gbms_main}  where each  column of  plots shows
relative  differences  in  the  central   values  of  the  stellar  parameters
determined with  two different GBM runs  that differ from one  another in just
one aspect. Differences  are taken in the  sense indicated at the  top of each
column. In  each plot,  red solid  and blue dashed  lines are  the statistical
uncertainties returned by the first  and second pipelines respectively, binned
in 60~K intervals.  Note that it is not within the scope of this paper to look
into detail on the origin of the  differences among GBM results, but rather to
present  typical cases  and  point  out some  possible  causes for  systematic
differences.

The  first  column  of  plots in  Fig.\,\ref{fig:gbms_main}  compares  results
obtained with the \texttt{YB} pipeline but with different evolutionary tracks,
\texttt{YREC2}  or  \texttt{BaSTI}.  For  $g$, $\mrho$,  $R$,  and  $M$  a
systematic offset is  seen between the sets of results.  A possible reason for
systematic offset  could be that  $\teff$ scales  or even luminosities  in the
models are  somewhat different. A  mismatch in  the $\teff$ scale  is possible
even if  both sets of tracks  are based on  a solar calibration of  the mixing
length parameter,  especially if  the solar  calibration and  the evolutionary
tracks have  been computed  with different physical  assumptions. This  is the
case of the  \texttt{BaSTI} tracks, that include microscopic  diffusion in the
solar   calibration  but   do   not   in  the   library   of  stellar   models
\citealp{pietrinferni:2004}, which leads  to a $\teff$ offset  as discussed in
Sect.\,\ref{sec:central}. Another source of differences might be the choice of
the critical ${\rm  ^{14}N(p,\gamma)^{15}O}$ rate, which is about  a factor of
two    smaller    in    \texttt{YREC2}    than    in    \texttt{BaSTI}    (see
Table~\ref{tab:grids1} and \ref{tab:grids2} in Appendix~\ref{app:models} for a
description  of  input  physics  used  in the  stellar  models  used  in  this
work). This rate affects  the $\teff$ scale of low mass  stars (typically $M <
1.3\msun$) close to  the turn-off, when the CNO-cycle takes  over pp-chains as
the dominant  H-burning process. It also  affects the whole main  sequence for
more massive  models and the luminosity  of the subgiant branch,  where models
with a larger rate are less luminous \citep{magic:2010}.

An   additional  difference   is  the   inclusion  (\texttt{YREC2}),   or  not
(\texttt{BaSTI} models  used in this  work), of overshooting, which  will also
affect the $\teff$ scale of stellar  models that have a convective core during
the  main sequence.  It  is also  interesting that  the  systematic offset  in
stellar  mass   is  not  translated  into   ages,  as  it  would   be  naively
expected. This  points towards \texttt{YREC2} predicting  longer main sequence
lifetimes than \texttt{BaSTI}, which is consistent (at least for stars with $M
\gtrsim 1.2\,\msun$ that have convective  cores during the main sequence) with
the fact  that \texttt{YREC2}  models include convective  overshooting whereas
the \texttt{BaSTI} models used here do not.  It is also possible that, even in
the absence of  convective cores during the main  sequence, \texttt{BaSTI} and
\texttt{YREC2} models do not predict the same evolutionary timescales at equal
stellar mass and composition.

Despite all the  above differences, it is reassuring that  in almost all cases
the  differences  are  smaller  or comparable  to  the  1$\sigma$  statistical
uncertainties returned  by the pipelines. The  exception is a small  number of
stars with  $\teff <  5400$\,K for  which $R$ and  $M$ have  larger fractional
differences.

The second  column compares results obtained  by two sets of  GBM results that
employ  the  same  underlying  set  of  \texttt{BaSTI}  tracks  but  that  are
implemented  differently. These  two GBMs,  \texttt{SFP/B} and  \texttt{MPS/B}
also determine  central parameters and  uncertainties by different  methods as
well. Here it  is very difficult to  single out reasons why  results from both
GBM pipelines show such dispersions. A clear difference is the mass resolution
because \texttt{SFP/B} uses an interpolated  grid with a finer mass resolution
than  the  original \texttt{BaSTI}  grid.  This  may  play an  important  role
especially when seismic uncertainties are small, which can lead to statistical
uncertainties well  below the  mass resolution  of the  grid. For  all stellar
quantities, the  dispersion of results  seen among pipelines sharing  the same
evolutionary tracks is  comparable to those seen in the  first column, showing
the  same  pipeline  with  different  stellar  models.  Again,  we  note  that
systematic differences  are almost  in all cases  not larger  than statistical
uncertainties returned by the pipelines.

The third column  compares results from the \texttt{BAS/G}  pipeline where the
only difference is  the use of $\dnuscl$  or $\dnuo$. This case  has also been
discussed in  the literature, as  described in Sect.\,\ref{sec:gbm},  but with
emphasis  in the  change in  $\mrho$, that  is most  directly affected  by the
choice  of  $\dnu$ in  the  stellar  models.  Clearly,  and as  expected,  the
systematic  effects follow  the inverted  shape of  the $\dnuo/\dnuscl$  ratio
(Fig.\,\ref{fig:dnucorr}) multiplied  by 2. This  clear trend with  $\teff$ is
also reflected in $R$  and $M$ estimates, and also in  $\logg$, which to first
order is  independent of  $\dnu$ according to  the scaling  relations. Stellar
ages are affected as well, but in  this case they reflect an almost one-to-one
correspondance with the mass variation, so changes are quite small.  Note that
$\dnuscl$ has been scaled such that the reference solar $\dnusun= 135.1 \mu$Hz
is  reproduced.  On the  other  hand,  the grid  using  $\dnuo$  has not  been
calibrated to match $\dnusun$.

Column    four   compares    results    from   \texttt{BeS/G}-$\dnuscl$    and
\texttt{BeS/G}-$\dnuo$ and in  this case both $\dnuscl$ and  $\dnuo$ have been
rescaled, as explained  in Sect.\,\ref{sec:central}, so that a  solar model is
reproduced properly.  The trends  seen in  the plots of  this column  are very
similar to those in  the previous column but with an  offset that reflects the
rescaling    of    $\dnuo$   by    the    factor    $f_{\dnu}$   defined    in
Sect.\,\ref{sec:central}.  It is  important  to point  out  that, despite  the
$f_{\dnu}$  correction  (or  calibration)   factor,  the  differences  between
\texttt{BeS/G}-$\dnuscl$ and \texttt{BeS/G}-$\dnuo$ are  not zero at the solar
$\teff$  because the  majority  of stars  in the  APOKASC  sample around  this
$\teff$ range  are past  the main  sequence or do  not have  a solar  mass and
composition   so   the   $\dnuo/\dnuscl$   is  different   from   solar   (see
Fig.\,\ref{fig:dnucorr}).

\subsubsection{Global GBM systematic uncertainties}

From the discussion  of the previous section it becomes  clear that a complete
determination of  systematic uncertainties  is not  possible. First,  it would
require surveying all physical inputs in stellar models (some of which are not
even properly modeled, such as overshooting or semiconvection) and quantifying
the resulting model uncertainties. Secondly,  it would be necessary to account
for differences in stellar properties due  to the statistical methods that can
be employed  in GBMs. Instead, we  follow the pragmatic approach  initiated in
C14  for  defining  and   computing  uncertainties  associated  to  systematic
differences arising from the use of various GBMs. The method is based on using
all available sets of GBM results described in Sect.\,\ref{sec:gbm} and relies
on  results returned  by GBMs  being robust  and that  a good  measure of  the
systematic uncertainty due to stellar  models and of the statistical inference
is given by the dispersion among all GBM results.

\begin{figure*}[ht!]
\centering
\includegraphics[scale=.38]{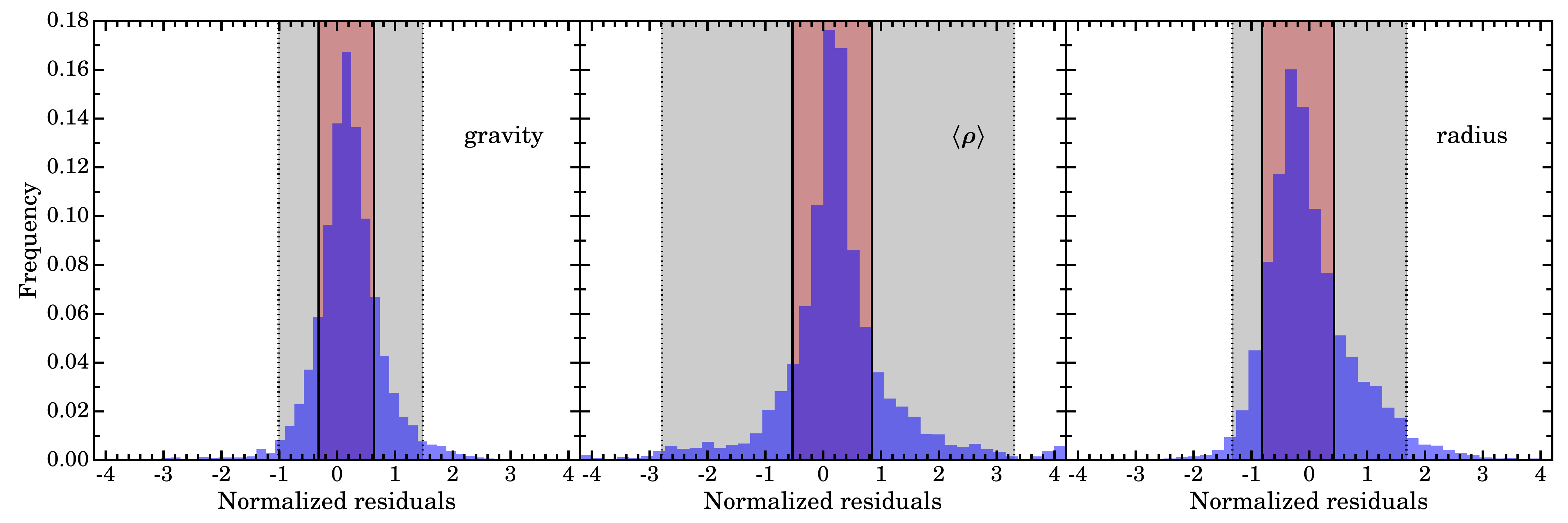}
\includegraphics[scale=0.38]{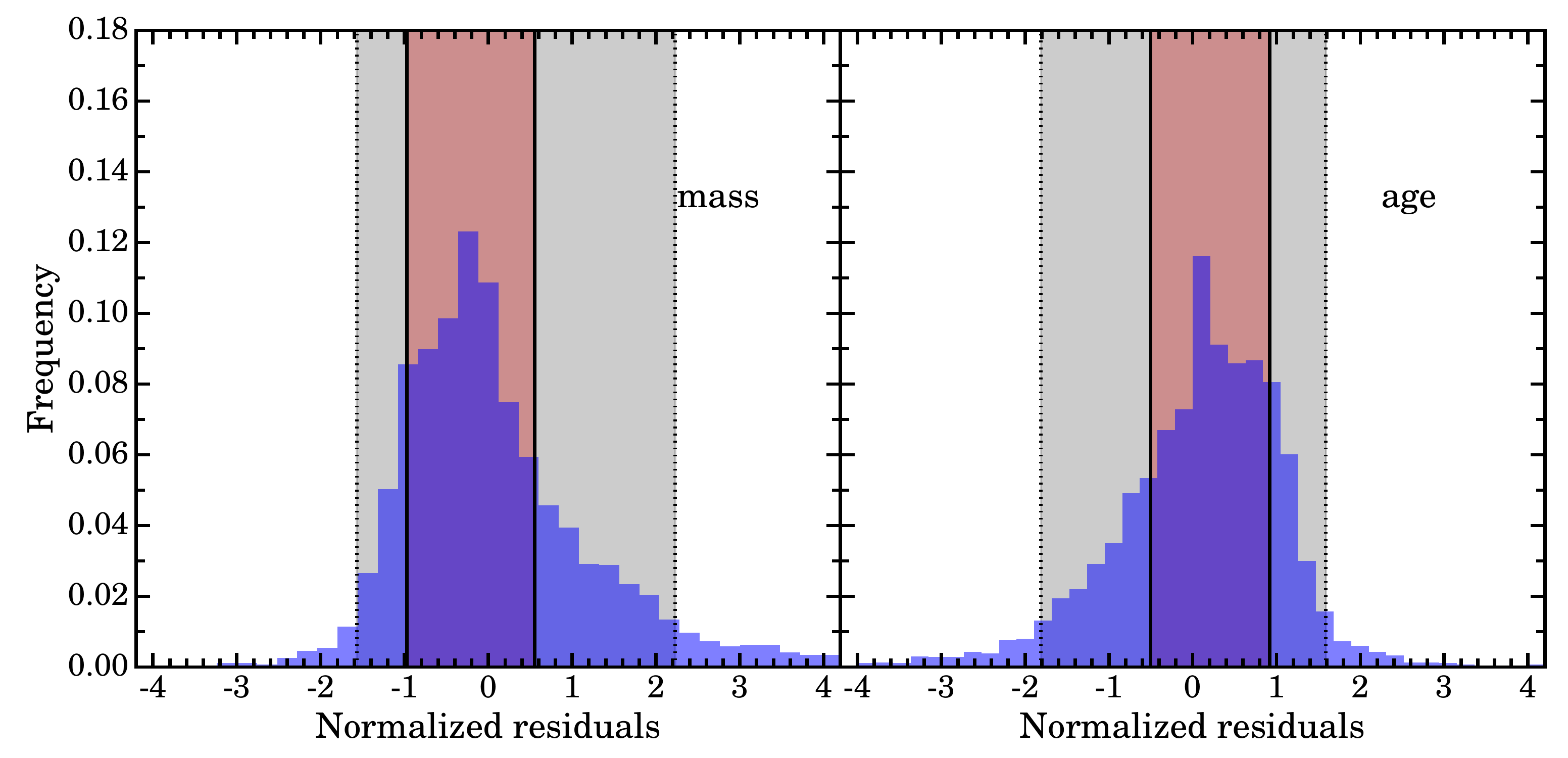}
\caption{Distribution of  normalized residuals  for all GBM  calculations. For
  each star and  property, each residual is computed as  the difference of the
  GBM  central  value minus  the  corresponding  value  in the  reference  GBM
  (\texttt{BeS/G}-$\dnuo$)  normalized by  the median  statistical uncertainty
  obtained from  the statistical uncertainties  returned by all  pipelines for
  the given  star and  property. Color  zones indicate  the 68.3\%  and 95.4\%
  confidence levels (computed over smoothed histograms). \label{fig:syst}}
\end{figure*}

A measure of the robustness of the results returned by the GBM can be obtained
by computing  the difference of the  central values returned by  each pipeline
with respect to reference values, which we  take here as those returned by the
combination   \texttt{BeSPP/GARSTEC}-$\dnuo$,  and   then  normalizing   these
differences  by the  median  of  the statistical  uncertainty  across all  GBM
pipelines. These normalized  residuals give an overall picture  of the scatter
across different sets of GBM results. The resulting distributions for $g$,
$\mrho$, $R$, $M$, and $\tau$ are shown in Fig.\,\ref{fig:syst}, including the
68.3\% and  95.4\% confidence levels  which, in  all cases except  for $\mrho$
correspond to narrower  than normal distributions of  standard deviation equal
to 1, i.e. when systematic and statistical uncertainties are equal.

\floattable
\begin{deluxetable}{cccc|ccc}
\tablecaption{Uncertainties\label{tab:mederrors}}
\tablehead{& \multicolumn{3}{c}{SDSS} & \multicolumn{3}{c}{ASPCAP} \\ \cline{2-7}
& Stat. & Syst. & Total & Stat. & Syst. & Total}
\startdata
$g$ & 0.025 & 0.009 & 0.028 & 0.025 & 0.016 & 0.030 \\
$\mrho$ & 0.032 & 0.012 & 0.034 & 0.032 & 0.014 & 0.035 \\
$R$ & 0.023 & 0.012 & 0.026 & 0.023 & 0.016 & 0.027 \\
$M$ & 0.041 & 0.030 & 0.051 & 0.045 & 0.044 & 0.063 \\
$\tau$ & 0.15 & 0.12 & 0.19 & 0.14 & 0.18 & 0.23 \\
\enddata
\tablecomments{Median values of  statistical, systematic and (quadratically) combined fractional
uncertainties in the catalog for both $\teff$ scales.}
\end{deluxetable}

The distribution  of $\mrho$ has  a slight shift  introduced by the  choice of
reference values.  The reason  of the shift  is related to  the fact  that the
reference  values have  been  computed with  a grid  of  stellar models  using
$\dnuo$, as  opposed to  $\dnuscl$ and  9 out of  the 12  GBM sets  of results
employ $\dnuscl$.  As discussed  by \citet{white:2011},  in the  $\teff$ range
between 5000 and 6300~K which comprises most of our sample, $(\dnuo/\dnusun) >
(\mrho/\mrhosun)^{1/2}$   whereas,   by   definition,   $(\dnuscl/\dnusun)   =
(\mrho/\mrhosun)^{1/2}$.  This  implies that  GBM  relying  on $\dnuscl$  will
introduce a (small)  systematic shift towards higher $\mrho$  values than GBMs
relying on  the more  physically correct  $\dnuo$, as it  is reflected  in the
distribution shown  in Fig.\,\ref{fig:syst}. The  second feature to  notice is
the  extended tails  of the  distribution towards  both positive  and negative
values that are  formed by results corresponding, in all  cases, to stars with
very small $\dnu$ uncertainties, below 0.4\%. For all cases where the absolute
value  of  the  normalized  residuals   is  larger  than  2,  the  statistical
uncertainty  of $\mrho$  is smaller  than 1.5\%,  a value  driven by  the high
precision of $\dnu$ for most of our sample.

The same arguments that explain the  shift of peak of the $\mrho$ distribution
are  behind   the  shifts   of  the   central  peaks  in   the  $R$   and  $M$
distributions. The  shifts here  take the  opposite sign  with respect  to the
change  in  $\mrho$  because  both   quantities  depend  inversely  on  $\dnu$
(Eqs.\ref{eq:mass}-\ref{eq:radius}).  Finally,  the  age  distribution  simply
mirrors differences in masses, as it is  apparent from the lower two panels in
Fig.\,\ref{fig:syst}. The similarity between  these two distributions is, once
again,  reassuring in  that  evolutionary timescales  obtained from  different
stellar  evolution codes  are in  reasonably good  agreement so  that a  given
fractional  difference  in mass  translates  into  a very  similar  fractional
difference in age.
  
Results  discussed  show  that  differences  among GBM  sets  of  results  are
typically consistent  with each  other when  measured against  the statistical
uncertainties, i.e. given the current  limitations imposed by uncertainties in
the seismic and spectroscopic data.

The systematic  GBM uncertainty for each  star and quantity is  defined as the
standard deviation $\sigma$  of the central values returned by  all twelve GBM
sets  after  applying  outlier  rejection.  Initially,  results  lying  beyond
3$\sigma$ of the  median of the results  are removed. Then, the  median of the
remaining sample is  recalculated and the process is iterated  until no values
beyond the  3$\sigma$ level  remain. The standard  deviation of  the remaining
values determine the systematic GBM uncertainty.

\subsubsection{ASPCAP $\teff$ scale}\label{sec:systteff}

We have considered  the presence of a systematic effect  in the calibration of
the  $\teff$ ASPCAP  scale. This  is  included in  the present  catalog as  an
additional systematic uncertainty in the derived stellar parameters. To obtain
this   uncertainty,  we   have  performed   two  additional   GBM  runs   with
\texttt{BeSPP/GARSTEC}-$\dnuo$  where  the  ASPCAP   $\teff$  scale  has  been
modified  by $\pm  100$~K. Then,  for each  stellar parameter,  the systematic
uncertainty has  been defined  as the  absolute value  of half  the difference
between those two sets of GBM results. The median fractional uncertainties for
the whole sample are 1.0\%\,($\logg$), 0.3\%\,($\mrho$), 1\%\,($R$), 3\%\,($M$)
and 13\%~($\tau$).  In the case of  $M$ and $\tau$ these  are smaller,
but  comparable,  with  median   statistical  uncertainties  as  discussed  in
Sect.\ref{sec:stat}  whereas  they are  substantially  smaller  for the  other
stellar parameters.

This systematic  component of the  total error budget  is not included  in the
catalog based on the SDSS \emph{griz}  $\teff$ scale. The reason, discussed in
Sect.~\ref{sec:teffscales}, is that  this $\teff$ scale is  more accurate than
the ASPCAP $\teff$ scale.

\subsection{Final uncertainties}\label{sec:errors}

Each  quantity  in  the  catalog  includes  both  statistical  and  systematic
uncertainties separately. These are assigned as follows:

\begin{itemize}
\item[-]   \textbf{Statistical}:   the    formal   uncertainty   returned   by
  \texttt{BeSPP/GARSTEC}-$\dnuo$,  the reference  GBM.  Positive and  negative
  errors are quoted individually.

\item[-] \textbf{Systematic}:  for the  stellar properties  based on  the SDSS
  $\teff$ scale,  the systematic  uncertainty is equal  to the  GBM systematic
  uncertainty  (Sect.\,\ref{sec:systgbm}). For  stellar properties  determined
  using  the ASPCAP  $\teff$  scale both,  GBM (Sect.\,\ref{sec:systgbm})  and
  $\teff$  scale  (Sect.\,\ref{sec:systteff})  systematic  uncertainties,  are
  added in quadrature to provide the final systematic uncertainty.

\end{itemize}

\begin{figure*}[ht!]
\centering
\includegraphics[scale=0.38]{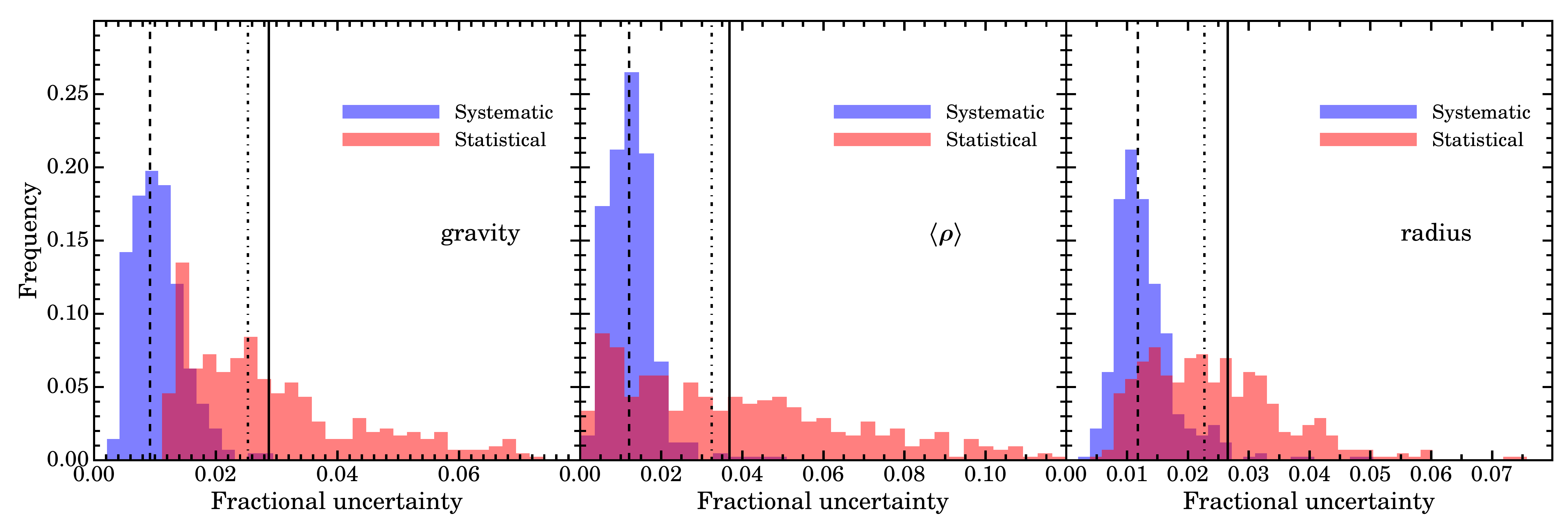}
\includegraphics[scale=0.38]{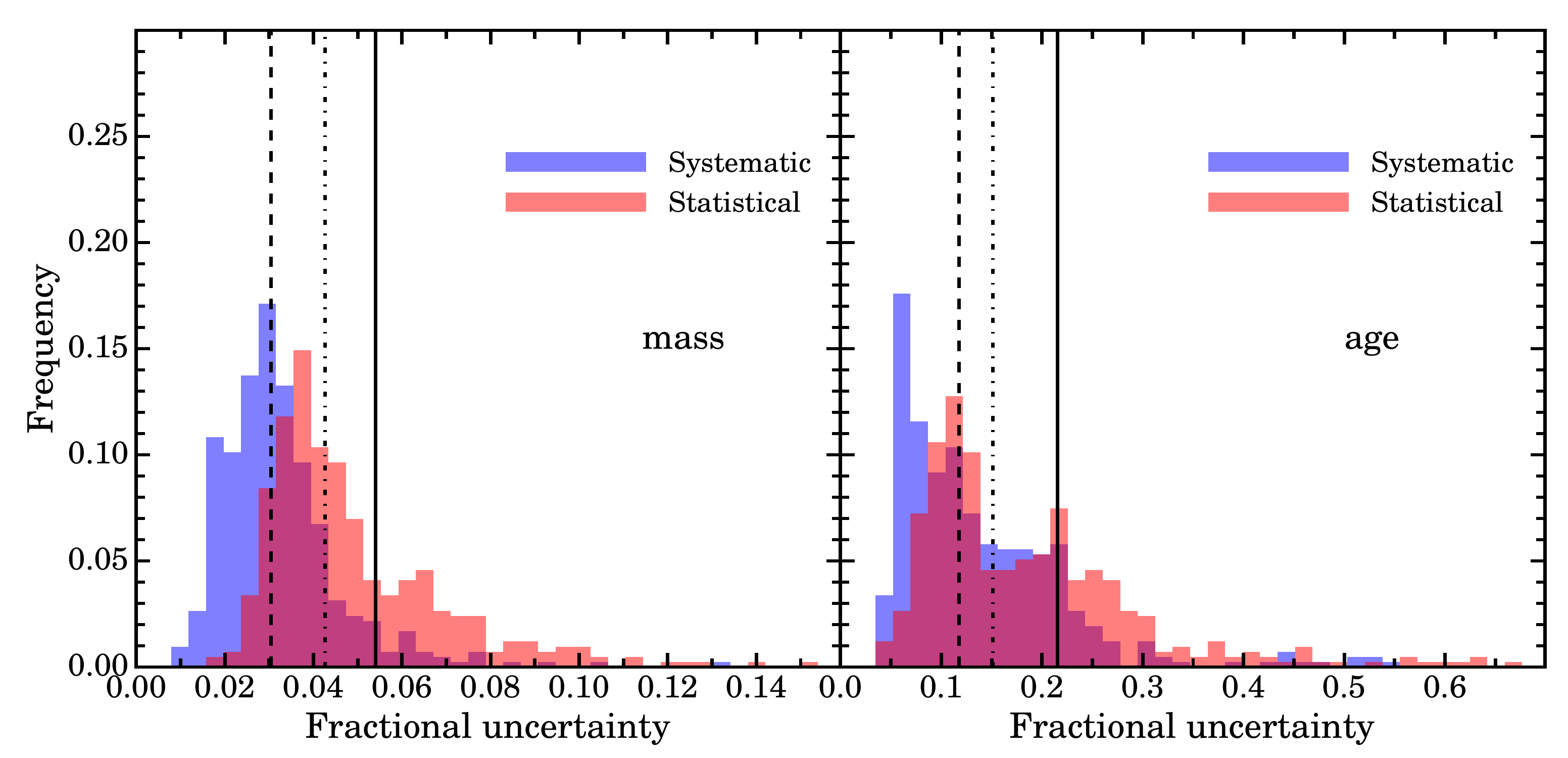}
\caption{Distributions of final errors in the catalog based on the SDSS $\teff$ scale. Statistical (average of positive and negative) and systematic components are given separately. Median values are shown in dashed, dotted-dashed and solid lines for the systematic, statistical and quadratically combined (total) uncertainty respectively. \label{fig:caterrsdss}}
\end{figure*}

\begin{figure*}[ht!]
\centering
\includegraphics[scale=.38]{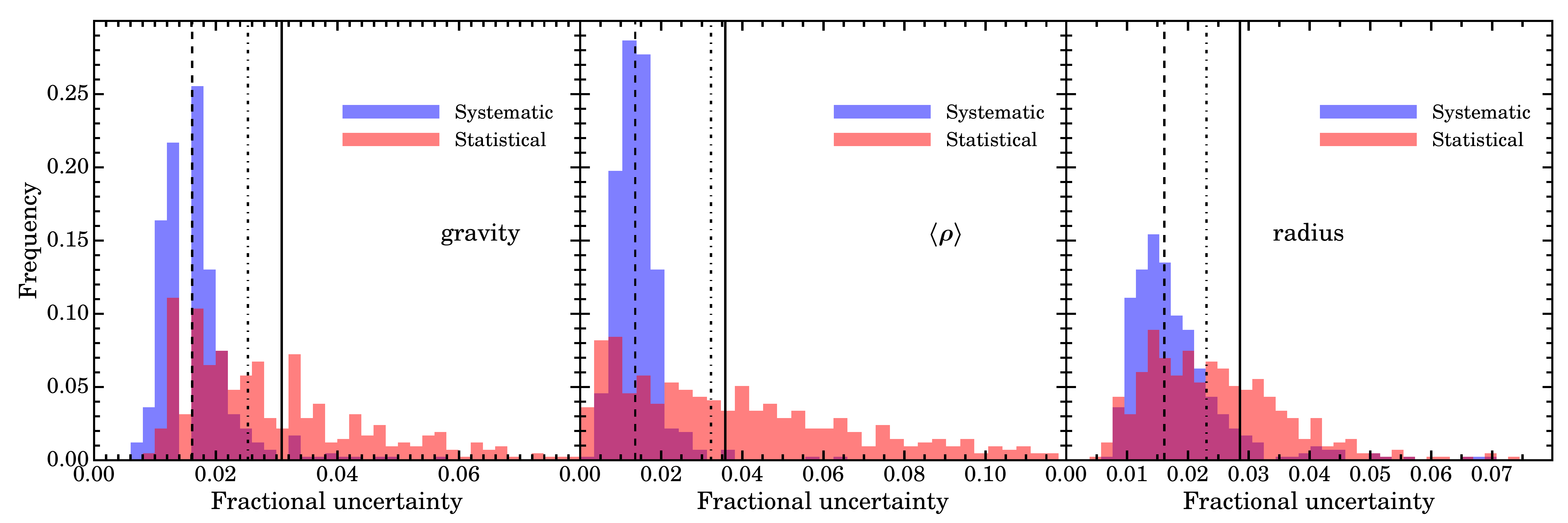}
\includegraphics[scale=0.38]{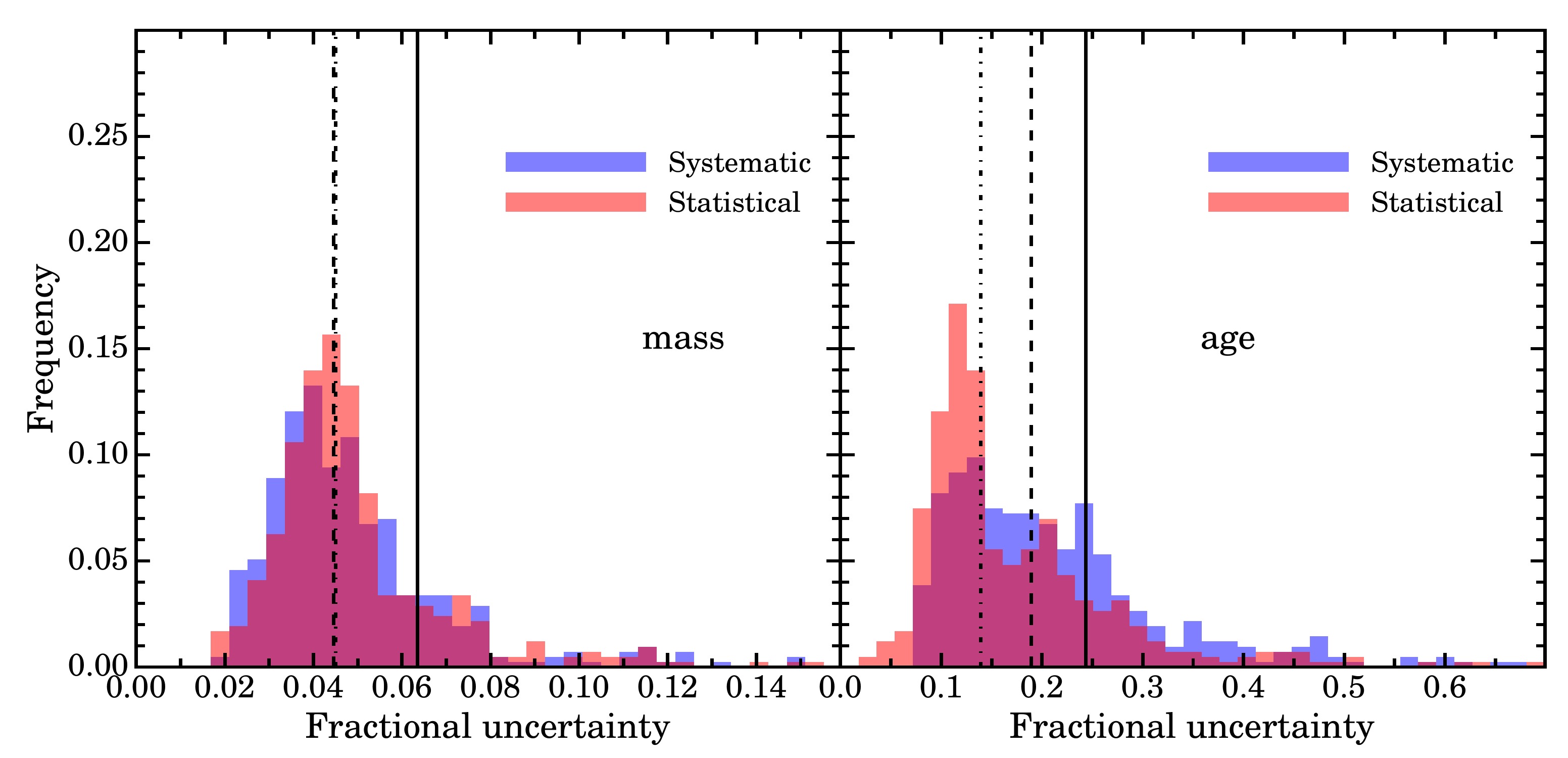}
\caption{Same as Fig.\,\ref{fig:caterrsdss} but for the ASPCAP $\teff$ scale. \label{fig:caterrasp}}
\end{figure*}

The   final  distributions   of   errors   in  the   catalog   are  shown   in
Fig.\,\ref{fig:caterrsdss}    for   the    SDSS   $\teff$    scale   and    in
Fig.\,\ref{fig:caterrasp}  for  the  ASPCAP   $\teff$  scale.  Vertical  lines
indicate median values  for the different uncertainty  components as indicated
in the figure. These values are also presented in Table~\ref{tab:mederrors}.

For the SDSS case, statistical uncertainties dominate over systematic ones for
$g$, $\mrho$, and $R$, and $M$, whereas they are comparable for $\tau$. In
the  case of  the ASPCAP  $\teff$  scale, statistical  uncertainties are  very
similar to those in the SDSS case, but systematic uncertainties are larger due
to     the      inclusion     of     the     $\teff$      systematic     error
(Sect.~\ref{sec:systteff}). This is particularly  relevant for $M$ and $\tau$,
for  which systematic  uncertainties are  comparable to  or dominate  over the
statistical uncertainties.

\section{APOKASC catalog} \label{sec:catalog}

\subsection{Presentation}

The final catalog contains 415  dwarfs and subgiant stars. Their asteroseismic
properties, reported  in Table~\ref{tab:seismic},  include the  global seismic
quantities $\numax$  and $\dnu$, the  length of the \emph{Kepler}  time series
used in their determination as well as the height-to-background ratio HBR (Sect.\,\ref{sec:seismicparam} and \citealt{mosser:2012}).

\floattable
\begin{deluxetable}{rcccc}
\tabletypesize{\footnotesize}
\tablecaption{APOKASC catalog - Seismic data \label{tab:seismic}}
\tablehead{\colhead{KIC ID} & \colhead{$\numax$\,[$\mu$Hz]} & \colhead{$\dnu$\,[$\mu$Hz]}& \colhead{Length\,[d]} & \phantom{1}HBR}
\startdata
   1435467 & $1382.31 \pm  19.04$  & $70.56  \pm  0.09$ & 938.1 & 1.66 \\
   2010607 & $\phantom{1}674.92 \pm 146.5$  & $42.48  \pm  2.18$ &  28.9 & 1.55\\
   2309595 & $\phantom{1}643.21 \pm 11.23$  & $39.03  \pm  0.72$ &  28.9 & 1.77\\
\enddata
\tablecomments{See text for details. The full table is available in a machine-readable form in the online journal. A portion is shown here for guidance regarding its form and content.}
\end{deluxetable}

The seismically determined stellar quantities  reported in the catalog are the
radius, mass, surface gravity, mean density and age. Properties obtained using
either  the SDSS  or  the ASPCAP  $\teff$ scales  are  reported separately  in
Tables~\ref{tab:catasdss}~and~\ref{tab:cataaspcap}   respectively.  For   each
quantity  asymmetric  statistical  uncertainties and  a  symmetric  systematic
uncertainty,  determined as  described in  Sect.\,\ref{sec:errors}, are  given
separately.

\floattable
\begin{deluxetable}{rccccccc}
\tabletypesize{\footnotesize}
\tablecaption{APOKASC dwarfs and subgiants catalog - SDSS $\teff$ scale \label{tab:catasdss}}
\tablehead{ \colhead{KIC ID} & \colhead{$\teff$ [K]}& \colhead{$\mh$ [dex]} & \colhead{Radius [$\rsun$]} & \colhead{Mass [$\msun$]} & \colhead{$\logg$ [dex]}& \colhead{$\mrho$ [solar units]} & \colhead{Age [Gyr]}}
\startdata   
1435467 & $ 6433 \pm 86 $ & $ -0.03 \pm 0.10 $ & $1.69^{+0.01 + 0.01}_{-0.02 -0.01} $ & $1.34^{+0.04 + 0.02}_{-0.04 -0.02} $ & $4.109^{+0.006 + 0.002}_{-0.006 -0.002} $ & $0.2777^{+0.0024 + 0.0015}_{-0.0024 -0.0015} $ & $2.60^{+0.30 + 0.21}_{-0.29 -0.21} $  \\
2110607 & $ 6361 \pm 71 $ & $ -0.07 \pm 0.10 $ & $2.41^{+0.10 + 0.03}_{-0.09 -0.03} $ & $1.40^{+0.07 + 0.05}_{-0.05 -0.05} $ & $3.819^{+0.027 + 0.003}_{-0.027 -0.003} $ & $0.0997^{+0.0099 + 0.0014}_{-0.0094 -0.0014} $ & $2.75^{+0.30 + 0.18}_{-0.30 -0.18} $ \\
2309595 & $ 5238 \pm 65 $ & $ -0.09 \pm 0.10 $ & $2.42^{+0.08 + 0.02}_{-0.08 -0.02} $ & $1.17^{+0.08 + 0.02}_{-0.08 -0.02} $ & $3.736^{+0.008 + 0.002}_{-0.008 -0.002} $ & $0.0818^{+0.0025 + 0.0007}_{-0.0025 -0.0007} $ & $5.46^{+1.35 + 0.42}_{-1.02 -0.44} $ \\
\enddata
\tablecomments{For seismically determined stellar parameters the first error term is the statistical error (Sect.\,\ref{sec:stat}) and the second one the total systematic error (Sect.\,\ref{sec:syst}). The full table is available in a machine-readable form in the online journal. A portion is shown here for guidance regarding its form and content.}
\end{deluxetable}

\floattable
\begin{deluxetable}{rccccccc}
\tabletypesize{\footnotesize}   \tablecaption{APOKASC  dwarfs   and  subgiants
  catalog   -  ASPCAP   $\teff$   scale  \label{tab:cataaspcap}}   \tablehead{
  \colhead{KIC   ID}  &   \colhead{$\teff$  [K]}&   \colhead{$\mh$  [dex]}   &
  \colhead{Radius  [$\rsun$]} &  \colhead{Mass  [$\msun$]} &  \colhead{$\logg$
    [dex]}& \colhead{$\mrho$ [solar units]}  & \colhead{Age [Gyr]}} \startdata
1435467 & $ 6096  \pm 69 $ & $ -0.03 \pm 0.10  $ & $1.68^{+0.01 + 0.01}_{-0.01
  -0.01}  $  &  $1.28^{+0.03  +  0.03}_{-0.03  -0.03}  $  &  $4.095^{+0.005  +
  0.004}_{-0.005 -0.004} $ & $0.2711^{+0.0011  + 0.0015}_{-0.0010 -0.0015} $ &
$3.75^{+0.35 + 0.50}_{-0.32  -0.50} $ \\ 2010607 &  $ 6013 \pm 69 $  & $ -0.07
\pm  0.10  $  &  $2.39^{+0.10  +   0.03}_{-0.09  -0.03}  $  &  $1.33^{+0.06  +
  0.04}_{-0.05  -0.04}  $  &  $3.802^{+0.026  +  0.006}_{-0.027  -0.006}  $  &
$0.0967^{+0.0097 +  0.0015}_{-0.0095 -0.0015} $ &  $3.54^{+0.40 + 0.33}_{-0.39
  -0.33} $ \\ 2309595 & $ 5000 \pm 69  $ & $ -0.09 \pm 0.10 $ & $2.30^{+0.08 +
  0.06}_{-0.07   -0.06}  $   &  $1.02^{+0.08   +  0.07}_{-0.07   -0.07}  $   &
$3.723^{+0.008   +    0.007}_{-0.008   -0.007}   $   &    $0.0839^{+0.0026   +
  0.0013}_{-0.0026  -0.0013}   $  &  $9.54^{+2.67  +   2.52}_{-2.19  -2.52}  $
\\ 
\enddata 
\tablecomments{For seismically determined stellar parameters the first error term is the statistical error (Sect.\,\ref{sec:stat}) and the second one the total systematic error (Sect.\,\ref{sec:syst}). The  full table is available  in a machine-readable
  form in the  online journal. A portion is shown  here for guidance regarding
  its form and content.}
\end{deluxetable}

\begin{figure*}
\centering\includegraphics[scale=.45]{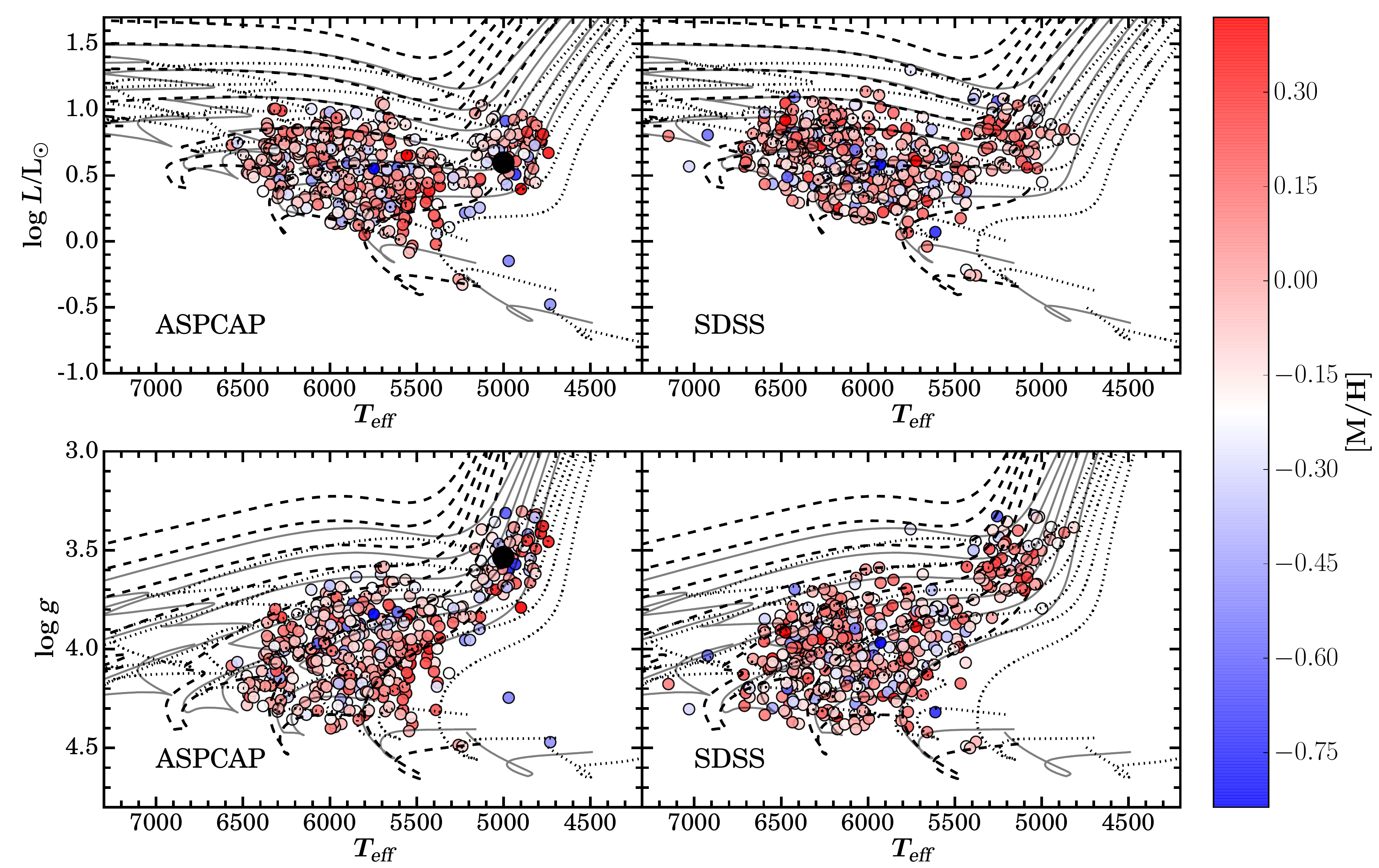}
\caption{Hertzsprung-Russell and Kiel diagrams of the full APOKASC sample (top
  and  bottom  panels  respectively),  for  the  spectroscopic  $\teff$  scale
  (ASPCAP;  left panels)  and  the  $\teff$ scale  based  on SDSS  \emph{griz}
  photometry  (right panels).  The black  symbol identifies  a star  with $\mh
  =-1.98$ (out of the color scale for $\mh$). \label{fig:hrd}}
\end{figure*}

Figure~\ref{fig:hrd} shows in the  top panels the Hertzsprung-Russell diagrams
and in the  bottom panels the Kiel  diagrams for the full sample  and for both
$\teff$  scales.  Stars  have  been  color  coded  according  to  their  $\mh$
value.  \texttt{GARSTEC}  evolutionary  tracks  for  masses  between  0.8  and
2~$\msun$, and $\mh =-0.5, 0.0, 0.5$ are overplotted.

\begin{figure*}
\centering\includegraphics[scale=.42]{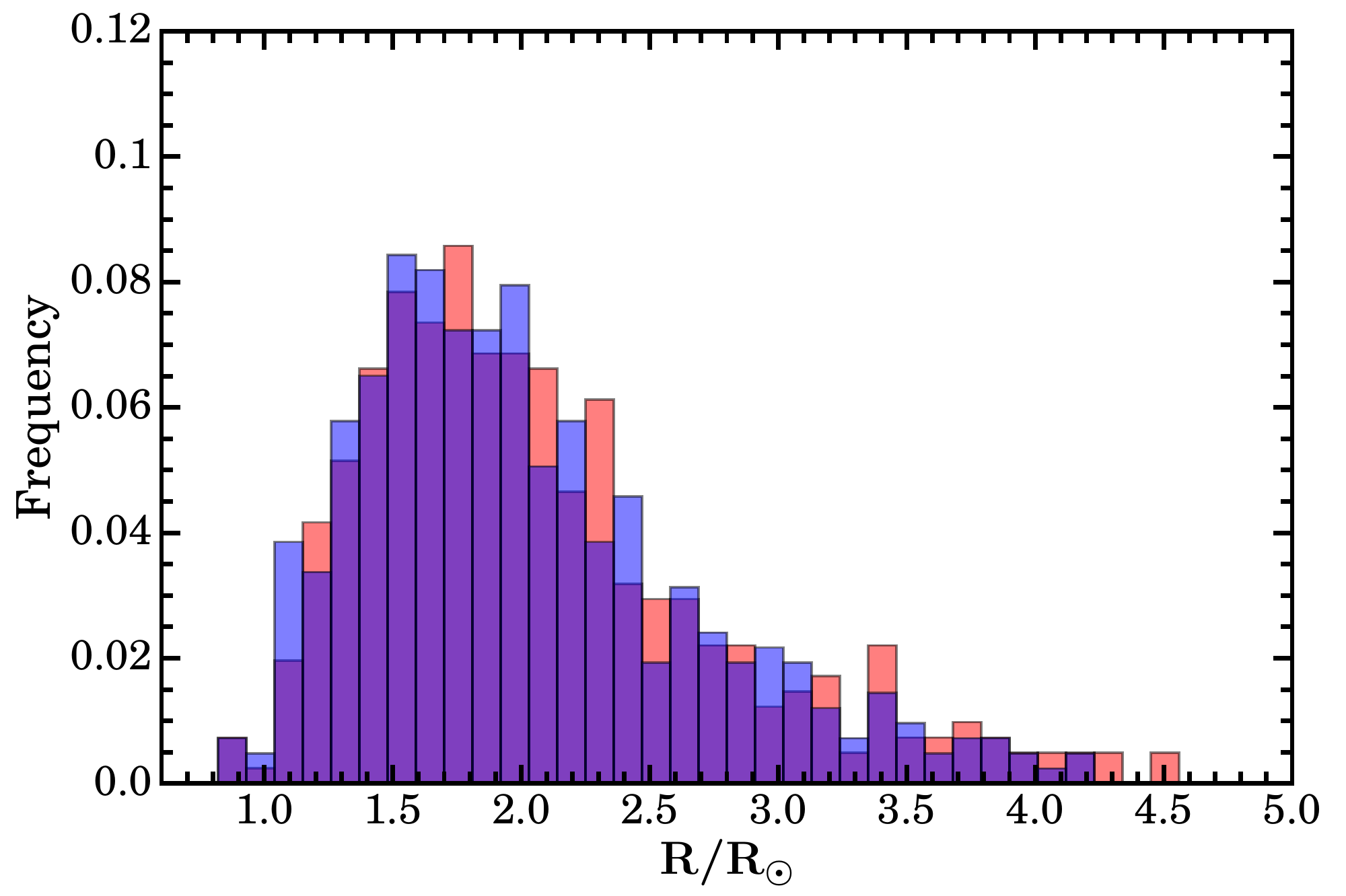}\includegraphics[scale=.42]{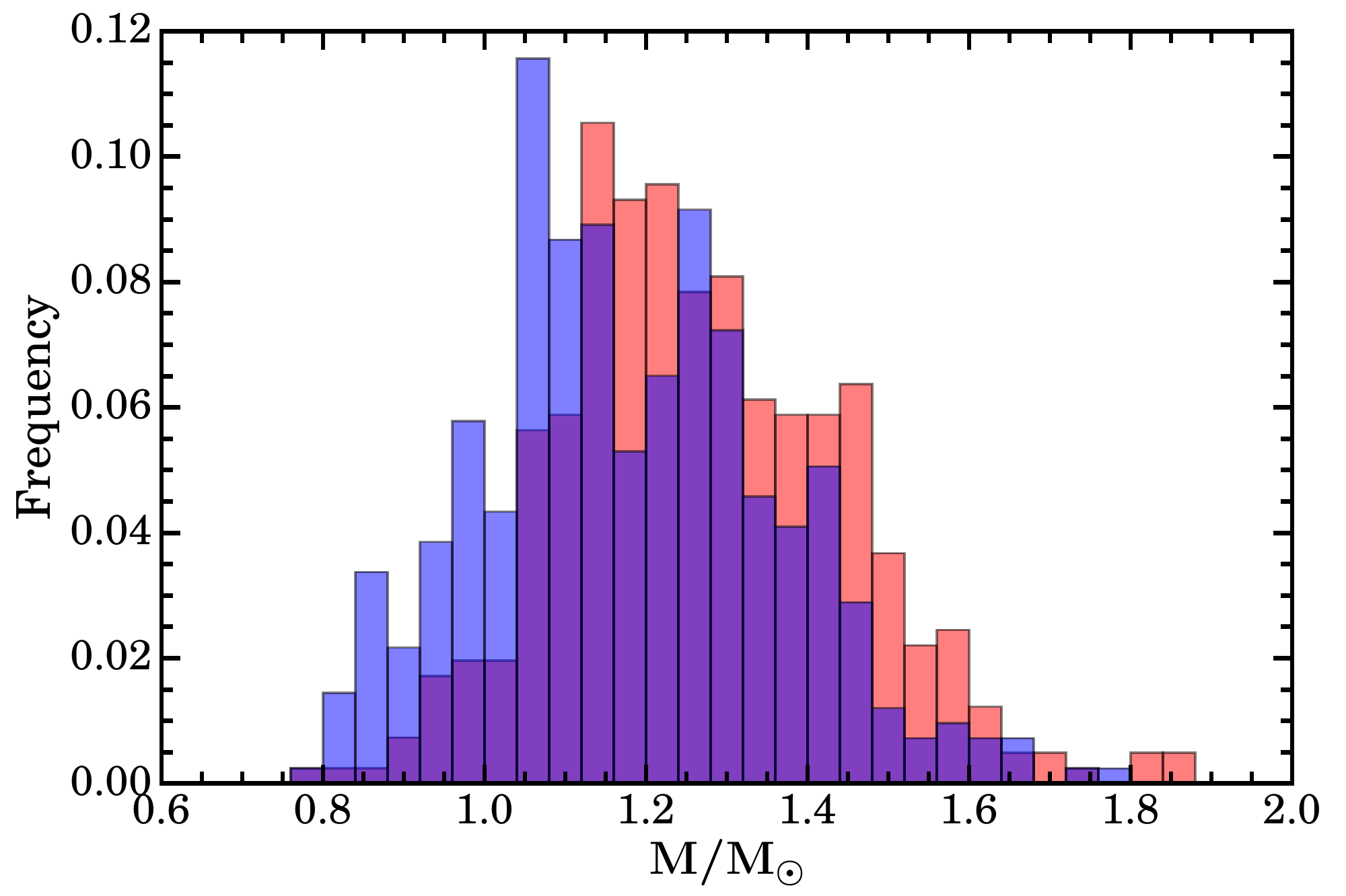}
\caption{Distributions of radii and masses for  the catalog for the SDSS (red)
  and ASPCAP (blue) $\teff$ scales.  \label{fig:mrpdf}}
\end{figure*}

Figure~\ref{fig:mrpdf} presents the distributions of  radii and masses for the
whole  sample and  both ASPCAP  and  SDSS $\teff$  scales. The  effect of  the
systematic offset between  the two scales is visible. This  is more noticeable
in the  mass distribution,  for which  the lower  ASPCAP temperatures  lead to
smaller stellar masses. This is discussed further in the next section.

\subsection{Impact of the $\teff$ scale}

The  systematic differences  between the  ASPCAP and  the SDSS  $\teff$ scales
(Sect.\,\ref{sec:teffscales}) lead to systematic variations in the seismically
determined stellar parameters that we discuss here.

\begin{figure*}[ht!]
\centering
\includegraphics[scale=.35]{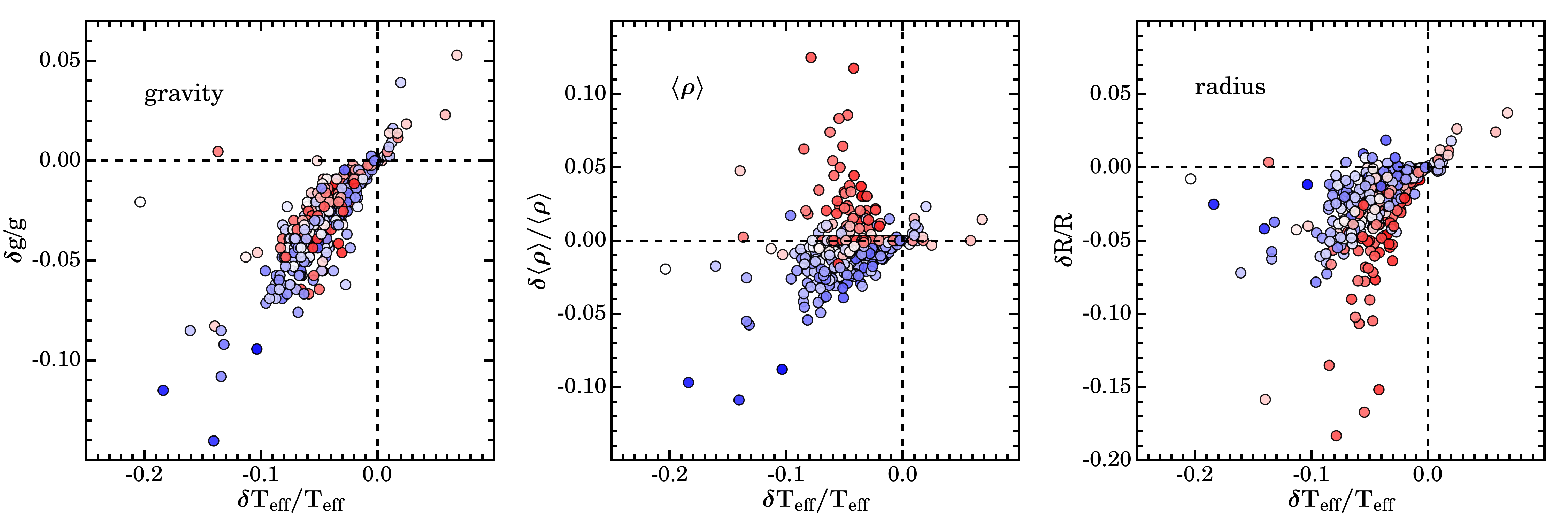}
\includegraphics[scale=.35]{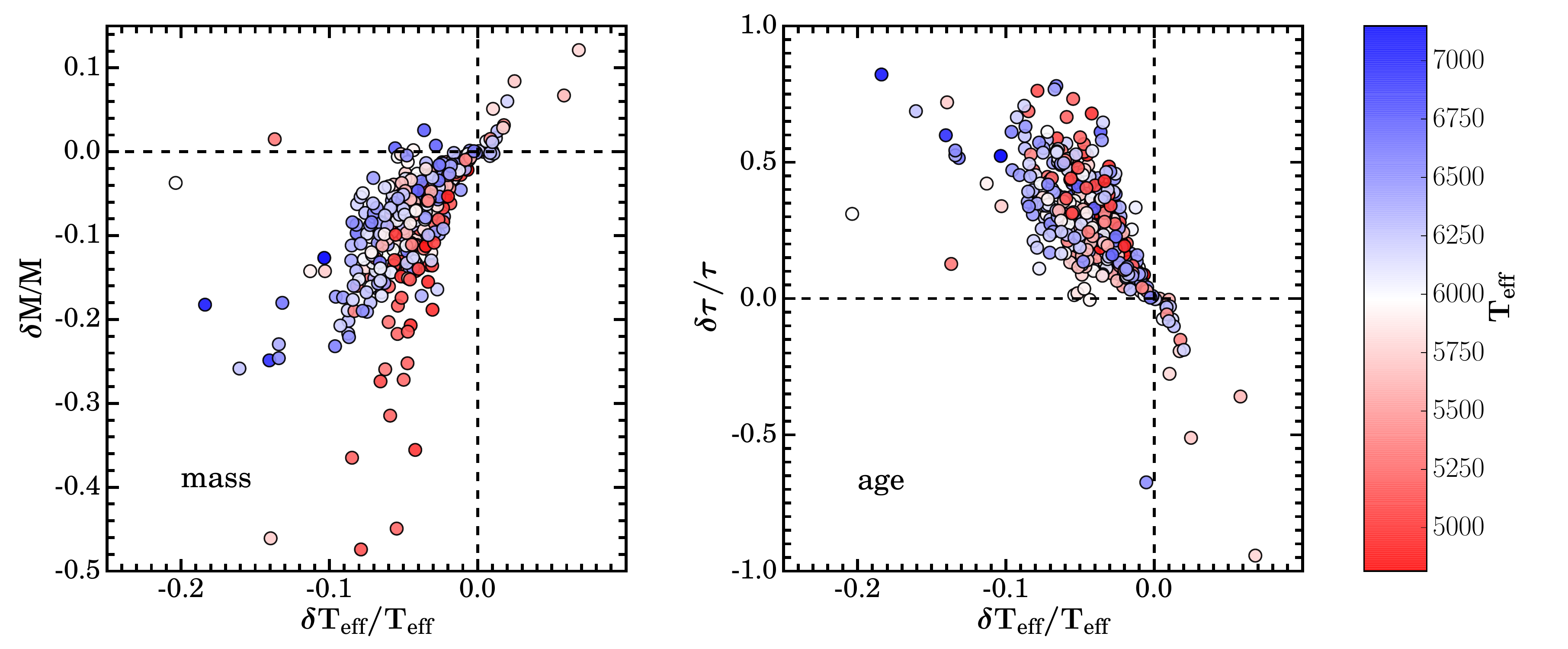}
\caption{Fractional   variations  in   the   seismically  determined   stellar
  parameters  as a  function of  the relative  $\teff$ difference  between the
  ASPCAP    and    SDSS    scales.    Differences    are    in    the    sense
  ASPCAP~$-$~SDSS. \label{fig:teffcomp}}
\end{figure*}

The  fractional   differences  in   the  stellar   parameters  are   shown  in
Fig.\,\ref{fig:teffcomp} as  a function  of the fractional  difference $\delta
\teff /\teff$.  Stars are  color coded  according to  their SDSS  $\teff$. For
$g$, the variations directly reflect the dependence of the seismic gravity
on $\sqrt{\teff}$  (Eq.\,\ref{eq:numax}). Differences  in $\mrho$ can  be both
positive and negative for a given  $\delta \teff$. To first order, the scaling
between  $\mrho$ and  $\dnu$ does  not include  $\teff$. However,  the central
values  in  the  catalog make  use  of  $\dnuo$,  not  of $\dnuscl$,  and  the
difference  between  the two  does  in  fact depend  on  $\teff$  as shown  in
Fig.\,\ref{fig:dnucorr}.  Due   to  the  systematic  nature   of  the  $\teff$
differences between  the ASPCAP and SDSS  scales, the sign of  $\delta \mrho /
\mrho$ therefore depends,  for a given star, on whether  its $\teff$ is cooler
or hotter  than the $\teff$  value for  which $\dnuo/\dnuscl$ has  its maximum
(Fig.\,\ref{fig:dnucorr}).   This   sets   the    general   trend,   seen   in
Fig.\,\ref{fig:teffcomp}, that  $\mrho_{\rm ASP}  > \mrho_{\rm SDSS}$  for the
cooler  stars of  the sample  and  the reverse  effect for  stars hotter  than
$\approx 5500$~K. For $\sim 97\%$ of the sample, changes are smaller than 5\%.
The median difference is $-2.3\%$ in $g$ and $-0.3\%$ in $\mrho$.

Changes  in  stellar radii  have  two  contributions  when the  $\teff$  scale
changes,  one through  its dependence  on  $\sqrt{\teff}$ and  the other  one,
indirect, through the  $\dnuo/\dnuscl$ dependence on $\teff$.  The two effects
partially cancel each other out for  stars with $\teff \gtrsim 5500$~K but are
added in the case  of the cooler stars, for which the change  from the SDSS to
the   ASPCAP   scale   leads   to   systematically   smaller   $\dnuo/\dnuscl$
corrections. This  is reflected  in Fig.\,\ref{fig:teffcomp} which  shows that
radii differences for cooler stars are stretched in comparison to hotter stars
with similar  relative $\teff$ variations.  The median difference  for stellar
radii is $-1.8\%$.

Stellar  masses show  an  analogous  response to  $\teff$  variations but  the
magnitude  of the  changes is  augemented  with respect  to radii  due to  the
steeper  dependence  of  seismic  mass   determinations  on  both  $\dnu$  and
$\teff$. In the case of the mass  it becomes even clearer than for radius that
there are  two separate  \emph{branches}, one  for hotter  and one  for cooler
stars. Stellar  ages, once again,  mostly reflect  the changes in  the derived
stellar masses.  The median  difference for  mass is $-6.2\%$  and for  age is
$+35\%$.

\subsection{Comparisons with other works} \label{sec:comparisons}

\subsubsection{Chaplin et al. 2014}  \label{sec:chap14}

The vast  majority of stars  in the catalog presented  in this work  were also
part of the C14 catalog. The most  important difference with respect to C14 is
that a  single value $\feh=-0.2\pm0.3$,  taken as representative of  the solar
neighborhood \citep{victor:2011},  was adopted  for the  whole sample  in that
work. The  large uncertainty  effectively implied that  $\feh$ actually  had a
relatively minor influence on the derived stellar properties. The SDSS $\teff$
scale used in this work is the same used in C14.

\begin{figure*}[ht!]
\centering
\includegraphics[scale=0.38]{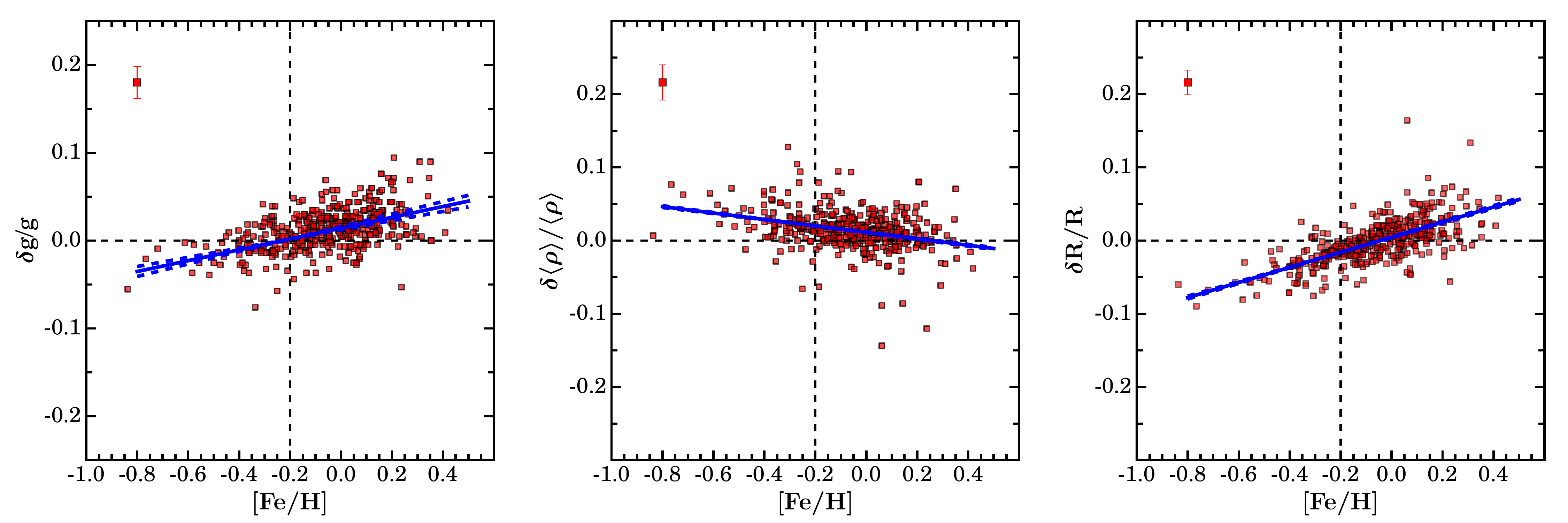}
\includegraphics[scale=0.38]{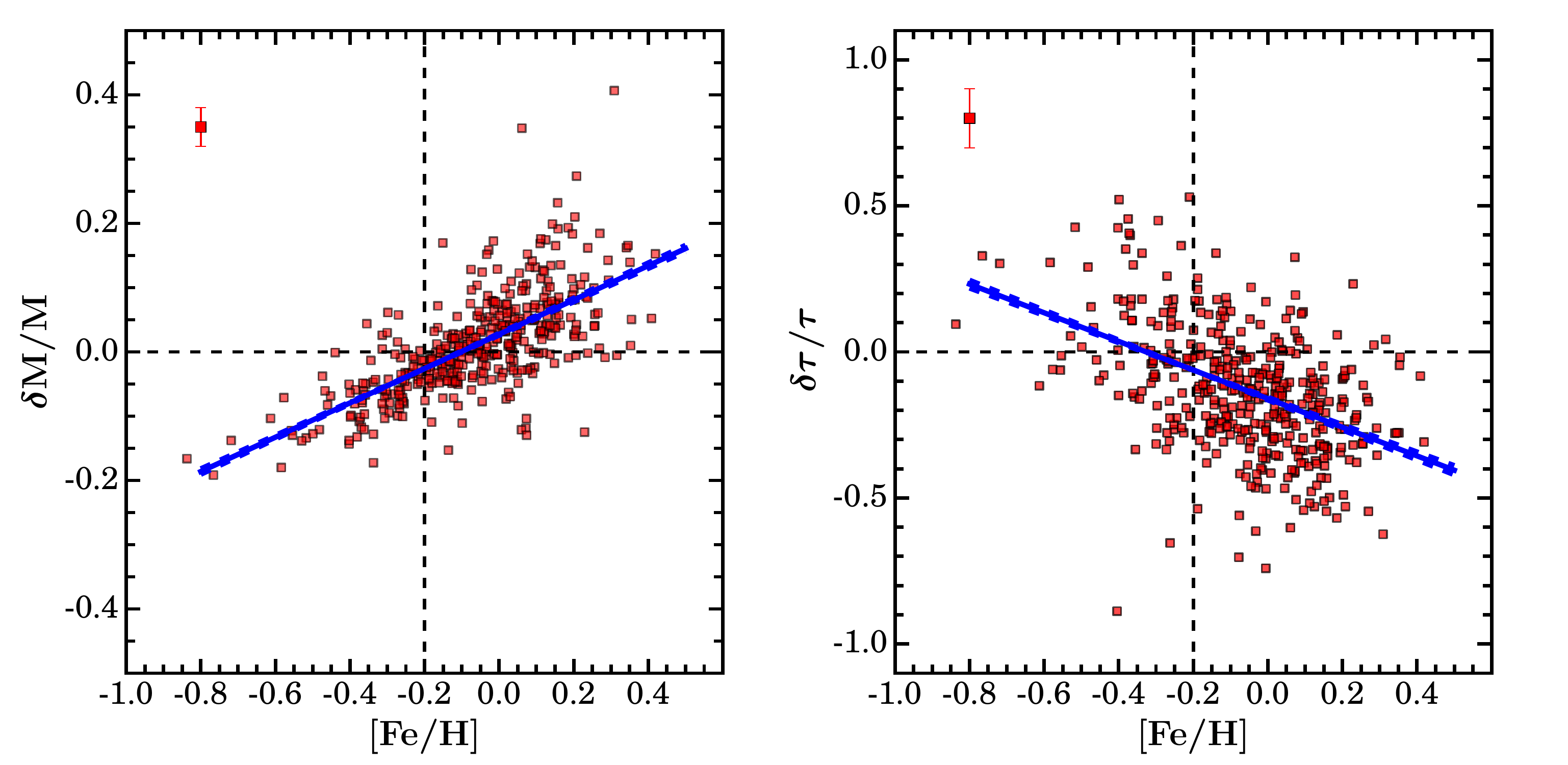}
\caption{Fractional   variations  in   the   seismically  determined   stellar
  parameters as  a function of  the ASPCAP metallicity between  the SDSS-based
  results  and  those in  the  \citet[C14]{chaplin:2014}  \emph{Kepler}  catalog. Vertical  line  at
  $\feh=-0.2$  denotes the  fiducial value  adopted  for the  whole sample  in
  C14. Differences are in the sense "this work - C14". Linear fits are shown in blue solid lines and 1$\sigma$ variations in dashed lines. Median uncertainties of each quantity are shown on top left corners. \label{fig:chapcomp}}
\end{figure*}

We can highlight the impact of having $\mh$ measurements by directly comparing
the stellar  properties derived in this  catalog with respect to  those in C14
when  the  SDSS $\teff$  scale  is  used in  both  cases.   This is  shown  in
Fig.\,\ref{fig:chapcomp}, where fractional differences are shown as a function
of the ASPCAP $\mh$  value. For reference, the $\feh$ adopted  in C14 is shown
as a vertical  dashed line. For all seismically  determined stellar quantities
there are  clear correlations in the  differences between the present  and the
C14 catalogs with the  actual $\mh$ of stars. A simple linear  fit $\delta = a
\cdot (\mh+0.2) + b$,  shown in each plot together with the 1$\sigma$ variation in solid and dashed lines respectively, is indicative  of the dependence of
each  stellar  parameter  on  $\mh$.  The   fits  have  slopes  equal  to  $a=
0.062\,(g), -0.044\,(\mrho), 0.103\,(R),  0.267\,(M), -0.493\,(\tau)$. The
offsets  at $\mh=-0.2$  for the  different  quantities are  $b= 0.004,  0.020,
-0.016, -0.026, -0.062$ in the same  order as above. Stellar parameters in C14
were   also   determined   using    \texttt{BeSPP},   models   computed   with
\texttt{GARSTEC}, and  $\dnuo$. The  reason for the  offset in  $\mrho$ arises
from the fact that  contrary to the procedure applied in  the present work, in
C14 $\dnuo$  in the stellar models  were not rescaled according  to the factor
$f_{\dnu}$ described in  Sect.\,\ref{sec:central}. This leads to  a 1\% offset
in $\dnuo$ between the two stellar  models grid that propagates as about twice
that value to the $\mrho$ determination, as expected, with a comparable offset
for $R$ and  a slightly larger one  for $M$ which, in turn,  propagates into a
$\tau$  offset. The  scatter seen  in the  relations can  have at  least three
sources: the  asteroseismic data used  here is  determined in many  cases from
longer time series and improved seismic analysis, and the grids of models have
not been  computed with  the exact  same assumptions,  the current  version of
\texttt{BeSPP} is an evolution of that used in C14, when it was still based on
a frequentist approach akin to that used by the \texttt{YB} pipeline.

As  a whole,  results in  this catalog  are consistent  with those  previously
published by  C14, but represent  a qualitative and quantitative  step forward
because  metallicity   information  is  now  included   individually  for  all
stars. The C14  catalog, however, includes about 100 more  stars that were not
observed with APOGEE (see Sect.~\ref{sec:sample}).

\subsubsection{The LEGACY Sample}  \label{sec:legacy}

The LEGACY Sample  is formed by 66  main sequence stars of  highest quality of
asteroseismic       data       from      the       \emph{Kepler}       mission
\citep{lund:2017}. Asteroseismic analysis  of these stars is  based on stellar
model  fitting  that  relies  on using  specific  combinations  of  individual
frequencies   that   allow   a   more   precise   determination   of   stellar
parameters. Detailed results have been presented in \citet{victor:2017}.

From the total  common sample between APOKASC and LEGACY,  we have removed two
stars because in one case the ASPCAP $\teff$ and in the other the SDSS $\teff$
values  differ from  the  LEGACY  $\teff$ value  (based  on SPC  spectroscopic
analysis by \citealt{buchhave:2015}) by more than 600~K. In total, we are left
with  a common  sample  of 45  stars  that can  be used  as  a benchmark.  Our
comparison of stellar parameters makes use of LEGACY results computed with the
\texttt{BASTA} pipeline, which are based  on \texttt{GARSTEC}. By using LEGACY
results based  on the  same evolutionary  code used  to determine  the central
values in the APOKASC catalogs, we minimize the systematic differences arising
from stellar models.

Note that,  in addition to  the different type of  input seismic data  used in
this  catalog  ($\dnu$  and  $\numax$)  and in  the  LEGACY  work  (individual
frequencies  or  frequency separation  ratios),  another  source of  deviation
between  the two  sets of  results is  likely  due to  the fact  that all  GBM
presented in  this work  are based  on libraries of  stellar models  where the
mixing  length parameter  is fixed,  whereas  in the  LEGACY work  this is  an
additional free  parameter in  the stellar models.  This precludes  a detailed
discussion  of  star-by-star  comparions.  Instead,   we  focus  on  a  global
comparison of mean differences and dispersions.

\begin{figure*}[ht!]
\centering
\includegraphics[scale=.42]{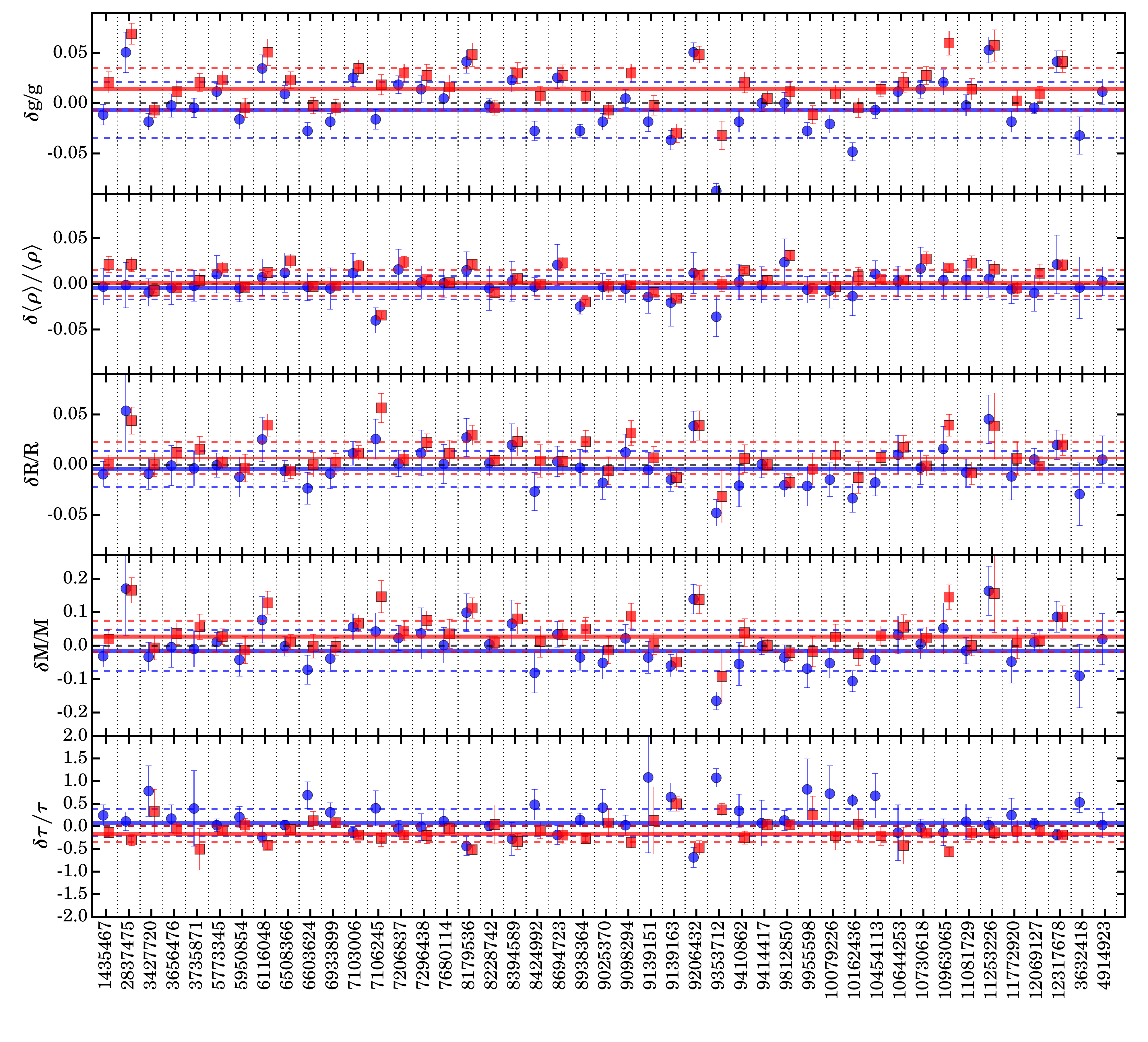}
\caption{Comparison of the  APOKASC results against the LEGACY  sample for all
  stars common  to both samples. Red  squares (blue circles) refer  to results
  obtained with the SDSS (ASPCAP) $\teff$ scale. Red and blue horizontal lines
  show the  weighted mean differences (solid) and rms (dashed) with respect  to the 
  LEGACY  sample for each $\teff$ scale (see Table~\ref{tab:legacycomp}). 
  \label{fig:legacy}}
\end{figure*}

\floattable
\begin{deluxetable}{lccccccc}
\tabletypesize{\small}     
\tablewidth{9cm}      
\tablecaption{APOKASC$-$LEGACY\label{tab:legacycomp}} 
\tablehead{  & \multicolumn{3}{c}{ASPCAP}  & &
  \multicolumn{3}{c}{SDSS} \\ \cline{2-4} \cline{6-8} & Weig. Mean & Weig. rms
  & Unw.  Mean &  & Weig.  Mean & Weig.  rms &  Unw. Mean}  
  \startdata 
  $\delta g/g$ & $-0.007$ & 0.028 & $-0.002$ & & \phantom{$-$}0.014 & 0.021 &
\phantom{$-$}0.016   \\  
$\delta   \mrho/  \mrho$   &  $-0.004$   &  0.013   &
$-5\times10^{-4}$  &   &  \phantom{$-$}0.001  &  0.014   &  \phantom{$-$}0.007
\\ 
$\delta R/R$ & $-0.004$ & 0.018 & $-0.001$ & & \phantom{$-$}0.007 & 0.016 &
\phantom{$-$}0.009  \\  
$\delta  M/M$  &  $-0.015$ &  0.061  &  $-$0.003  &  &
\phantom{$-$}0.027  & 0.047  &  \phantom{$-$}0.036 \\  
$\delta  \tau/ \tau$  &
\phantom{$-$}0.078  &  0.300 &  \phantom{$-$}0.213  &  &  $-0.167$ &  0.183  &
$-0.116$ \\ 
\enddata 
\tablecomments{Comparison of APOKASC  and LEGACY results
  shown as  APOKASC--LEGACY. Weighted mean and rms fractional differences are  shown for
  both $\teff$  scales used  in this work.  Additionally, the  unweighted fractional 
  mean differences are also given. }
\end{deluxetable}

Figure~\ref{fig:legacy} shows  the comparison for  all stars and  both $\teff$
scales, identifying in  blue (circles) results for the SDSS  $\teff$ scale and
in red (squares) those for the ASPCAP $\teff$ scale. Horizontal lines indicate
the weighted  mean difference with  respect to  the LEGACY results.  For three
stars only ASPCAP $\teff$ values are available.

The  weighed mean and the dispersion (rms) of the  fractional differences
 of the  two
APOKASC sets of  results with respect to the LEGACY  results are summarized in
Table~\ref{tab:legacycomp}.  For all stellar parameters, ASPCAP results show a
smaller  mean  difference  than  SDSS.  Overall, it  is  reassuring  that  the
systematic  offsets  between APOKASC  and  LEGACY  results  are in  all  cases
substantially smaller  than the median error  of each quantity in  the APOKASC
catalog (Figs.\,\ref{fig:caterrsdss} and \ref{fig:caterrasp}).

In the following  we discuss the comparison in more  detail, in particular for
stellar   mass.  The   mean   fractional  difference   $\delta  M/M$   between
APOKASC/ASPCAP and LEGACY  is $-1.5\%$ and between APOKASC/SDSS  
and LEGACY is
2.7\%. The mean $\teff$  in ASPCAP is 1.4\% cooler than  in the LEGACY sample,
and the SDSS  $\teff$ scale is 1.3\% hotter. By  propagating these differences
through the  scaling relation  Eq.~\ref{eq:mass}, we infer  that asteroseismic
masses from  global seismology show a  systematic mean deviation of  about 2\%
with respect  to the LEGACY  sample. We have  confirmed this by  performing an
additional GBM run with \texttt{BeS/G-$\dnuo$} using as inputs the combination
of our global  seismic parameters and the $\teff$ and  $\feh$ values used from
LEGACY sample \citep{victor:2017}. In this case, we find that the mean $\delta
M/M=0.6\%$, in  agreement with  the above  estimate, and  an rms  $\delta M/M=
6\%$. In this case, the mean and rms  values for $\delta \tau / \tau$ are -4\%
and  26\%,  respectively.   These  numbers  can be  taken  as  a  first  order
estimation  of the  systematic differences  in  mass and  age results  between
global seismology and more detailed analysis based on modeling many individual
mode  frequencies.  However, a  more  careful  analysis  required to  be  more
conclusive is  beyond the  scope of  this paper.  For example,  global seismic
parameters reported for the LEGACY  sample \citep{lund:2017}, are not the same
as the ones  in our work ($\numax$ in  particular is different by up  to a few
percent for some  stars) and we have not tested  if this introduces systematic
effects. Also, despite that stellar  models are computed with \texttt{GARSTEC}
in both cases, they are not exactly the same.

From Table~\ref{tab:legacycomp}  we also  see that  the mass  and age  rms are
larger  for the  ASPCAP results  than for  SDSS. For  the ASPCAP  results, the
larger rms  is driven by  three stars with  $\teff$ values different  from the
LEGACY values by more than 300~K. By removing these stars the rms decreases to
6.3\%, around the 6\% found between global and detailed seismic analysis. Mass
differences (global$-$detailed seismology) do  not show clear correlation with
the $\teff$ differences in the analysis. However, age differences do correlate
with  the $\teff$  scale.  This  explains in  part  the  variation in  $\delta
\tau/\tau$ from almost 8\% (ASPCAP) down to -17\% (SDSS), which is larger than
expected from  the differences found  for $\delta M/M$. A  detailed comparison
between global and detailed seismic analysis is worth further investigation.

\begin{figure*}
\centering\includegraphics[scale=.42]{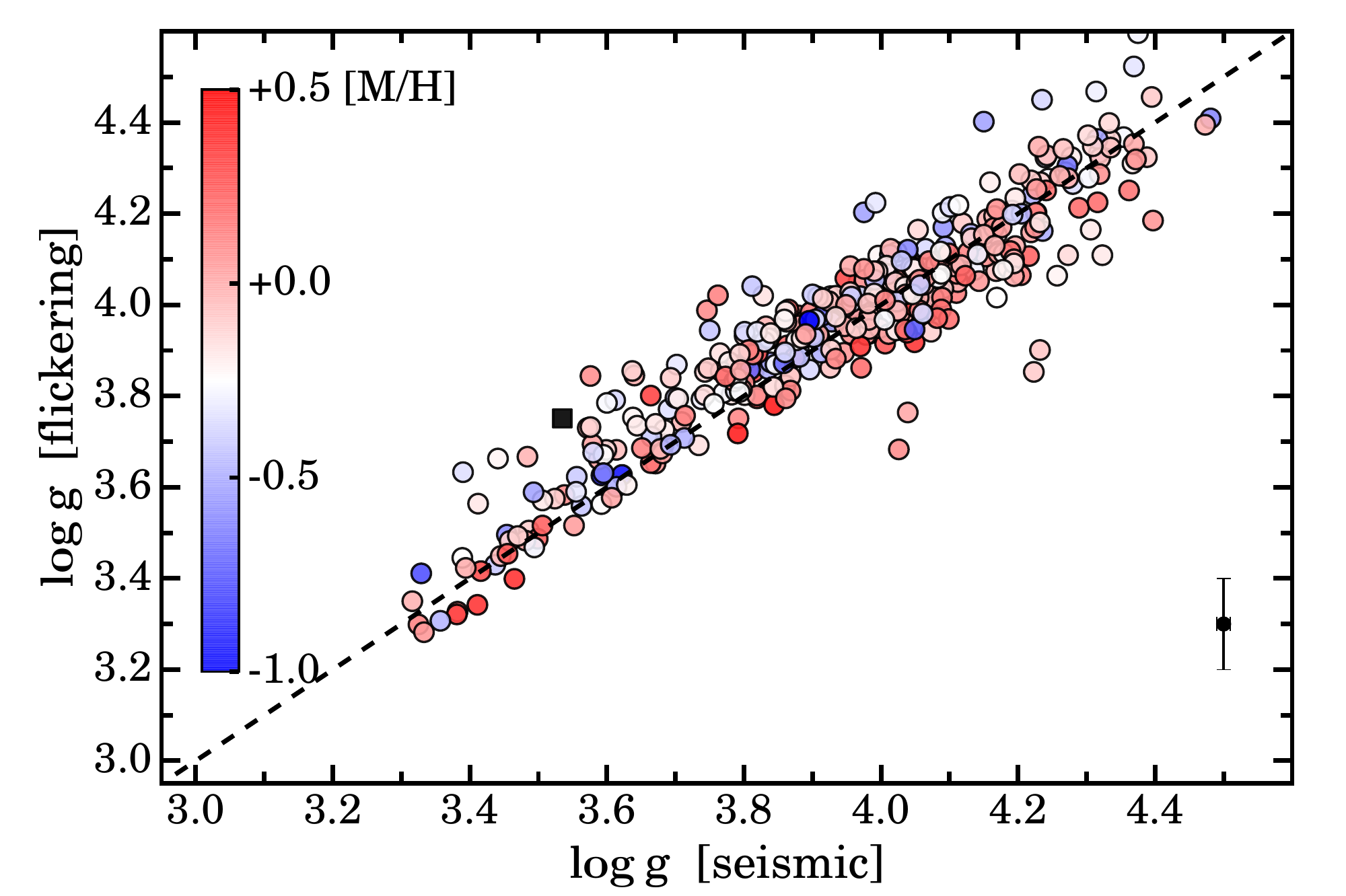}\includegraphics[scale=.42]{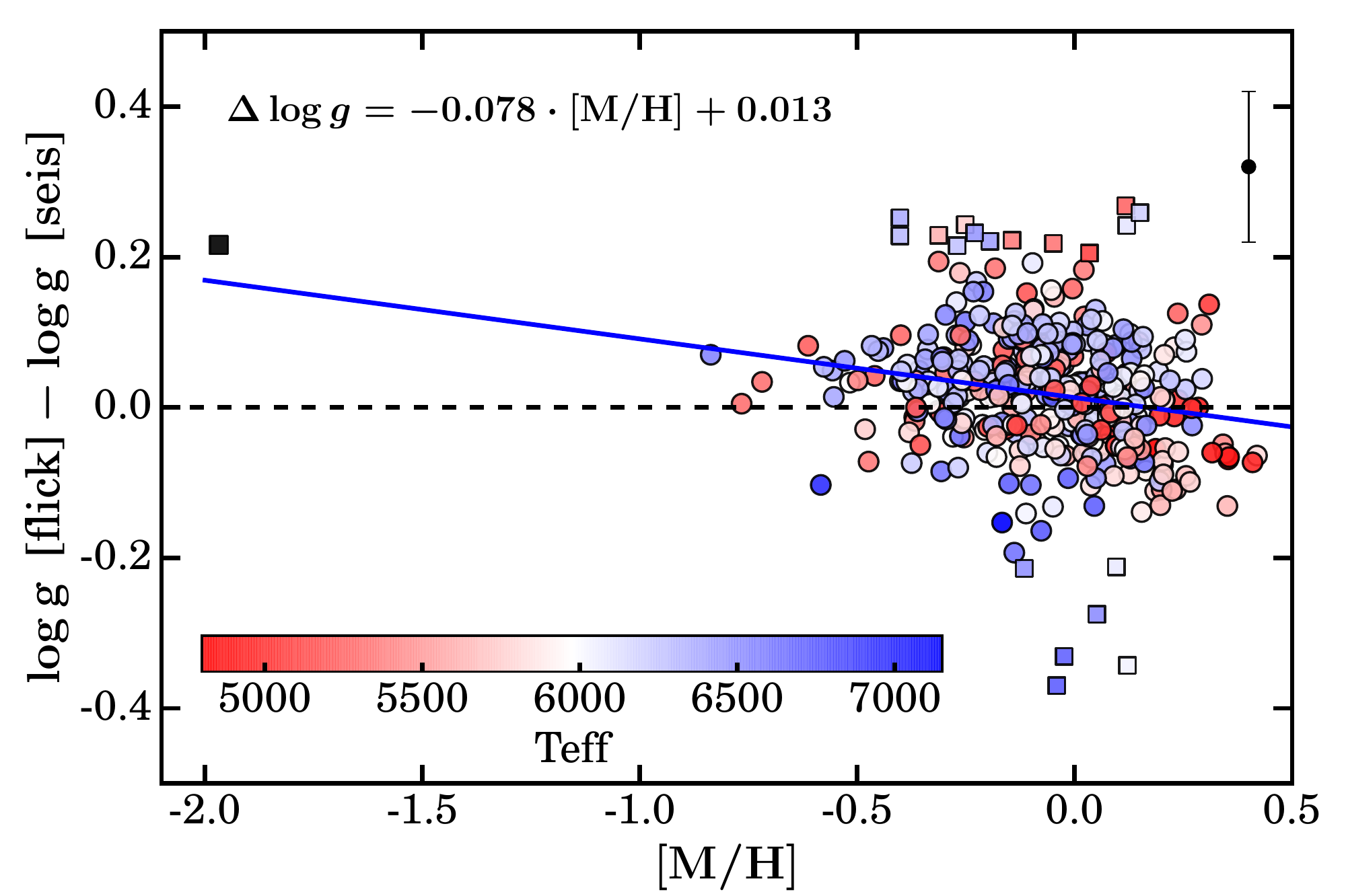}
\caption{Left panel: comparison between  seismic and flickering gravities from
  \citet{bastien:2016}.  A typical  error bar  is shown  for orientation.  The
  black square is  KIC~7341231, with $\mh=-1.97$. Right panel:  residuals as a
  function of \mh\  and a linear fit signaling a  possible, albeit mild, trend
  in flickering gravities with \mh. Squares indicate stars for which residuals
  are larger than 0.2~dex and are excluded from the fit. \label{fig:flicker}}
\end{figure*}

A  final and  useful comparison  between the  APOKASC catalog  and the  LEGACY
results  is that  of  the  typical errors  with  which  the different  stellar
quantities can be  determined based on global seismic  parameters (APOKASC) or
on  using  individual  frequencies.   The  error  distributions  presented  in
\citet{victor:2017} (see their Figure~2) are the equivalent of the statistical
errors  in the  APOKASC catalog.  Following  the publication  of the  original
LEGACY work \citep{lund:2017},  an error was identified in  the preparation of
the frequency ratios (\citealp{roxburgh:2017};  Lund et al., submitted), which
caused an  overestimation of  uncertainties in  the LEGACY  results determined
with some  of the  pipelines used in  \citet{victor:2017}. Taking  this factor
into account  (Silva Aguirre private  comm.), typical errors  embracing LEGACY
results  obtained  with  all  pipelines range  between  0.5-1\%  for  $\mrho$,
0.8-1.8\%  for $R$,  2-4\% for  $M$, and  5-10\% for  $\tau$. The  statistical
errors in  the catalog presented  in this paper  (see Tab.~\ref{tab:mederrors}
and Fig.\,\ref{fig:stat}) are  about 3\%\,($\mrho$), 2.3\%\,($R$), 4\%\,($M$),
and 15\%\,($\tau$). The  better precision of the detailed  seismic analysis is
most  evident for  $\mrho$, while  for  $R$ and  $M$ results  are closer.  For
$\tau$,  the precision  of global  seismology is  on average  a factor  of two
larger than  detailed seismic analysis. We  note that the final  errors in the
present catalog  include the  systematic component  to the  total uncertainty,
whereas  in  \citet{victor:2017} this  has  not  been  included in  the  error
estimate because results for all seismic pipelines were reported.

According  to our  results, the  precision of  global seismology  is typically
larger  than that  of detailed  seismology  by factors  from about  1.5 to  3,
depending on  the stellar property  considered. Note that in  this discussion,
the APOKASC results include stars with timeseries of all available lengths. On
the  other  hand,   the  LEGACY  sample  is  formed  by   the  best  available
\textit{Kepler} data.

The  overall comparison  of the  APOKASC  catalog, with  both $\teff$  scales,
against the  LEGACY sample  shows there are  no strong  systematic differences
present at levels larger than the mean uncertainties in the catalog. Moreover,
the typical  precision with  which GBM  pipelines estimate  stellar parameters
using  only  $\dnu$  and  $\numax$  is  comparable  to  those  obtained  using
individual oscillation  frequencies, at  least when stars  without mixed-modes
are considered. More detailed work is  however desirable to understand in full
the limitations of global seismology of dwarf and subgiant stars.

\subsubsection{Gravities from flickering}

Timescales  and  amplitudes of  granulation  are  tightly related  to  surface
gravities  \citep{kjeldsen:2011},  which was  first  observed  for red  giants
\citep{mathur:2012b} and later applied to  measure surface gravities of dwarfs
and   giants   from   long-cadence   data.  The   so-called   flicker   method
\citep{bastien:2013} is based  on a correlation between the  rms variations of
the stellar brightness on timescales shorter  than 8 hours, the so-called 8-hr
flicker $F_8$, and the surface gravitity  of stars. The relation between $F_8$
and $\logg$ was further calibrated and analyzed in \citet{bastien:2016}, using
in part seismic $\logg$ from C14. 

As discussed in Sect.\,\ref{sec:chap14}, the  availability of \mh\ from APOGEE
leads to changes in the determined seismic $\logg$ values with respect to C14,
although these corrections are small in comparison with the typical $0.10$~dex
uncertainties   in    $\logg$   obtained    from   $F_8$.   Left    panel   in
Figure~\ref{fig:flicker} compares  the newly  determined seismic  $\logg$ with
the $F_8$  $\logg$ from \citet{bastien:2016}  with symbols coded  according to
stellar metallicity. The agreement of  the flickering gravities with the newly
determined seismic  gravity is  excellent overall, with  a mean  difference of
0.017~dex (flickering$-$seismic) and  a dispersion of 0.068~dex  for the whole
APOKASC  sample. If  the C14  $\logg$  values are  used instead  for the  same
sample, the mean and dispersion of  the difference are 0.023~dex and 0.071~dex
respectively.  Including  the  \mh\  information brings  both  $\logg$  scales
(slightly) closer.  

\begin{figure*}
\centering\includegraphics[scale=.42]{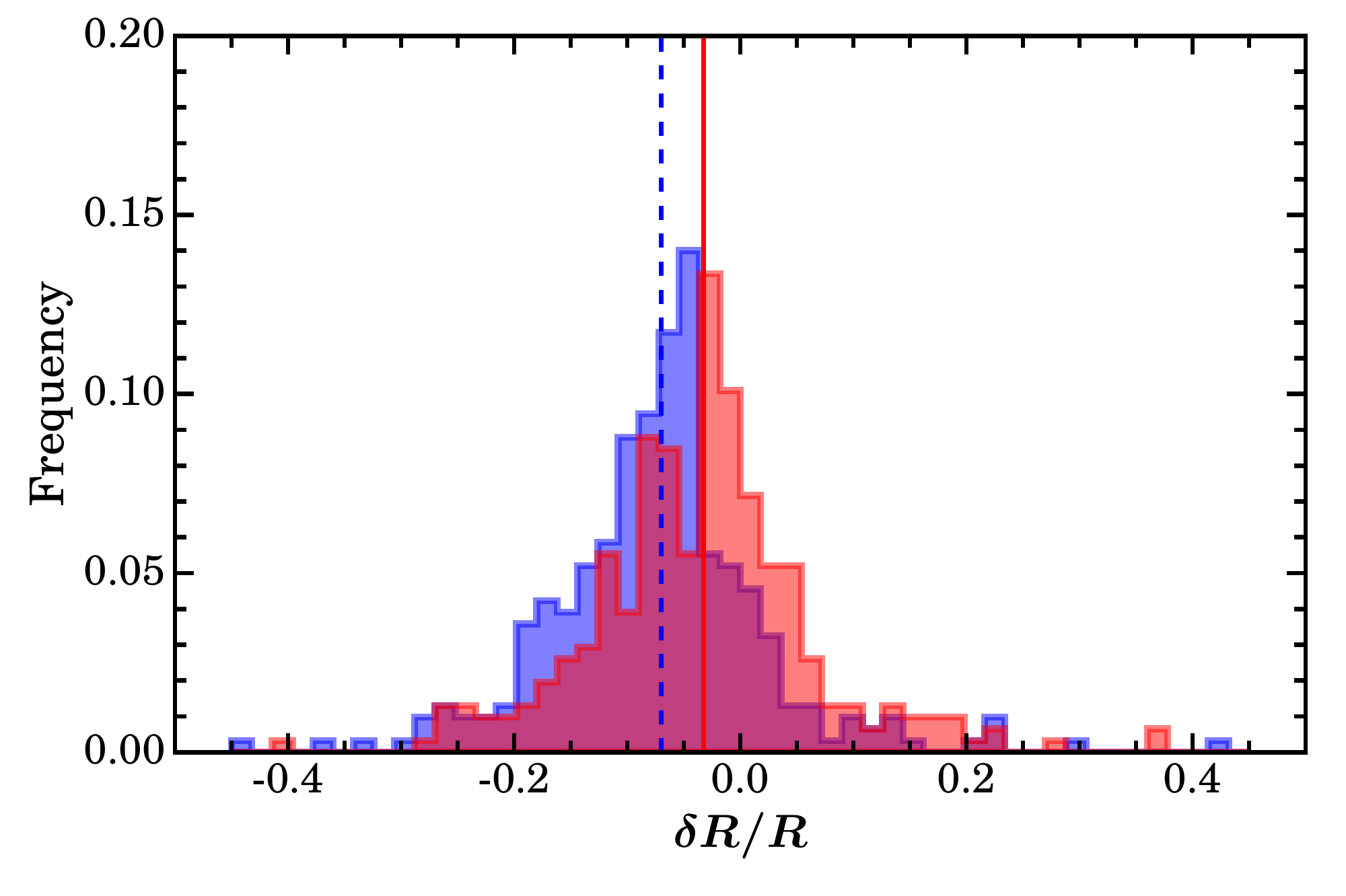}\includegraphics[scale=.42]{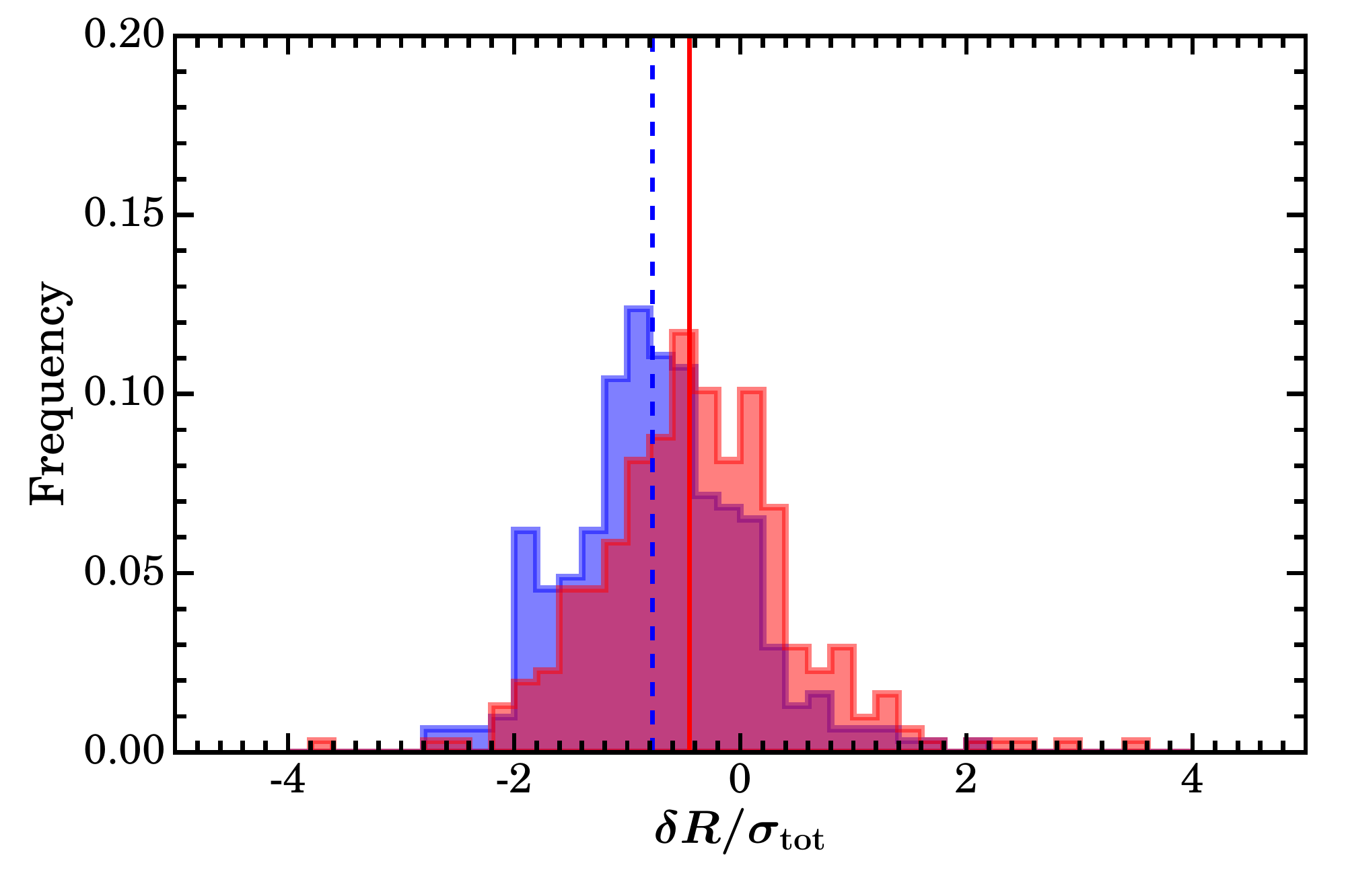}
\caption{The left panel shows fractional differences of the catalog radii with
  respect to  TGAS radii \citep{huber:2017}  for ASPCAP (blue) and  SDSS (red)
  $\teff$  scales  (in  the  sense  APOKASC$-$TGAS).  The  right  panel  shows
  differences  normalized to  the combined  uncertainties, $\sigma_{\rm  tot}=
  \sqrt{\sigma^2_{cat}+\sigma^2_{TGAS}}$.  Vertical lines  show median  values
  for ASPCAP (dashed) and SDSS (solid) samples. \label{fig:tgas}} 
\end{figure*}

More interesting is perhaps the difference between the flicker and the seismic
gravities ($\Delta \logg$)  as a function of  \mh\ that is shown  in the right
panel in  Fig.\,\ref{fig:flicker}, where colors  are the  same as in  the left
panel. Here, stars showing $|\Delta \logg| > 0.2$~dex ($\sim 2\sigma_{\rm F_8}$)
are plotted with squares. The larger  negative differences are related to some
of the hotter stars  in the sample, above $\approx 6500$~K;  as pointed out in
\citet{bastien:2016}, the  tight relation between  $F_8$ and $\logg$  seems to
break down around this or slightly  higher $\teff$. We have performed a linear
fit to $\Delta \logg$  as a function of \mh, excluding from  the fit all cases
where $|\Delta \logg| > 0.2$~dex. The  metal-poor star KIC 7341231, shown with a
black square, is also  excluded from the fit. The result is  plotted as a blue
solid line. We have also carried out linear fits to binned data. This was done
by  sorting stars  by increasing  \mh\, computing  the mean  \mh\ and  $\Delta
\logg$ in batches of  10 stars and then performing the  linear fit. The result
is    virtually    unchanged    compared    to    the    one    reported    in
Fig.\,\ref{fig:flicker}. We  have checked  that results are  not significantly
affected by  the bin size. Interestingly,  the linear fit reproduces  well the
result  for KIC  7341231, although  the typical  scatter in  flicker gravities
prevents us from drawing firm conclusions. 

The significance of the correlation between  $\Delta \logg$ and $\mh$ has been
tested  by   computing  the  Kendall's  $\tau_K$   non-parametric  correlation
test. Applying the  cut $|\Delta  \logg|  > 0.2$~dex  we obtain  for 381  stars
$\tau_K=  0.177$, leading to $z_K=5.16$ and a p-value=  $1.2\times  10^{-7}$. 
If stars are grouped in 12 bins of 30 stars each, then  $\tau_K=0.522$,   i.e. 
$z_K= 2.47$ or a p-value= $6.7\times  10^{-3}$). A  highly
significant correlation between $\Delta \logg$ and $\mh$ exists in both cases.

The result above seems to suggest  there is some degree of correlation between
the level  of flickering activity  $F_8$ and the  metallicity of the  star for
dwarfs and subgiants. As \mh\ decreases, the  tendency is for $F_8$ to lead to
larger estimates of $\logg$ than the  seismic values. Based on the $F_8-\logg$
relation  in  \citet{bastien:2016},  this  impies  lower  levels  of  activity
associated with lower  \mh. This is in qualitative  agreement with theoretical
models \citep{samadi:2011} linking F8 and mode amplitudeswith $\mh$ and recent
detailed analysis of red giant  stars carried out by \cite{corsaro:2017} based
on open star clusters. The relation between stellar activity, in particular as
measured  by $F_8$,  metallicity, and  $\logg$ deserves  further and  detailed
study. 

\subsubsection{Comparison with TGAS radii}

Recently,  \citet{huber:2017} have  used  the Gaia  DR1  (TGAS) parallaxes  in
combination with  the 2MASS-K band  and $\teff$ from  \citet{buchhave:2015} to
determine stellar  radii. Almost all stars  in the catalog in  this paper have
been  analyzed  in  \citet{huber:2017}  and  form  their  dwarf  and  subgiant
sample. Figure~\ref{fig:tgas} shows histograms comparing stellar radii for our
two $\teff$ scales with radii determined from TGAS by \citet{huber:2017} using
the same $\teff$ scales. The  median differences, indicated by vertical lines,
are  3.3\%  and 7.0\%  for  the  SDSS and  the  ASPCAP  $\teff$ based  results
respectively, with a dispersion in both cases of 10\%. This offset was already
highlighted by  \citet{huber:2017}, who  concluded that  TGAS results  favor a
hotter  $\teff$ scale  such  as the  SDSS  scale rather  than  the cooler  one
provided by ASPCAP. The right  panel shows differences normalized with respect
to  the combined  APOKASC  +  TGAS uncertainty,  where  the  offsets are  also
visible. The dispersion of uncertainties  is 0.9$\sigma$ and 1$\sigma$ for the
ASPCAP and SDSS scales respectively, and it is dominated by errors in the TGAS
results.

Our  results confirm  those  in \citet{huber:2017},  that asteroseismic  radii
agree  with  those  determined  from  Gaia parallaxes.  However,  there  is  a
systematic offset present, and its nature is not fully understood. It might be
tempting to present  this as a limitation of global  seismology (i.e. based on
$\dnu$ and  $\numax$), stellar models  (i.e. shortcomings in  determination of
$\dnuo$), or on the scaling relation  of $\numax$. But, our radii compare very
well with  those from the LEGACY  sample (Sect.\,\ref{sec:legacy}), determined
from a  seismic analysis  based on oscillation  frequencies. In  fact, seismic
parallaxes   of    the   LEGACY   sample   computed    by   \citet[see   their
  Figure~13]{victor:2017} show a systematic  offset of $\sim 0.25$~mas towards
larger  values than  those from  TGAS. Analogous  results were  also found  by
\citet{davies:2017} using asteroseismic  parallaxes of red clump  stars and by
\citet{stassun:2016} using  binaries. Because seismic parallaxes  are based on
computing  the   total  stellar  flux,  results   in  \citet{victor:2017}  are
equivalent to  having seismic radii  smaller than TGAS radii,  consistent with
our findings.  About 2\% of  the offset  in radius can  be accounted for  by a
systematic offset in TGAS parallaxes. Still, further study is required to have
a  better understanding  of  whether  the reasons  for  the  offset lies  with
asteroseismic analysis, $\teff$ scale, or a systematic underestimation of TGAS
parallaxes. Gaia DR2 will certainly help shed light on this.  

\begin{figure*}
\centering \includegraphics[scale=.42]{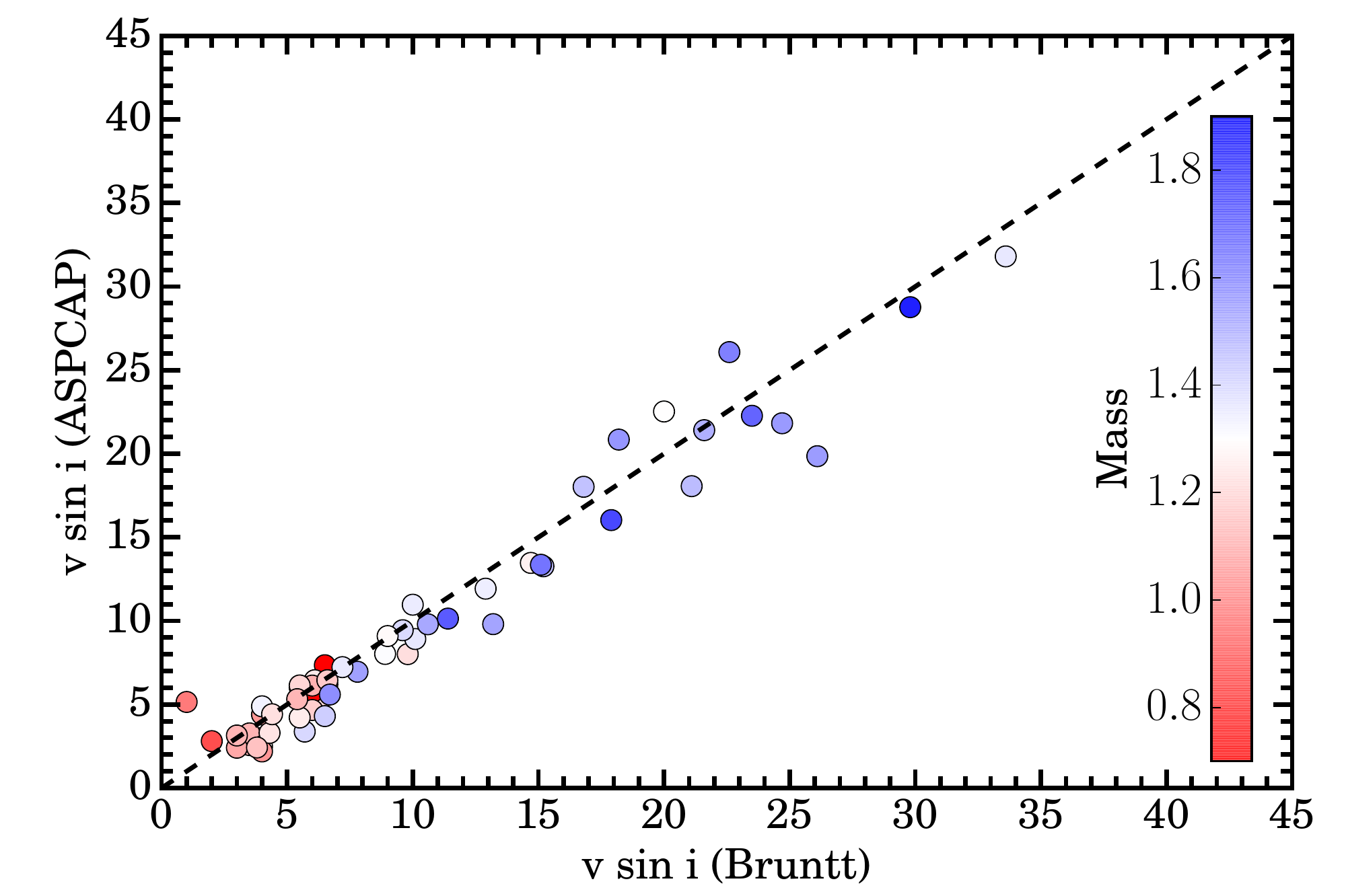}\includegraphics[scale=.42]{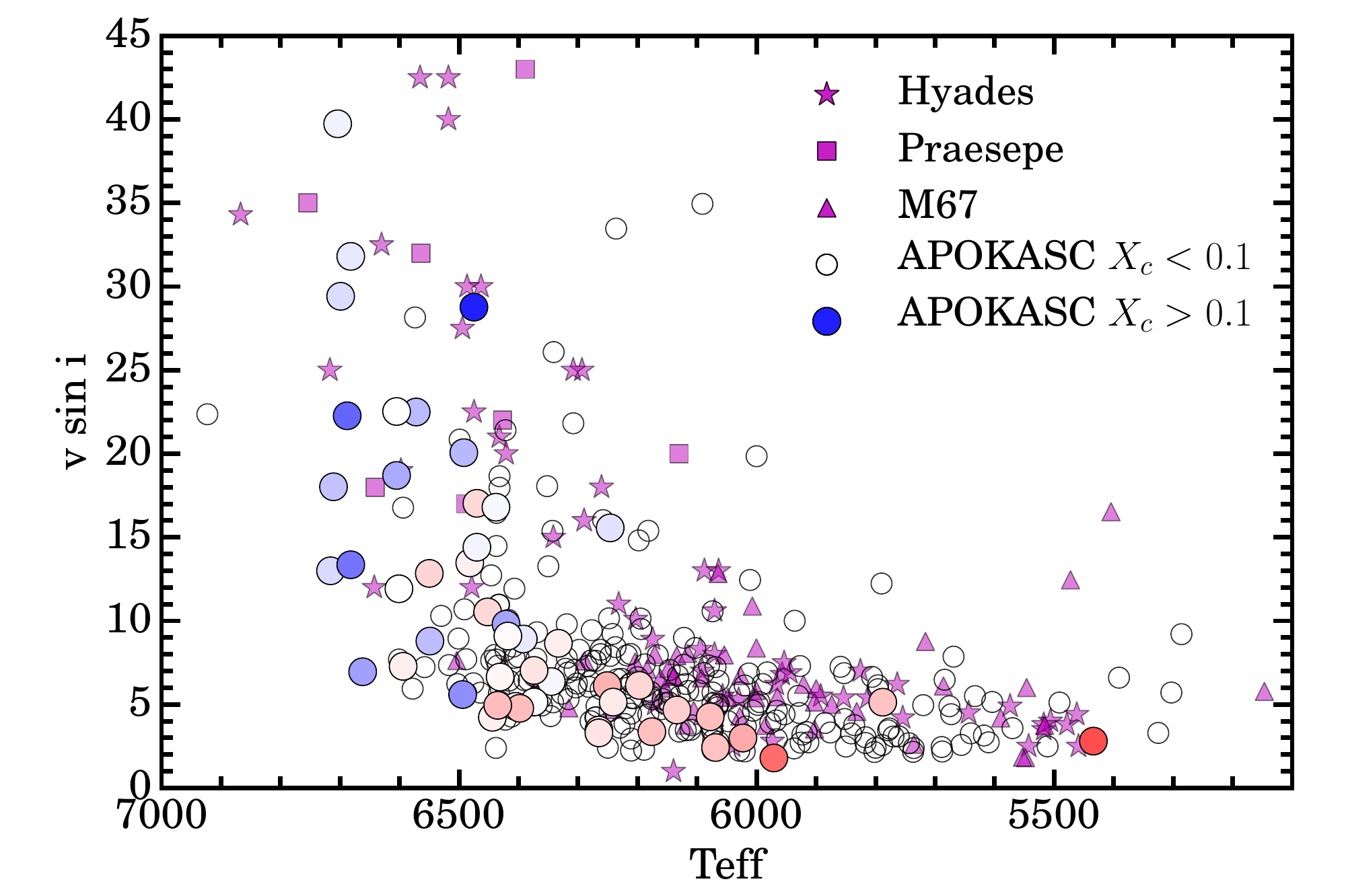}
\caption{Left  panel: ASPCAP  vs  \citet{bruntt:2012}  $\vsini$. Right  panel:
  $\vsini$ vs $\teff$ for the whole  APOKASC sample, M67 stars, and Hyades and
  Praesepe stars from \citet{vansaders:2013}.  Main sequence APOKASC stars are
  color coded according to their mass. \label{fig:vsini}} 
\end{figure*}

\subsubsection{Projected rotational velocities} \label{sec:vsini}

The synthetic grid of spectra  in ASPCAP incorporates rotational broadening as
an additional dimension in the analysis  since DR13. The possibility of having
a large sample  of stars with $\vsini$ determinations  and seismic parameters,
even  if   the  inclination   $i$  remains  unknown,  opens   up  interesting
possibilities  for  studies  of  gyrochronology, at  least  in  a  statistical
sense. In  order to assess the  quality of ASPCAP $\vsini$  determinations, we
use  stars in  common with  \citet{bruntt:2012} for which  $\vsini$ has  been
determined from optical  spectroscopy taken at resolution R$\sim 80000$ and S/N $>200$. Results are shown in the  left panel of
Figure~\ref{fig:vsini},  where  stars  are  color  coded  according  to  their
mass. There is a very good  agreement between both datasets, with a dispersion
of  $1.8\  \hbox{km/s}$  between  the  two  datasets  and  a  mean  offset  of
$-0.5\ \hbox{km/s}$ (in the sense ASPCAP$-$Bruntt).

The  comparison above  lends strong  support to  $\vsini$ determinations  with
ASPCAP. There are 331 stars in the catalog with ASPCAP $\vsini$ determinations
and they are shown in the right panel of Figure~\ref{fig:vsini} where $\vsini$
is shown as a function of $\teff$ (SDSS). We have excluded from the plot stars
that have been  flagged by ASPCAP with problematic $\vsini$.  These are mostly
stars with low ($\sim 2~\hbox{km/s}$)  $\vsini$. Main sequence stars are shown
as large filled circles, color coded as in the left panel of the figure. Stars
that are close to the end of the  main sequence or have already evolved off it
are shown as  small open circles. This is an  approximate classification based
on  whether the  central  hydrogen  mass fraction  $X_c$,  estimated from  the
\texttt{BeS/G-$\dnuscl$} GBM results, is larger  or smaller than 0.1. The plot
includes ASPCAP  values of $\vsini$ for  M67 stars. Hyades and  Praesepe stars
from \citet{vansaders:2013} are also included. For all three clusters, $\teff$
have  been determined  using  $(B-V)$ colors  and the  \citet{casagrande:2010}
transformation. The  general trend in  the APOKASC  sample is similar  to that
seen in cluster stars. Cool stars ($\teff < 6200$~K) show slow rotation in all
but a few cases. Above this temperature, as surface convective envelopes thin,
magnetic braking becomes less effective and higher rotation rates are present,
as in cluster stars \citep{vansaders:2013}. At $\teff > 6200$~K there seems to
be an  excess of  low $\vsini$  in comparison  to cluster  stars. This  may be
indicative that there  is a relative lack of fast  rotators in seismic samples
\citep{tayar:2015} and the particular exclusion  of close binary stars, which,
at  least for  red giant  stars, tend  to have  rapid rotation  but suppressed
oscillations  \citep{gaulme:2014}. Ongoing  work  includes  the comparison  of
estimated  $\vsini$ distributions  obtained  for stars  with known  rotational
periods  to further  confirm  the reliability  of  $\vsini$ determinations  by
ASPCAP (Simonian et al. in prep.).

\section{Summary} \label{sec:summary}

This  work  extends  the  first   APOKASC  red  giants  catalog  presented  in
\citet{pinsonneault:2014} to include 415 dwarf and subgiant stars. It is based
on global seismic  parameters $\dnu$ and $\numax$  obtained from \emph{Kepler}
short cadence asteroseismic  data with length of lightcurves  spanning from 30
up to  1055 days. Adopted  seismic values  are from the  \texttt{SYD} pipeline
\citep{huber:2009},  and  four other  pipelines  have  been used  to  estimate
systematic uncertainties in $\dnu$ and  $\numax$. Two $\teff$ scales have been
used throughout this work, a spectroscopic scale based on the $H$-band spectra
taken  by  APOGEE  \citep{majewski:2015}  and  the other  one  based  on  SDSS
\emph{griz}  photometry  \citep{pinsonneault:2012}.   APOGEE  data  have  been
reduced  using  ASPCAP, in  the  version  corresponding  to the  SDSS-IV  DR13
\citep{albareti:2017}, from which we have also taken the metallicity \mh\ used
in combination  with both  $\teff$ scales.  Typical $\teff$  uncertainties are
about 70~K  both for ASPCAP  and SDSS scales.  The formal \mh\  uncertainty in
ASPCAP is 0.03~dex, but we have expanded it to 0.1~dex to account for possible
systematic errors.

The catalog includes  the stellar quantities, $g$, $\mrho$,  $R$, $M$, and
$\tau$ determined with GBM techniques. In total, twelve different combinations
of GBM pipelines and sets of  stellar evolution tracks have been used. Central
values     in     the      catalog     are     from     \texttt{BeS/G-$\dnuo$}
(Table\,\ref{tab:gbms}).   Each   quantity   is  accompanied   by   assymetric
statistical errors (those  coming from uncertainties in the input  data) and a
symmetric  systematic  uncertainty that  captures  the  dispersion of  results
across GBM pipelines.  In the case of ASPCAP results  an additional systematic
component, computed  by redetermining  all stellar  quantities using  as input
$\teff$ the  ASPCAP values  $\pm100$~K, has  been added  in quadrature  to the
systematic component from the GBM pipelines. The median total uncertainties in
the catalog are (SDSS/ASPCAP): 2.3/2.3\%\,($g$), 2.8/2.7\%\,($\mrho$),
2.1/2.4\%\,($R$), 4.4/5.5\%\,($M$), and 17/21\%\,($\tau$).

Stellar properties in the catalog compare well with other published data, most
notably   with    the   asteroseismic    results   of   the    LEGACY   sample
\citep{victor:2017}, that are based on  a more detailed asteroseismic analysis
which   employs   individual  frequencies   rather   than   just  two   global
quantities.  Our  results  show  small  systematic  offsets  for  all  stellar
quantities and the ASPCAP $\teff$ scale and,  for the SDSS scale, a 3.6\% mean
offset  for $M$  and a  $-15.6$\% offset  for $\tau$,  still smaller  than the
typical uncertainties in  the catalogs. We have checked that  about half these
offsets are due to  the SDSS $\teff$ scale being 1.4\%  hotter than the LEGACY
$\teff$ scale.  The overall agreement  between our global seismic  results and
the  more  detailed  LEGACY  work  is remarkable  considering  how  much  less
information   is   contained  in   global   seismology   than  in   individual
frequencies.  Prospects for  accurate determination  of stellar  properties of
dwarfs and subgiants  with the TESS mission, that will  observe most stars for
27~days and in  many cases will allow measuring only  $\dnu$ and $\numax$, are
excellent.

We have  also compared  the statistical uncertainties  of our  derived stellar
parameters with those in the LEGACY  sample and find them comparable. In fact,
our  statistical   uncertainties  are   comparable  to  the   most  pesimistic
uncertainties in \citet{lund:2017,victor:2017} for $R$  and $M$, and larger by
factors between  1.5 and  3 for  $\mrho$ and $\tau$.  The good  performance of
global  seismology is  due to  the  rather small  fractional uncertainties  in
$\dnu$ and $\numax$ in comparison to  those of the frequency separation ratios
used  by  several  pipelines  in LEGACY.  Error  propagation  from  individual
frequencies, even  they are  measured to  levels of 0.1\%  or better,  lead to
uncertainties in  frequency ratios that  range typically  between a few  up to
20\%.  It  is  then  apparent  that a  better  understanding  of  the  surface
correction term  continues to be searched  for, because only by  being able to
perform seismic modeling based on individual frequencies will unleash the full
power of detailed seismology.

The importance  of $\mh$ for  determination of stellar properties  from global
seismic measurements  is made explicit in  the comparison of our  results with
those in C14.  We have derived approximate linear relations  between $\mh$ and
fractional variations  in stellar properties such that $\delta  \logg =
0.062 \,\delta\mh$, $\delta \mrho/\mrho =  -0.044 \, \delta\mh$, $\delta R/R
 = 0.103 \,  \delta\mh$, $\delta  M/M = 0.267  \, \delta\mh$, and  $\delta \tau 
 / \tau = -0.493 \,\delta \mh$.

Finally, our  stellar radii have  been compared  with those inferred  from the
TGAS parallaxes  \citep{huber:2017}. We confirm  previous results and  find an
overall  good  agreement between  asteroseismic  and  TGAS  radii but  with  a
systematic shift of TGAS radii being  $\sim 3\%$ larger than seismic radii for
the  SDSS  $\teff$   scale.  This  is  in  agreement  with   the  analysis  by
\citet{victor:2017}   using  different   seismic   information.  Assuming   no
systematic  errors  in asteroseismic  radii  and  $\teff$ scales,  this  would
indicate that TGAS parallaxes are systematically underestimated by $\sim 3\%$.

We have  presented the first large-scale  catalog of dwarf and  subgiants with
seismically   inferred   stellar   parameters   and   individual   metallicity
measurements.  The detailed  information  on the  composition  of these  stars
obtained and released as part of DR13 makes this APOKASC catalog a unique data
source for stellar studies.

\acknowledgements

We thank the anonymous referee for a very careful reading of the manuscript and the many comments tha have helped improving the presentation of results.

Funding for the Sloan Digital Sky Survey IV has been provided by the
Alfred P. Sloan Foundation, the U.S. Department of Energy Office of
Science, and the Participating Institutions. SDSS acknowledges
support and resources from the Center for High-Performance Computing at
the University of Utah. The SDSS web site is \url{www.sdss.org}.

SDSS is managed by the Astrophysical Research Consortium for the Participating
Institutions of  the SDSS Collaboration including  the Brazilian Participation
Group, the Carnegie  Institution for Science, Carnegie  Mellon University, the
Chilean    Participation    Group,    the    French    Participation    Group,
Harvard-Smithsonian  Center for  Astrophysics, Instituto  de Astrof\'isica  de
Canarias, The  Johns Hopkins University,  Kavli Institute for the  Physics and
Mathematics of  the Universe (IPMU)  / University of Tokyo,  Lawrence Berkeley
National  Laboratory,   Leibniz  Institut   für  Astrophysik   Potsdam  (AIP),
Max-Planck-Institut f\"ur Astronomie  (MPIA Heidelberg), Max-Planck-Institut f\"ur
Astrophysik (MPA Garching),  Max-Planck-Institut f\"ur Extraterrestrische Physik
(MPE),  National  Astronomical  Observatories   of  China,  New  Mexico  State
University,  New  York  University,  University of  Notre  Dame,  Observatório
Nacional /  MCTI, The  Ohio State  University, Pennsylvania  State University,
Shanghai  Astronomical   Observatory,  United  Kingdom   Participation  Group,
Universidad Nacional  Aut\'onoma de México, University  of Arizona, University
of  Colorado   Boulder,  University  of  Oxford,   University  of  Portsmouth,
University  of  Utah,  University   of  Virginia,  University  of  Washington,
University of Wisconsin, Vanderbilt University, and Yale University. 

A.S.  is partially  supported by  ESP2015-66134-R (MINECO).  M.H.P., J.T.  and
J.A.J. acknowledge support from NASA grant NNX15AF13G. Funding for the Stellar
Astrophysics Centre  is provided  by The  Danish National  Research Foundation
(Grant  DNRF106). V.S.A.  acknowledges  support from  VILLUM FONDEN  (research
grant  10118). M.N.L.  acknowledges  the  support of  The  Danish Council  for
Independent  Research|Natural Science  (Grant DFF-4181-00415).  We acknowledge
funding  from the  European Research  Council under  the European  Community's
Seventh Framework  Programme (FP7/2007-2013) /  ERC grant agreement  no 338251
(StellarAges). D.S. is the recipient  of an Australian Research Council Future
Fellowship  (project  number  FT1400147).  R.A.G.   acknowledges  the  funding
received from  the CNES. D.A.G.H. was  funded by the Ram\'on  y Cajal fellowship
number RYC-2013-14182. D.A.G.H. and O.Z.   acknowledge support provided by the
Spanish  Ministry   of  Economy  and  Competitiveness   (MINECO)  under  grant
AYA-2014-58082-P. 

\software{ASPCAP \citep{zamora:2015,garciaperez:2016}, FERRE \citep{allende:2006}, A2Z \citep{mathur:2010}, COR \citep{mosser:2013}, FITTER \citep{davies:2016}, OCT \citep{hekker:2010}, SYD \citep{huber:2011},  BASTA \citep{victor:2015}, BeSPP \citep{serenelli:2013}, GARSTEC \citep{weiss:2008}, GOE \citep{hekker:2014}, CESTAM2k \citep{marques:2013}, MESA \citep{paxton:2013}, MPS \citep{hekker:2014}, RADIUS \citep{stello:2009b}, BaSTI \citep{pietrinferni:2004}, ASTEC \citep{astec:2008}, SFP \citep{kallinger:2012}, YB \citep{basu:2010, gai:2011}.}

\appendix
\section{Description of stellar models}\label{app:models}

This appendix  gives details of  the physical inputs  in the grids  of stellar
models that have been used in the  GBM pipelines used in this work. Overall, 7
different  grids  of models  have  been  used  based  on 6  stellar  evolution
codes. Two grids  have been computed with \texttt{GARSTEC}  but with different
physical  inputs  and  assumptions   regarding  chemical  composition  of  the
grid. The  implementation of a grid  of stellar models might  be different for
each GBM pipeline, e.g. some pipelines interpolate the original grid to have a
denser coverage of  the parameter space; these details  have been discussed
in the  main body of  the paper (Sect.\,\ref{sec:gbm}).  The  basic references
for   each   stellar   code  are:   \texttt{GARSTEC}   -   \citet{weiss:2008},
\texttt{CESTAM2k} -  \citet{morel:2008}, \texttt{MESA}  - \citet{paxton:2013},
\texttt{BASTI}      -     \citet{pietrinferni:2004},      \texttt{ASTEC}     -
\citet{astec:2008}, \texttt{YREC2} - \citet{demarque:2008}.

The     most    important     information    has     been    summarized     in
Tables~\ref{tab:grids1}~and~\ref{tab:grids2}. The first table lists quantities
related to  calibrations. All grids use  variants of the Mixing  Length Theory
(MLT; \citealp{bohm:1958,kippen:1990})  except \texttt{CESTAM2k}  that employs
the Full Spectrum Theory (FST;  \citealp{canuto:1996}). In both cases there is
a free parameter  that relates the mixing length to  the pressure scale height,
$\Lambda = \alpha_c H_P$, with $c$  denoting either MLT or FST. This parameter
is determined in all grids from a solar calibration. The MLT is not a uniquely
defined  theory   because,  in  addition   to  the  mixing   length  parameter
($\alpha_{\rm    MLT}$)   other    free    parameters    are   present    (see
e.g. \citealt{salaris:2008}).  Moreover, a  solar calibration  of $\alpha_{\rm
  MLT}$ depends on  other physical inputs of the models  such as the inclusion
and  efficiency of  microscopic  diffusion,  low-temperature opacities,  model
atmosphere  and the  reference  solar composition.  Therefore, the  $\alpha_c$
values are  given mostly for  reference but  not for direct comparison  among each
other. 

Each set  of models  also use  a different relation  to determine  the initial
composition of models, although they all  assume a linear relation between the
initial metal and  helium mass fractions, i.e. $Y_{\rm  ini}= (\Delta Y/\Delta
Z) \cdot  Z_{\rm ini}  + Y_{\rm R}$.  For each grid,  $Y_{\rm R}$  and $\Delta
Y/\Delta Z$ are  included in the table. With the  exception of \texttt{ASTEC},
these choices  lead to  a small range  of $Y_{\rm ini}$  values for  any given
$Z_{\rm ini}$. The  distribution of metals is determined by  the adopted solar
mixture, and all  grids are based  either on GN93 or GS98,  two solar mixtures
with  more similarities  than differences.  Finally, the  definition of  \feh=0
associated with each grid  is also listed. This is necessary because
stellar models depend  on the absolute $Z$ and a zero point  relating $Z$ (or
$Z/X$) with  $\feh$ is needed to  place stellar models onto  the observational
plane.

\floattable
\tabletypesize{\footnotesize}
\begin{deluxetable}{lcccccc}
\tablecaption{Reference quantities\label{tab:grids1}}
\tablehead{Grid & \multicolumn{6}{c}{Reference values} \\
& Conv.\,Theory & $\alpha_c$ & \yr & $\Delta Y/\Delta Z$ & Mixture &  \feh=0}
\startdata   
\texttt{GARSTEC/BASTA} & MLT & 1.791 & 0.248 & 1.4 & GS98 & $Z/X=0.0230$ \\
\texttt{GARSTEC/BeSPP} & MLT & 1.798 & 0.2485 & 1.17 & GN93 & $Z/X=0.02439$\\
\texttt{CESTAM2k} & FST & 0.707 & 0.245 & 1.45 & GS98& $Z/X=0.02293$ \\
\texttt{MESA} & MLT & 1.908 & 0.245 &  1.45 & GS98& $Z/X=0.02293$\\
\texttt{BASTI} & MLT & 1.913 & 0.245 & 1.4 & GN93 & $Z/X=0.0245$ \\
\texttt{ASTEC} & MLT & 1.800 & 0.300 &  $-$1 & GN93 & $Z=0.0188$ \\
\texttt{YREC2} & MLT & 1.826 & 0.245 &  1.54 & GS98 & $Z/X=0.0230$ \\
\enddata
\tablecomments{See text for detailed meaning of each column.}
\end{deluxetable}

Table~\ref{tab:grids2}  lists inputs  to  stellar  models describing  physical
inputs. Microscopic diffusion is included  only in the \texttt{GARSTEC} models
used in combination  with \texttt{BeSPP}. The basic prescription  is that from
\citet{thoul:1994}, but  these models  also include extra  mixing (generically
linked to  turbulent mixing) below  the convective envelope that  moderate the
efficiency  of gravitational  settling of  helium and  metals as  suggested by
solar   models  \citep{delahaye:2006,villante:2014}   and  required   for  low
metallicty  stars  \citep{richard:2002}.  This  is modeled  according  to  the
parametrization  of   \citet{vandenberg:2012},  that  links   the  macroscopic
diffusion coefficient  to the density at  the base of the  convective envelope
and the mass thickness of the convective envelope. 

\floattable
\tabletypesize{\footnotesize}
\begin{deluxetable}{lcccccc}
\tablecaption{Physical inputs\label{tab:grids2}}
\tablehead{Grid & Diffusion & Core OV & Nucl. rates & High/low-$T$ opac. &  EoS }
\startdata   
\texttt{GARSTEC/BASTA} & No & $f=0.016$+geom.cut & NACRE+LUNA & OPAL/F05 & OPAL05  \\
\texttt{GARSTEC/BeSPP} & Yes+extra mixing & $f=0.016$+geom.cut & A11 & OPAL/F05 & FreeEOS  \\
\texttt{CESTAM2k} & No & No & NACRE & OPAL/F05 & OPAL05  \\
\texttt{MESA} & No &  No & NACRE & OPAL/F05 & OPAL05  \\
\texttt{BASTI} & No & No & NACRE & OPAL/AF94 & FreeEOS  \\
\texttt{ASTEC} & No & No & BP95 & OPAL/K91 & E73  \\
\texttt{YREC2} & No & 0.2~$H_P$ & A98+LUNA & OPAL/F05 & OPAL05  \\
\enddata
\tablecomments{See text for detailed meaning of each column. NACRE: \citet{angulo:1999}, 
LUNA: \citet{formicola:2004,marta:2008}, A11: \citet{adelberger:2011}, BP95: \citet{bahcall:1995}, 
A98: \citet{adelberger:1998}, OPAL: \citet{opal:1996}, F05: \citet{ferguson:2005},
AF94: \citet{alexander:1994}, K91: \citet{kurucz:1991}, OPAL05: \citet{rogers:2002},
FreeEOS: \citet{cassisi:2003}, E73: \citet{eggleton:1973}. 
Note that A11 recommended  value for ${\rm ^{14}N(p,\gamma)^{15}O}$ is
  that from LUNA \citealp{marta:2008}}.
\end{deluxetable}

Convective  core  overshooting  (core OV) is  included   in  some  sets  of  models.  
In
\texttt{YREC2} models, this is done as  an extension of the convective core by
a fixed fraction  of the pressure scale height.  In \texttt{GARSTEC}, chemical
mixing is treated as a diffusive process and overshooting at all boundaries of
convective regions is applied by  means of an exponentially decaying diffusive
coefficient \citep{freytag:1996}. The free parameter $f$ is adjusted such that
it is roughly equivalent to 0.2-0.25~$H_{\rm  P}$ for convective cores of main
sequence stars. Note  that $H_{\rm P}$ tends to infinity  for convective cores
approaching zero  extension. In  order to avoid  such unphysical  situation, i.e. 
a tiny small convective  core with an infinite overshooting  region, a geometric
suppression  is added  in the  calculation of  the diffusion  coefficient (see
\citealt{magic:2010} for details).  

Nuclear reaction rates  come from a variety  of sources. But, in  this work we
are  only concerned  about  hydrogen  burning processes  and  among all  rates
involved in the pp-chains and CNO-bicycle the only relevant difference between
the  stellar   codes  is   the  adoption,   or  not,   of  the   LUNA  results
\citep{formicola:2004,marta:2008}   for  the   ${\rm  ^{14}N(p,\gamma)^{15}O}$
rate.  This is  done in  both sets  of models  computed with  \texttt{GARSTEC}
(\citealt{adelberger:2011}  rate for this reaction is taken from the LUNA value available at that time, \citealt{marta:2008}) and  also
with \texttt{YREC2}. The LUNA rate for this  reaction is roughly a factor of a
half    compared   to    the   NACRE    \citep{angulo:1999}   or    the   BP95
\citep{bahcall:1995} values  and the  most relevant  impact on  stellar models
used in this work is that  the transition from pp-chain dominated evolution to
CNO-bicycle dominated evolution is shifted  towards slightly larger masses. As
a result, the stellar mass at which convective cores start to develope is also
slightly larger, by 0.07~$\msun$ approximately (e.g. \citealp{magic:2010}).  

Radiative opacities  in stellar interiors  are from OPAL  \citep{opal:1996} in
all   cases.    Low-temperature   opacities    are   in   most    cases   from
\citet{ferguson:2005},    but    those   from    \citet{alexander:1994}    and
\citet{kurucz:1991} are also used. It must be noted, however, that most of the
differences that  might arise  from using different  low-temperature opacities
are absorbed  by using an $\alpha_c$  value that is solar  calibrated. This is
true because we are concerned with  the modeling of main sequence and subgiant
stars, i.e. with  stars that are not  very cool and far  from solar conditions
such that different combinations  of low-temperature opacities and calibration
of convection start to produce strongly deviant stellar models.  

Finally, the  equation of state  (EoS) is also  different. Some codes  use the
OPAL EoS \citep{rogers:2002}  in its 2005 release, some an  updated version of
the                                                                    FreeEOS
\citep{cassisi:2003}\footnote{\url{http://freeeos.sourceforge.net/}}       and
\texttt{ASTEC} employs the simpler EoS developed by \citet{eggleton:1973}. 

It is not  our goal here to  delve into the detailed changes  in the estimated
stellar properties  that occur  due to  variations in  the physical  inputs of
models.  Instead,  we  just  present  the results  obtained  for  all  stellar
quantities  from all  GBM pipelines  and assume  the scatter  as a  measure of
systematic uncertainties  associated, at least partially,  to uncertainties in
the   physical   inputs   of   stellar   models  as   it   is   discussed   in
Sect.\,\ref{sec:systgbm}.

\section{GBM Comparisons}\label{app:gbm}

This appendix shows detailed comparisons of results for different pairs of GBM
calculations for the five  stellar quantities seismically determined: $g$,
$\mrho$, $R$, $M$, and $\tau$. There  are twelve different sets of GBM results
that, as  described in Sect.\,\ref{sec:syst},  are used to  derive systematics
uncertainties. Results  shown here  are based  on the  SDSS $\teff$  scale but
those based on the ASPCAP $\teff$ scale portrait the same picture. 

\begin{figure*}
\centering
\includegraphics[scale=.42]{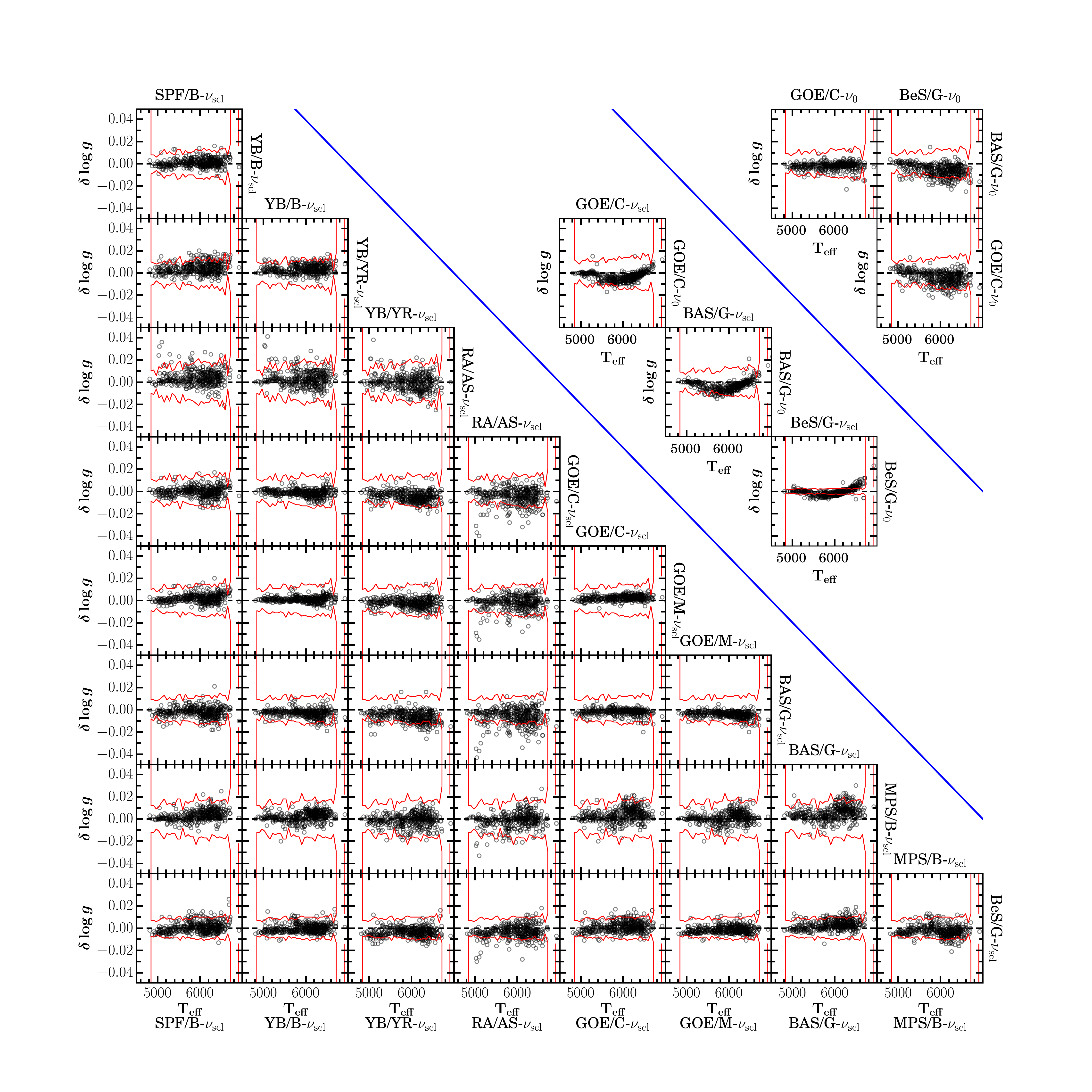}
\caption{One-to-one comparisons of different  GBM sets of results. Differences
  are computed in the sense ($\delta \logg= \logg_r - \logg_c$), where $c$ and
  $r$ denote, respectively, the GBM identifying the column and the row of each
  subplot. Red lines show the median formal  error of the pipeline $r$ in 60~K
  $\teff$  bins.   Results  shown   here  are  based   on  the   SDSS  $\teff$
  scale. GBM labels are given in Table~\ref{tab:gbms}. \label{fig:gbmlogg}}
\end{figure*}

\begin{figure*}
\centering
\includegraphics[scale=.42]{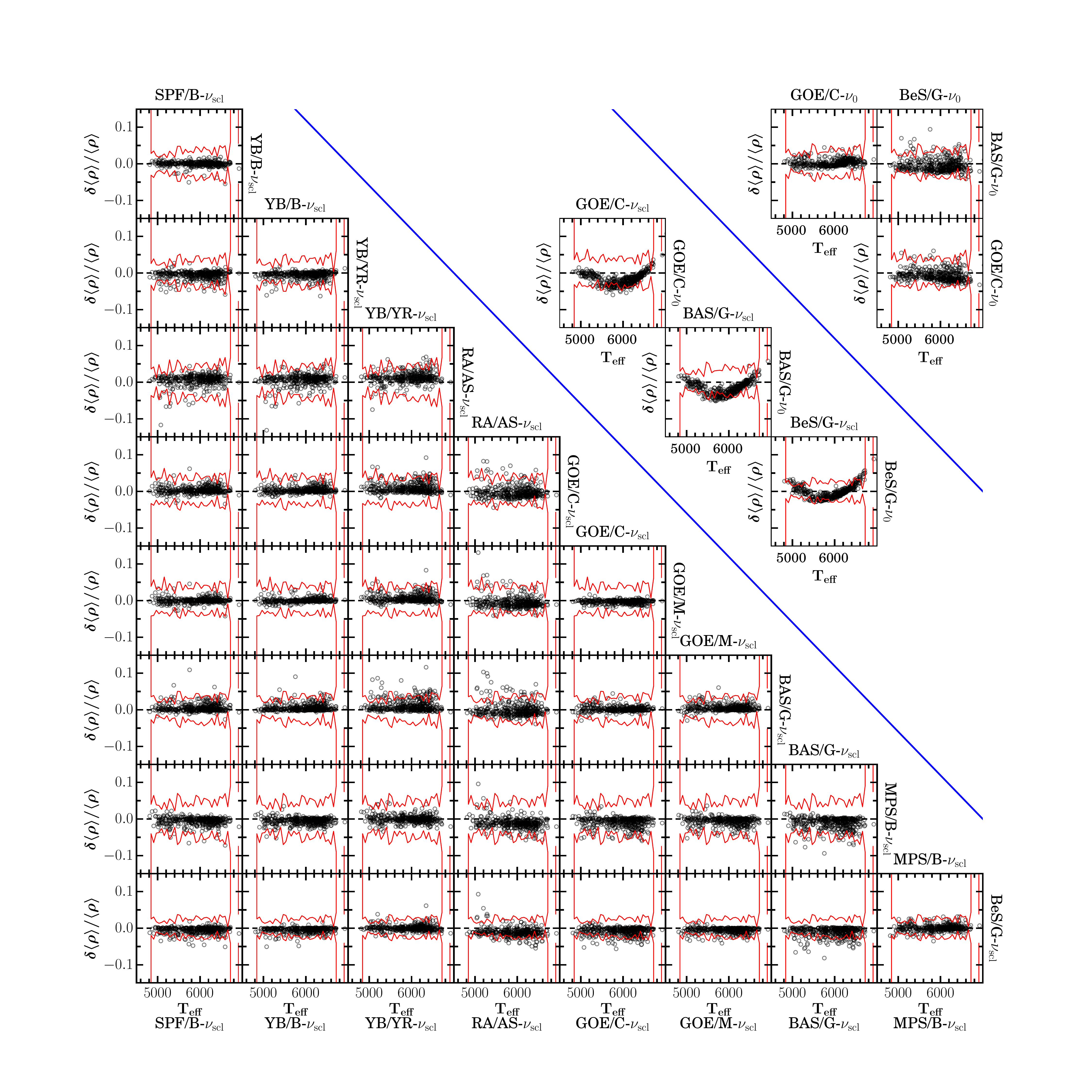}
\caption{Same as Fig.\,\ref{fig:gbmlogg} but for   mean density $\mrho$. Fractional differences are computed as 
$\delta \mrho = (\mrho_r - \mrho_c)/\mrho_c$. GBM labels are given in Table~\ref{tab:gbms}. 
\label{fig:gbmrho}}
\end{figure*}

\begin{figure*}
\centering
\includegraphics[scale=.42]{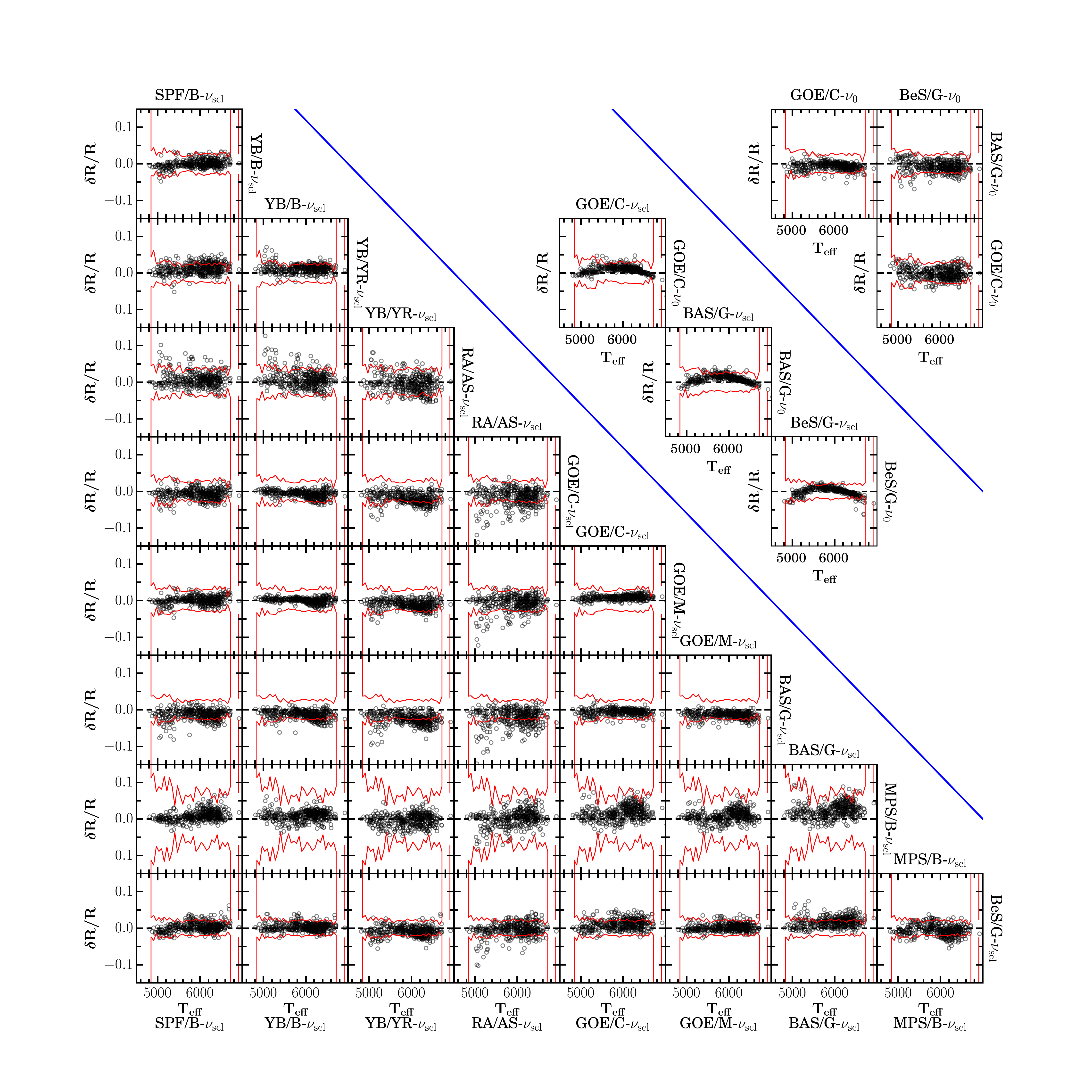}
\caption{Same    as   Fig.\,\ref{fig:gbmrho}    but    for   stellar    radius
  $R$. GBM labels are given in Table~\ref{tab:gbms}.  \label{fig:gbmrad}} 
\end{figure*}

\begin{figure*}
\centering
\includegraphics[scale=.42]{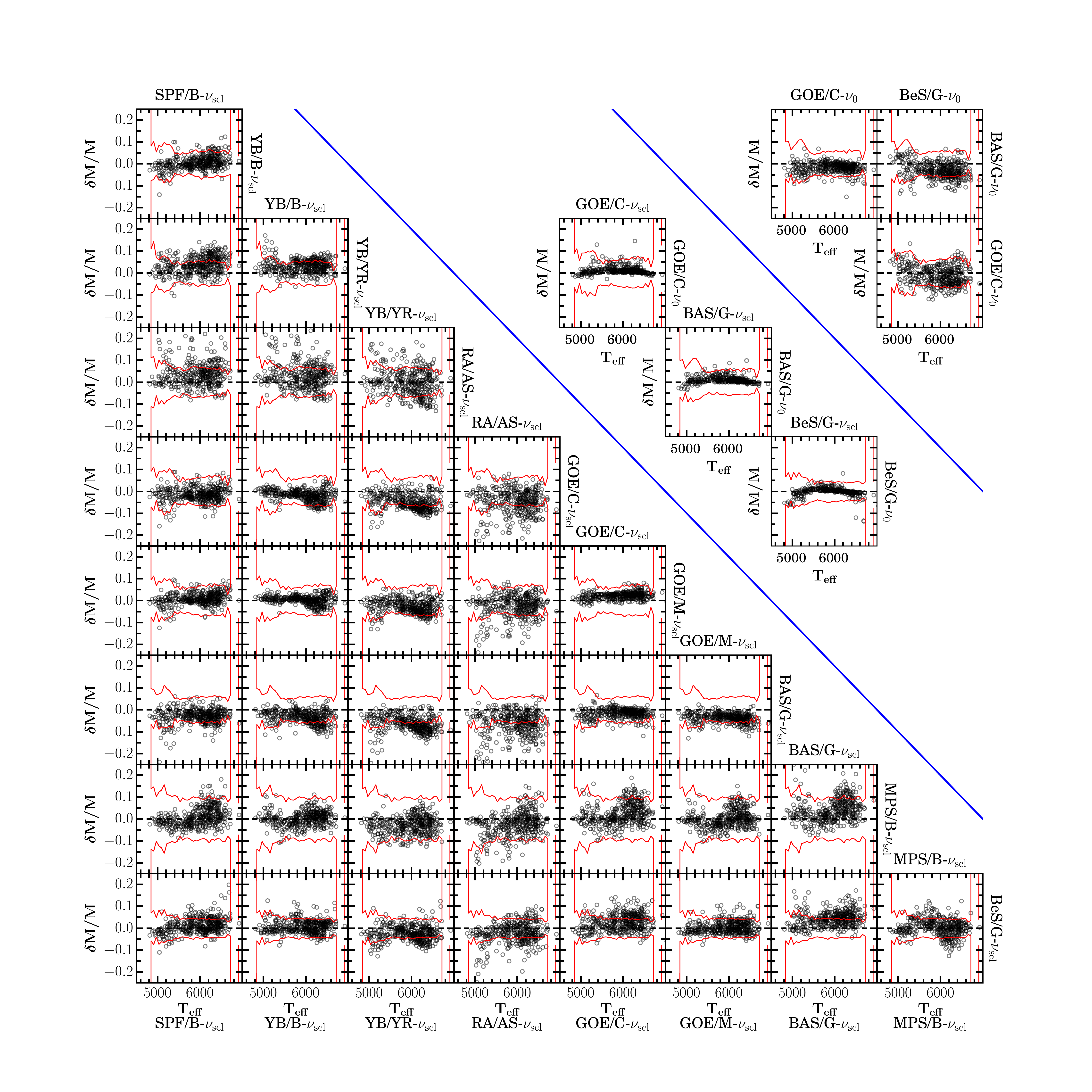}
\caption{Same    as    Fig.\,\ref{fig:gbmrho}    but    for    stellar    mass
  $M$. GBM labels are given in Table~\ref{tab:gbms}. \label{fig:gbmmass}} 
\end{figure*}

\begin{figure*}
\centering
\includegraphics[scale=.42]{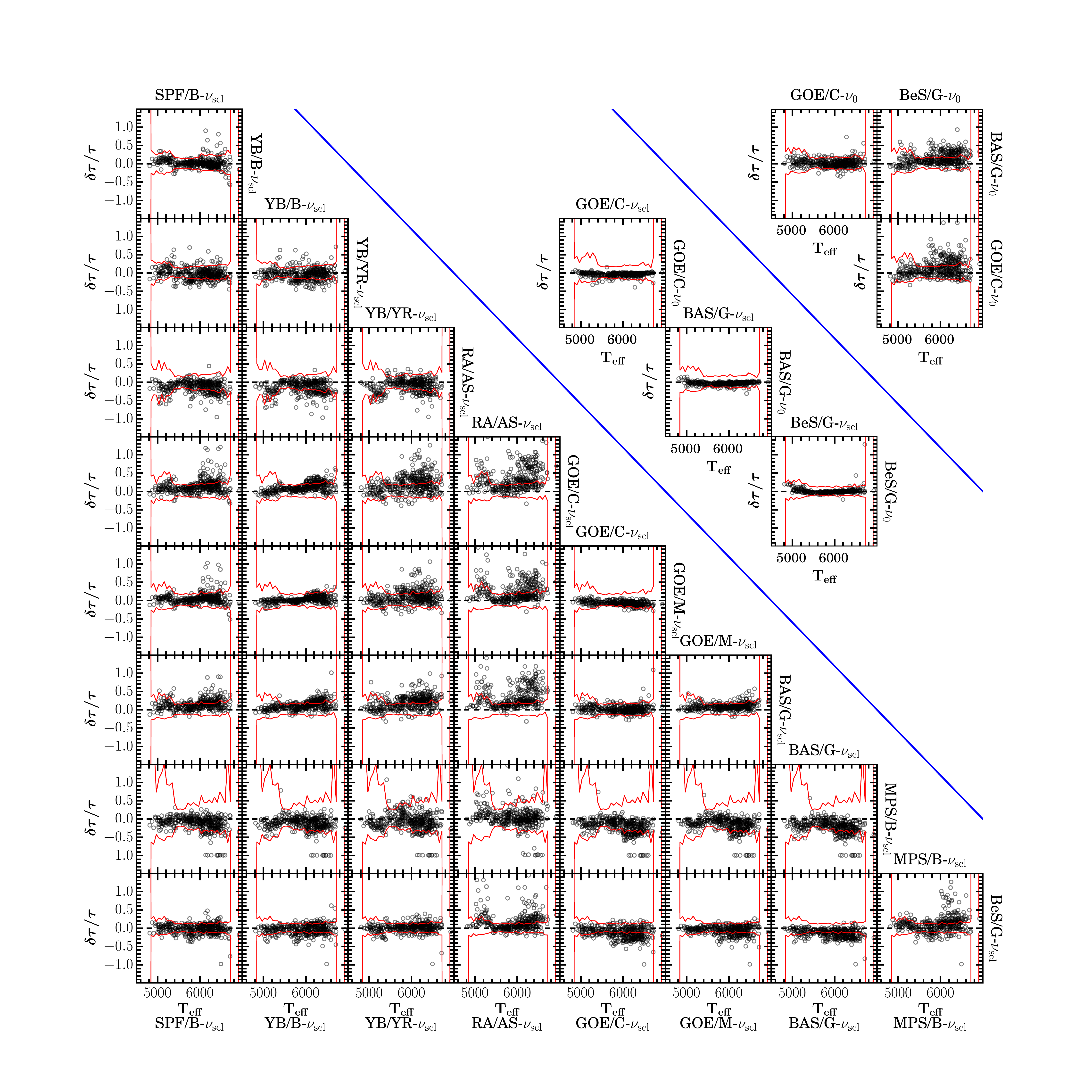}
\caption{Same    as    Fig.\,\ref{fig:gbmrho}     but    for    stellar    age
  $\tau$. GBM labels are given in Table~\ref{tab:gbms}. \label{fig:gbmage}} 
\end{figure*}

Figures~\ref{fig:gbmlogg}-\ref{fig:gbmage}  compare   one-to-one  GBM  results
where fractional differences (in dex for $\logg$) are plotted as a function of
$\teff$.  In each  subplot,  differences are  computed in  the  sense $(q_r  -
q_c)/q_c$, where  $q$ is any seismic  quantity (except $\logg$, for  which the
denominator is obviously omitted) and the subindices $r$ and $c$ represent the
GBM identifying  the row or column  respectively, e.g. the top  leftmost panel
shows differences in the sense \texttt{YB/B}$-$\texttt{SPB/B}. The list of all
sets of GBM results  is given in Table~\ref{tab:gbms} in the  main body of the
article. 

GBM comparisons  are shown in  three groups  separated by diagonal  lines. The
first and largest group compares results  of the nine GBMs that use $\dnuscl$.
In the second  group, the middle diagonal band, each  subplot compares results
of a given GBM against itself, in  one case with $\dnuscl$ and with $\dnuo$ in
the other. Finally, the third group in the right upper corner compares results
from GBMs using  $\dnuo$ against each other.  In all cases, red  lines are the
median formal uncertainty returned by  the pipeline identifying the row, taken
over 60~K temperature  bins. Below, we discuss general  characteristics of the
comparisons. Some relevant  aspects have been commented upon in  the main body
of this work (Sect.\,\ref{sec:systgbm}).  

The general  and most important lesson  from these comparisons is  that in the
majority  of cases  the  differences in  the central  values  returned by  the
different GBMs  are in agreement  within the formal uncertainties  returned by
the pipelines, even for the  modest 67~K $\teff$ uncertainty characteristic of
the SDSS $\teff$ scale. This justifies the choice made in the determination of
systematic uncertainties, in  particular that central values  returned by GBMs
that are  away by more  than 3 times the  standard deviations from  the median
value are  considered outliers and removed  from the sample used  to determine
the  final systematic  uncertainty (Sect.\,\ref{sec:errors}).  The most  clear
examples of these outliers can be  seen, for example, in comparisons involving
\texttt{RA/AS} or  \texttt{MPS/B}. But plots also  show that out of  the total
number of stars and stellar parameters,  the fractional number of results that
are rejected is small.

Focusing on  the first group  of comparisons, there  are specific sets  of GBM
results  that  help to  understand  the  impact  of  using different  sets  of
evolutionary  tracks/isochrones with  the same  statistical inference  tool or
vice   versa.   Our   discussion   here   extends   the   one   presented   by
C14. \texttt{SFP/B}, \texttt{YB/B}, and \texttt{MPS/B} rely on the same set of
isochrones,  so the  differences  among  them should  be  related (mostly)  to
differences in  the statistical  methodology. In  fact, the  \texttt{SFP/B} vs
\texttt{YB/B} typically shows very good  agreement for all seismic quantities,
with some of the smallest dispersions.  On the other hand, when \texttt{MPS/B}
is compared against them, the  dispersion is somewhat larger, particularly for
$M$  and $\tau$.  Similarly, \texttt{BAS/G}  and \texttt{BeS/G}  results agree
well. It has to  be noticed in this case that,  although the same evolutionary
code has been  used, the sets of models have  been computed independently with
somewhat  different  assumptions  (see  Table~\ref{tab:grids2}).  Results  for
$\mrho$ in particular show that the presence of systematic differences between
GBM  results  is not  uncommon,  because  in  many cases  differences  between
pipelines show  a preferential sign difference  with most results for  a given
comparison been  either positive  or negative. The  effect is  typically small
compared to statistical  uncertainties, and it is beyond our  scope here to go
into the  reasons why  it occurs.  We remark, however,  that our  procedure to
estimate systematic uncertainties  accounts for it. For  other quantities this
systematic   difference  between   GBMs   is  smeared   out  because   $\teff$
uncertainties play a more dominant role and scatter around the results. 

The  comparison  of  \texttt{GOE/C}  vs \texttt{GOE/M}  and  \texttt{YB/B}  vs
\texttt{YB/YR} are a measure of the  impact of using different sets of stellar
models  with  the same  statistical  inference.  In fact,  \texttt{GOE/C}  and
\texttt{GOE/M}  compare very  well with  each other  for all  quantities, with
typically the smallest dispersion even for stellar ages. For \texttt{YB/B} and
\texttt{YB/YR}  the comparison  does not  yield any  remarkable results,  with
dispersions that are comparable to most other GBM comparisons. The latter case
has been discussed more in extent in Sect.\,\ref{sec:systgbm}.  

Overall, it  does seem  that using  different sets  of evolutionary  tracks or
different  statistical  inference  tools  yields similar  dispersions  on  the
results. In both cases, as stated  above, differences are typically within the
median  formal uncertainties,  which  is  a good  sanity  check of  systematic
uncertainties.  

Results in the second group directly  reflect differences induced by using the
more physically  accurate $\dnuo$ instead  of $\dnuscl$,  as this is  the only
difference in the  results within each subplot. Differences are  shown here in
the  sense $G(\dnuo)  - G(\dnuscl)$,  where G  denotes any  of the  three GBMs
included here.  The effect of using  $\dnuo$ is more clearly  seen in $\mrho$,
where as expected the trend reflects that shown in Fig.\,\ref{fig:dnucorr} but
with the  opposite sign  and about  twice as  large in  amplitude. There  is a
systematic offset between  results based on \texttt{BeS/G} and  those based on
\texttt{GOE/C} and \texttt{BAS/G}.  The reason is that  in \texttt{BeSPP}, the
grids  of models  are always  rescaled such  that a  solar model  in the  grid
reproduces    $\dnusun$    and    $\numaxsun$,   even    when    $\dnuo$    is
used. \texttt{GOE/C-$\dnuo$} and \texttt{BAS/G-$\dnuo$}, on the other hand, do
not include this  additional scaling, leading to the  larger systematic offset
seen at  the solar $\teff$. Aside  from this scaling, the  overall changes are
the same  for the  three GBMs,  which is  reassuring that  differences between
$\dnuo$ and $\dnuscl$  in stellar models are quite independent  of the stellar
models.  The impact  of  using $\dnuo$  or $\dnuscl$  also  affects all  other
quantities, but in comparison to the formal uncertainties, it has a decreasing
impact from $R$ to $\logg$, $M$ and $\tau$.  

The    final   set    of    comparisons   includes    the   combinations    of
\texttt{GOE/C}-$\dnuo$,               \texttt{BAS/G}-$\dnuo$,              and
\texttt{BeS/G}-$\dnuo$. The dispersion  between the first two  is very similar
to that between \texttt{GOE/C} and  \texttt{BASTA/G} in the first group (based
on  $\dnuscl$). Comparisons  with \texttt{BeS/G}-$\dnuo$  show however  larger
dispersions.   This  is   due   to  the   additional   scaling  performed   in
\texttt{BeS/G}-$\dnuo$  and discussed  in  the previous  paragraph, that  puts
results from this  GBM aside. Recommended values in the  catalog, however, are
nevertheless  based   on  \texttt{BeS/G}-$\dnuo$  for  reasons   explained  in
Sect.\,\ref{sec:central}.

\end{document}